\documentclass[screen,review=false]{acmart}

\usepackage{marvosym}
\usepackage{braket}

\newcounter{nresearch}
\stepcounter{nresearch}

\usepackage{algorithm}
\usepackage[noend]{algpseudocode}

\usepackage{float}
\usepackage{lscape}

\usepackage{tabularx}
\usepackage{rotating} 
\usepackage{booktabs} 
\usepackage{makecell} 

\usepackage{changepage}

\usepackage{hyperref}
\usepackage{adjustbox}
\usepackage{tikz}
\usepackage{tabularx}
\usepackage{amsmath}
\usepackage{amsthm}
\usepackage{mathtools}
\usepackage{physics}
\usepackage{graphics}
\usepackage{graphicx} 
\usepackage{comment}

\usepackage{color}
\definecolor{brandeisblue}{rgb}{0.0, 0.44, 1.0}

\newtheorem{problem}{Problem}

\def\EXP{\mathit EXP}
\def\QMA{\textit{QMA}}
\def\QAOA{\textit{QAOA}}

\acmBooktitle{Quantum Complexity May Help Classical Complexity}

\begin{document}

\title{Quantum Computational Complexity vs Classical Complexity: A Statistical Comprehensive Analysis of Unsolved Problems and Identification of Key Challenges}

\author{Arash Vaezi}
\authornotemark[1]
\authornote{Corresponding Author.}
\email{avaezi@sharif.edu; avaezi@ipm.ir}
\orcid{0000-0003-4798-0029}

\author{Ali Movaghar}
\authornotemark[2]
\email{movaghar@sharif.edu}
\orcid{0000-0002-6803-6750}

\author{Mohammad Ghodsi}
\authornotemark[1]
\authornotemark[2]
\email{ghodsi@sharif.edu}
\orcid{0000-0002-1588-5337}

\author{Seyed Mohammad Hussein Kazemi}
\authornotemark[2]
\email{hussein.kazemi75@sharif.edu}

\author{Negin Bagheri Noghrehy}
\authornotemark[2]
\email{bageri.negin@physics.sharif.edu}

\author{Seyed Mohsen Kazemi}
\authornotemark[1]
\email{smohsen.kazemi84@sharif.edu}

\affiliation{
 \institution{Institute for Research in Fundamental Sciences (IPM)}
  \country{}
}

\affiliation{
  \institution{Sharif University of Technology}
  \country{}
}

\renewcommand{\shortauthors}{A. Vaezi, A. Movaghar, M. Ghodsi, N. Bagheri, SMH. Kazemi}

\begin{abstract}
Scientists have demonstrated that quantum computing has presented novel approaches to address computational challenges, each varying in complexity. Adapting problem-solving strategies is crucial to harness the full potential of quantum computing. Nonetheless, there are defined boundaries to the capabilities of quantum computing. This paper concentrates on aggregating prior research efforts dedicated to solving intricate classical computational problems through quantum computing. The objective is to systematically compile an exhaustive inventory of these solutions and categorize a collection of demanding open problems that await further exploration.
Through statistical analysis, we help the researchers with their further investigations. 
\end{abstract}


\maketitle
\twocolumn
\newpage
\setcounter{tocdepth}{3}
\tableofcontents
\newpage
\section{Introduction}
\label{sec:introduction}
Quantum computing and classical computing represent distinct fundamental approaches to conducting computations.

Classical Computing employs conventional binary units known as bits, which represent information as either 0 or 1. It processes instructions sequentially, executing tasks one by one. Grounded in the principles of classical physics, classical computing employs Boolean logic for operations. Data is stored in classical memory using these bits. However, as tasks become larger, certain problems experience exponential complexity, hindering efficient processing.

Quantum Computing utilizes unique units called qubits, which can be in states of 0, 1, or a combination of both due to quantum properties. It harnesses quantum phenomena, such as superposition and entanglement, to conduct multiple computations simultaneously, potentially leading to exponential acceleration for specific problems. Quantum computing is rooted in the principles of quantum mechanics, allowing qubits to exist in multiple states and interact through entanglement. Information is stored in qubits, but their susceptibility to quantum decoherence poses a challenge. Quantum computing has the potential to deliver exponential speedup for particular problems through parallelism and quantum interference, although this benefit is not universal. Presently, quantum computing is in its nascent stage, predominantly focusing on specialized tasks. Practical quantum computers are not as widespread as classical computers. Anticipations arise regarding the capacity of quantum computers to effectively tackle complex problems involving a multitude of variables and potential outcomes at a much swifter pace compared to classical computers.

Although classical computing is firmly established and appropriate for a wide range of tasks, quantum computing presents the opportunity for substantial acceleration in particular areas of problem-solving. The distinct characteristics of quantum computing, including superposition and entanglement, introduce fresh avenues for resolving intricate problems. However, obstacles related to stability, error correction, and the practical realization of quantum computing still require resolution.




In contrast to classical systems, where the influence of the environment on the system is minimal, quantum systems can rapidly and significantly become entangled with their environment, leading to disturbances. Complete isolation of a quantum system from its environment is nearly impossible, making interaction between the two unavoidable. The phenomenon of quantum decoherence poses a challenge in describing the dynamics of quantum systems. Essentially, decoherence characterizes the effects of entangling interactions on the future measurement statistics of the system~\cite{SCHLOSSHAUER20191}.

Quantum decoherence greatly hinders quantum information processing, such as quantum computation. Decoherence can impact the superposition states of quantum systems, leading to errors. To address this, qubit systems have been engineered to minimize the influence of environmental interactions on state preparation and longevity. Various strategies have been developed to mitigate the adverse effects of decoherence, ranging from preventing or minimizing errors (e.g., encoding information in a decoherence-free subspace) to correcting undesired changes (referred to as error correction)~\cite{SCHLOSSHAUER20191}.

The quantum error correction (QEC) process is vital for addressing the decoherence problem. However, classical error correction algorithms cannot be easily applied to quantum systems due to some obstacles. The no-cloning theorem, as described in~\cite{wootters2009no}, states that no machine can copy any given quantum state (note that some but not all states can be cloned). Additionally, measuring a quantum system disturbs the state and might destroy superposition. This means reading out information without destroying it is more complex in quantum systems than in classical ones. Quantum Noise has several continuous parameters, similar to some classical noises, but its probabilistic nature makes it harder to describe than the deterministic evolution of classical noise~\cite{lidar2013quantum}.

Over time, researchers developed many QEC codes to mitigate the detrimental effects of decoherence. One of the famous codes is Peter Schor's 9-qubit code, first introduced in 1995 in~\cite{1995PhRvA..52.2493S}, which encodes $1$ logical qubit in $9$ qubits, capable of correcting arbitrary errors in a single qubit. However, there exist other codes with a higher rate, such as those developed by Calderbank, Shor~\cite{PhysRevA.54.1098}, and Steane~\cite{1996} (referred to as CSS code) that were inspired by the classical Hamming code [4,7](paper~\cite{hamming1950errordetecting}. Furthermore, a more general class of code known as stabilizer codes was discovered by Gottesman~\cite{gottesman1997stabilizer}, and by Calderbank, Rains, Shor, and Sloane~\cite{calderbank1997quantum}.

The measurement of quantum systems differs fundamentally from that in classical physics. In quantum systems, observables are measured experimentally, not the vector states. Consequently, without knowing the exact vector state, we cannot predict the outcome of the measurement. The state's evolution during measurement is different from the evolution between measurements, contributing to the peculiar nature of this process. When an experimenter attempts to apply a specific observable operator on the system, the wave function collapses, meaning the state unpredictably jumps to one of the eigenstates of the observable operator~\cite{Susskind:2014qoa}.

To achieve scalable quantum computation despite noise-process variations, quantum fault-tolerant methods are essential. Although there are similarities in the basic concepts of fault tolerance in classical and quantum computation, such as creating gadgets that encode information and correct errors simultaneously, and protecting the system from noise, implementing fault-tolerant computation in quantum systems is significantly more challenging owing to the coherent implementation of superposition states and the management of noise that can alter both the bit and the phase of the state~\cite{lidar2013quantum}.

The distinction between quantum processors and classical ones lies in their utilization of quantum information instead of classical information. Quantum information possesses intrinsic features such as randomness, uncertainty, and entanglement, which are absent in classical systems. Quantum randomness alludes to an inherent randomness that does not stem from a lack of knowledge, unlike the everyday randomness we encounter. Uncertainty in the quantum system arises from the fact that certain observables cannot be measured simultaneously without interfering with each other, meaning that one measurement inevitably affects subsequent measurements. Furthermore, entanglement enables complete knowledge of the entire quantum system while imparting zero information about the individual subsystems~\cite{preskill2023quantum}.




Quantum computing allows achieving exponential speed-up compared to classical computers by harnessing superposition, where a quantum system can simultaneously exist in multiple states. This enables operations to be carried out in parallel on all states within the superposition. While measuring the result collapses the system to a single state, smart algorithms and initial state selections allow for efficient extraction of information. One can see a practical example in Deutsch's Algorithm (see~\cite{nielsen2010quantum}). In this algorithm, a 2-qubit unitary operator denoted as $U_f$, is created to compute a function $f(x)$ that maps inputs from $\{0, 1\}$ to $\{0, 1\}$. By applying $U_f$ to a superposition initial state, such as $\left| +, - \right> = \left| 0 \right> \pm \left| 1 \right> / \sqrt{2}$, and observing the resulting state evolution, one can effectively determine whether $f(0)$ equals $f(1)$ or not. The projection of the first qubit onto the $ \left| \pm \right>$ basis allows for extracting the desired information, showcasing the power of quantum superposition in computational tasks. This demonstration highlights the efficiency and effectiveness of quantum algorithms in processing information through superposition states, paving the way for exploring more complex quantum computational problems~\cite{chapman2022quantum}.

One may explain quantum computational complexity as the measure of the difficulty of a quantum computation task, quantified by the number of operations required to transform an initial state into a final state. Specific initial states are often necessary for efficient computation, as illustrated by the example where additional operations were needed to prepare the desired state $ \left| \psi 0 \right>$ starting from a canonical state $ \left| 00 \right>$. Additionally, transitioning between bases, e.g. from the computational basis to the $\left| \pm \right>$ basis, necessitates additional operators. This formalization of quantum computation consists of a sequence of operations on qubits, with the complexity of the task reflected by the number of operations. The concept of computational complexity closely connects to resource requirements for problem-solving to find the fastest algorithm. In the domain of quantum mechanics, the time required for computation is bound by physical limits, as indicated by the Margolus and Levitin~\cite{margolus1998maximum} bounds and the Aharonov et al.~\cite{anandan1990geometry,aharonov1961time,mandelstam1991uncertainty} bounds, which establish the minimum time needed to evolve a state into an orthogonal state based on the energy expectation value. These bounds provide insights into the fundamental limits imposed by quantum mechanics on the time required for quantum operations, emphasizing the intricate relationship between quantum computational complexity and physical implementation constraints~\cite{chapman2022quantum}.

Different concepts of complexity in quantum computing involve considering the number of qubits used, analogous to storage in classical complexity, and the Kolmogorov complexity~\cite{berthiaume2001quantum}, which measures the information content of a string and its compressibility without losing information. Quantum computational complexity can be centred on the number of operations, particularly significant in holographic duality and black hole interiors, offering a unique perspective distinct from traditional quantum computing. As the comprehension of the relationship between geometry and information advances, other complexity notions like Kolmogorov complexity in holography may also become pertinent~\cite{chapman2022quantum}.


Numerous prior studies have focused on addressing challenging problems in classical computing by quantum approaches. Several others aimed to study such results, as Dalzell et al.~\cite{dalzell2310quantum} did. They extensively discussed the areas of applications of quantum solutions. Further, they thoroughly provided details about quantum algorithmic primitives, namely the building blocks to achieve quantum speedups. For each algorithmic primitive, they sketched its basic idea, example use cases and important caveats. They further evaluate potential quantum speedups by comparing quantum solutions to state-of-the-art classical methods and complexity-theoretic limitations. ZHU et al.~\cite{zhu2022brief} provided a brief survey of quantum architecture search methods that address the challenges of designing quantum circuit architectures for practical tasks such as combinatorial optimization. These methods are inspired by techniques from classical neural network architecture search and machine learning, such as reinforcement learning. The methods may also apply to other fields, such as multi-objective optimization. Valerio et al.~\cite{massoli2022leap} pointed out, in their survey work, that quantum neural networks and algorithms are still far from proving decisive supremacy over classical ones, yet it is an ongoing quest. Quantum technologies may later exploit phenomena such as superposition and entanglement to obtain improvements whatsoever. The authors discussed the latest achievements and state-of-the-art approaches to Quantum Perceptrons and Quantum Neural Networks. It is further elaborated that the Quantum machine learning algorithms, such as Quantum Neural Networks, accept quantum states as input, requiring the translation of classical data into quantum states through \emph{state preparation}.

\subsection{Overview of this Paper}

In short, we have compiled a collection of articles to identify unresolved and significant problems worthy of attention. For numerous well-known problems in computer science, we have demonstrated a clear contrast between the quantum solutions' runtimes and the classical ones. We explored the potential of quantum resources in achieving exponential speedup in computational tasks and the challenges of error correction and scalability in quantum systems. Moreover, we revisited hybrid quantum-classical algorithms that leverage the speed and efficiency of quantum processors while tackling the constraints associated with quantum systems. This emphasizes the importance of interdisciplinary collaboration in pioneering innovative solutions for classical computational challenges. We then clustered the directions possible for future investigations into several categories.

In section~\ref{sec:stat_analysis}, as a descriptive analysis, we illustrated how bringing several classical problems to the quantum side showed theoretical and/or practical speedups in the literature. Further, we provided illustrations about where several well-known quantum computation techniques, e.g. Quantum Walk~\cite{apers2019unified, chailloux2021lattice, laarhoven2016search, marsh2019quantum, belovs2012learning,bernstein2013quantum} and Quantum Search~\cite{unsal2023faster, zhang2020procedure, srinivasan2018efficient, gheorghica2021gr3, miyamoto2020quantum, bonnetain2020improved}, are applied. We specifically referenced results that applied these techniques to solve well-known classical problems such as the Hamiltonian Cycle Problem, the 3-SAT Problem, and others. Furthermore, we conducted an inferential analysis to illuminate more nuanced details of the current results and explore potential future directions to address existing open problems.

In section~\ref{sec:previousworks}, we reviewed the existing findings on models of quantum complexity growth. We also revisited current results in the literature that illustrated the relationship between quantum randomness and classical complexity, showing how quantum randomness influences problem-solving. Furthermore, we delved into the approximation techniques in the literature for quantum problems, demonstrating how quantum algorithms can offer approximate solutions for complex computational tasks. Additionally, we covered quantum query complexity, that is, the efficiency of quantum algorithms in terms of query operations. We also revisited the concept of quantum random walk, highlighting its fundamental role in quantum algorithms and solving classical computational problems.

In Section~\ref{sec:quantumSolToClassProb}, we reviewed various well-known computational problems that can be efficiently solved using quantum computing techniques. These include problems such as the Satisfiability Problem, Maximum Independent Set, Graph Traversal Problems, and Maximum Matching on Graphs. We revisited the existing results that showed how quantum algorithms can improve computational efficiency by leveraging quantum principles like superposition and entanglement. This further opens up new possibilities for effectively addressing challenging computational tasks.

In Section~\ref{sec:openproblems}, we provided several open problems and potential areas for further exploration in the field of quantum computing. We covered various critical topics, including Quantum Approximate Optimization Algorithm (QAOA), Quantum Annealing, Quantum Search, Quantum Walk, and Mean Estimator. We also revisited Quantum String Matching Algorithms, Quantum Merging Trees, Quantum Congested Clique, Verification and Certification, Machine Learning, Complexity, Classification, Security, and related subjects. By highlighting these open problems, we aimed to assist future research endeavours toward enhancing quantum computing capabilities and addressing challenging computational problems.


In section~\ref{sec:summary}, the contributions of the earlier studies, including their essential methodologies and findings, were summarized. As will be further discussed in this paper, prior studies have explored the application of quantum algorithms to a wide variety of combinatorial optimization problems. These include domains such as graph theory~\cite{de2022quantum, mariella2023quantum, siraichi2019qubit, li2020qubit, bravyi2022hybrid, bochniak2023quantum, salehi2022unconstrained, Ge_2020, cocchi2021graph, kim2022rydberg, pelofske2019solving, kerger2023asymptotically, pelofske2019solving, ramanujan2021approximate, kimmel2021query, witter2022query, mcleod2022benchmarking, vert2020revisiting, bravyi2022hybrid, bravyi2020obstacles}, as well as chemistry simulations~\cite{kassal2011simulating}.

\section{Statistical Analysis}
\label{sec:stat_analysis}

In this section, we provide a \emph{descriptive analysis} of the quantum versus classical complexity of well-known computer science problems. We also demonstrate where different quantum approaches and solvers have been applied so far and could contribute to exploring future research directions. Later in the section, we draw remarks as an \emph{inferential analysis} to help the researchers with their further investigations. 

\subsection{Descriptive Analysis}

In this subsection, we provide a comparison, based on the existing results, between the complexity of quantum and classical approaches for well-known problems in computer science. We also demonstrate how various quantum approaches and solvers have been used for several problems, and refer to their potential future applications.\\

\begin{table*}[htp]
\centering
\adjustbox{max width=\textwidth}{
\begin{tabular}{|c|c|c|c|c|c|} 
 \hline
 \textbf{Problem} & \textbf{Classic C.} & \textbf{Quantum C.} & \textbf{Improvement} & \textbf{Problem Number(s)} \\ [0.5ex] 
 \hline \hline
 k-distinctness & $\Theta(n \log n)$ & $\tilde{\mathcal{O}}(n^{\frac{3}{4} - \frac{1}{4}\frac{1}{2^k - 1}})$ & Polynomial & -\\ 
 (Element Distinctness)& (In decision tree model~\cite{10.1145/800061.808735}) & ~\cite{jeffery2022multidimensional} &  & \\ \hline
 k-SAT ($k \geq 3$) & $\mathcal{O}(2^{1-\frac{\Theta(1)}{k}})$~\cite{10.1145/3313276.3316359} & $\mathcal{O}((2(1 - 1/k))^n)$~\cite{schoning1999probabilistic}& - & \ref{ab-QAOA1}, \ref{p:Scalability:1}, \ref{p:QAlgorithm:3}, \ref{p:QAlgorithm:4}, \ref{p:QAlgorithm:5}, \ref{p:QComplexity:6} \\
 &  & &  & \ref{p:Simulation:1}, \ref{p:Simulation:2}, \ref{SATtoIsing}, \ref{p:Scalability:6}, \ref{bian}\\ \hline
 Maximum Independent Set & $\mathcal{O}(1.1996^n \cdot n^{\mathcal{O}(1)})$~\cite{XIAO2017126} & $\mathcal{O}(1.1488^n)$~\cite{etde_21241168} & Polynomial & \ref{p:QAlgorithm:10}, \ref{prob:problemname2}, \ref{p:QSupremacy:3} \\ \hline
 Hamiltonian Cycle & $\mathcal{O}^{*}(1.657^n)$~\cite{bjorklund2014determinant} & $\mathcal{O}(n^{2n/(n+1)})$~\cite{dorn2007quantum} & Exponential & \ref{p:QSupremacy:0},  \ref{cocchi1}, \ref{cocchi2}, \ref{p:Hybrid:1}\\ \hline
 Eulerian Tour & $\mathcal{O}(|E|)$~\cite{fleischner1990eulerian} & $\mathcal{O}(\sqrt{n})$~\cite{dorn2007quantum} & Polynomial & -\\ \hline
 Travelling Salesman Problem & $\mathcal{O}(n^2 \cdot 2^n)$~\cite{doi:10.1137/0110015} & $\mathcal{O}({1.728}^n \cdot \text{poly}(n))$~\cite{doi:10.1137/1.9781611975482.107} & - &  \ref{p:Hybrid:1} \\ \hline
 Maximum Matching & $\mathcal{O}(|E| \sqrt{|V|})$~\cite{micali1980v} & $n^{3/2} \log^{2}  n$~\cite{beigi2022time} & Polynomial & \ref{p:QComplexity:7}, \ref{p:QSecurity:6}, \ref{p:QComplexity:8}, \ref{p:QComplexity:81}, \ref{p:QAlgorithm:11} \\ \hline
 Maximum Cut & $\text{exp}(n)$~\cite{garey1997computers} & $\sqrt{2^n / r}$~\cite{chang2023quantum} & - &  \ref{p:Scalability:3}, \ref{p:Error:1} \\
  &  & s.t. $r$ is num of max-cuts &  &  \\ \hline
 Graph Isomorphism & $2^{\mathcal{O}(\log n)^c}$~\cite{grohe2020graph} & NA~\cite{alagic2007quantum} & NA & \ref{mariella}, \ref{de2022}, \ref{li2022}, \ref{li2022-2} \\ \hline
 Maximum Clique & $2^{n/4}$~\cite{robson2001finding} & $\mathcal{O}(\log_2 n \cdot \sqrt{2^{n}})$~\cite{haverly2021comparison} & - &  \ref{prob:problemname7}, \ref{prob:problemname8} \\ \hline
 String Edit Distance & $\mathcal{O}(s^2 + n)$~\cite{landau1998incremental} & $\mathcal{O}(n^2 / \log n)$~\cite{equi2021quantum} & Poly-logarithmic & \ref{strEditDist1}, \ref{strEditDist2}, \ref{strEditDist3}\\ 
   & s.t. $s$ is a max. edit distance &  &  & \\ \hline 
 Longest Common Subsequence & $\mathcal{O}(n \log^2 n)$~\cite{bhowmick2019approach} & $\tilde{\mathcal{O}}(\sqrt{n})$~\cite{le2023quantum} & Polynomial & \ref{p:QComplexity:1} \\ \hline
 Protein Folding & $\omega(n^c), \ \forall c > 0$ & $\mathcal{O}(2^{n/2})$~\cite{wong2022fast} & - & \ref{pf1}, \ref{pf2}, \ref{p:Simulattion:1}\\ \hline
 Mixed-Integer Programming & $(\log n)^{^{\mathcal{O}(n)}} \cdot \text{poly}(n)$~\cite{reis2023subspace} & $\exp(n)$ & - & \ref{MIP1}\\ \hline
 Knapsack Problem & $\mathcal{O}(2^{n/2})$~\cite{doi:10.1137/0210033} & $\mathcal{O}(2^{n/2})$~\cite{fujimura2010quantum} & - & \ref{p:Hybrid:2}, \ref{p:QAlgorithm:7}, \ref{subsec: machine learning}, \ref{lai2020} \\ \hline
 Subset-Sum & $\tilde{\mathcal{O}}(2^{0.291n})$~\cite{becker2011improved} & $\tilde{\mathcal{O}}(2^{0.236n})$~\cite{bonnetain2020improved} & - & \ref{p:Error:4},  \ref{p:QSecurity:1}, \ref{p:QSecurity:2}, \ref{naya} \\ \hline
 Discrete Logarithm & $\omega(n^c), \ \forall c > 0$ & $\mathcal{O}(\log |\mathcal{G}|)$~\cite{hhan2023quantum} & Superpolynomial & - \\
 &  &  for a cyclic group $\mathcal{G}$ &  & \\
 &  & of prime order &  & \\ \hline
 Integer Factorization & $\exp(1.9(\log n)^{\frac{1}{3}}(\log \log n)^{\frac{2}{3}})$~\cite{buhler1993factoring} & ${\mathcal{O}} (b^3)$~\cite{shor1999polynomial} & Superpolynomial & - \\ 
 & $n$ is the number to factor & s.t. $b$ is num of bits of $n$ &  & \\ \hline
 Crypto. Certified  Random Bits & - & $\exp(n)$~\cite{aaronson2023certified} & -  & \ref{p:QSupremacy:2}, \ref{p:QSecurity:4} \\ [1ex]
 \hline
\end{tabular}
}
\caption{ Quantum vs. Classical complexity comparison for some known problems in computer science.}
\label{table:1}
\end{table*}

\begin{table*}[htp]
\centering
\adjustbox{max width=\textwidth}{
\begin{tabular}{|c|c|c|} 
 \hline
 \textbf{Approach/Solver}  & \textbf{Areas Used} & \textbf{Problem Number(s)} \\ [0.5ex] 
 \hline \hline
 &  Tail Assignment Problem~\cite{vikstaal2020applying}, &  \\  
 &  Approximate Solutions for Problems in NPO PB~\cite{marsh2019quantum}, & \\
 QAOA  & 3-SAT and Max-3-SAT~\cite{yu2023solution}, & \ref{ab-QAOA1},\ref{p:Scalability:1},\ref{RQAOA} \ref{RQAOA1},\ref{p:Scalability:2},\ref{p:QAlgorithm:1},\ref{QAOAMaxCut},\ref{p:Scalability:3} \\ 
 &   Combinatorial Search Problems on Graphs~\cite{farhi2020quantum,farhi2020quantum2}, & \\ 
 &   Max-Cut~\cite{boulebnane2021predicting,basso2021quantum,crooks2018performance,guerreschi2019qaoa}, & \\ 
 &  Maximum k-Vertex Cover~\cite{cook2020quantum,bravyi2022hybrid} & \\ \hline
 &  Theory of Cuts~\cite{cruz2019qubo}, &  \\  
 &  3-SAT Problems~\cite{10071022,nusslein2023solving}, & \\
 &  UnitDisk Maximum Independent Set Problem~\cite{Serret_2020}, & \\ 
 &  Maximum Independent Set Problems~\cite{8477865}, & \\ 
 &  Graph Partitioning into Subgraphs Containing Hamiltonian & \\
 Quantum Annealing &  Cycles of Constrained Length~\cite{cocchi2021graph}, & \ref{p:QSoftware:1},\ref{cocchi1},\ref{cocchi2}, \ref{p:Error:2},\ref{MIP1},\ref{p:Hybrid:1} \\ 
 &  TSP~\cite{salehi2022unconstrained}, & \\
 &  Maximum Cardinality Matching Problems~\cite{mcleod2022benchmarking}, & \\
 &  Minimum Vertex Cover~\cite{pelofske2019solving}, & \\
 &  Set Cover in Graphs~\cite{hamilton2017identifying}, & \\
 &  Graph Minor-Embedding~\cite{bernal2020integer}, & \\ 
 &  Multi-Knapsack Problems~\cite{awasthi2023quantum} & \\ \hline
 & Concentration Assignments~\cite{unsal2023faster},  & \\
 & 3-SAT Problem~\cite{zhang2020procedure}, & \\
 Quantum Search & TSP~\cite{srinivasan2018efficient}, & \ref{p:QAlgorithm:3},\ref{p:QAlgorithm:4},\ref{p:QAlgorithm:5},\ref{p:QSupremacy:0},\ref{p:QSupremacy:1},\ref{decomposition},\ref{prob:problemname4}\\
 & Subgraph Isomorphism~\cite{gheorghica2021gr3}, & \\ 
 & Minimum Steiner Tree Problem~\cite{miyamoto2020quantum}, & \\
 & Subset-Sum~\cite{bonnetain2020improved} & \\ \hline
 
 & Lattice Sieving~\cite{apers2019unified},  &  \\ 
 & Shortest Vector Problem~\cite{chailloux2021lattice,laarhoven2016search}, & \\
 Quantum Walk & Bounded NP optimization Problems~\cite{marsh2019quantum}, & \ref{quantumrandomwalk},\ref{p:QAlgorithm:6},\ref{Johnsongraph},\ref{quantumRandomWalk},\ref{p:Error:3},\ref{pf1}\\
 & $k\text{-distinctness}$~\cite{belovs2012learning}, & \\
 & Hamiltonian Cycle Problem~\cite{mahasinghe2019solving} & \\
 & Subset-Sum~\cite{bernstein2013quantum,bonnetain2020improved} & \\ \hline

 Mean Estimator & Approximating Partition Functions~\cite{cornelissen2023sublinear}  & \ref{problem::ME1},\ref{problem::ME3} \\ \hline
 Qubit Mapping & It is suggested it might bring enhancements in the future  & \ref{li2022},\ref{p:Simulation:1},\ref{p:Simulation:2},\ref{SATtoIsing},\ref{p:Scalability:6},\ref{bian},\ref{p:Error:4} \\
  & (see the related open problems) &  \\ \hline
 Quantum Particle Swarm Optimisation & Multidimensional Knapsack Problem~\cite{haddar2016hybrid,lai2020diversity}  & \ref{p:QAlgorithm:7},\ref{p:Hybrid:2}, \ref{quantumParticle},\ref{lai2020} \\ \hline
 Quantum Merging Trees & $k$-list Problem~\cite{naya2020optimal}  & \ref{p:QSecurity:1},\ref{p:QSecurity:2},\ref{naya} \\ \hline
 Quantum Machine Learning & Knapsack Problem~\cite{garcia2021knn} & \ref{knapsackML},\ref{li2022-2} \\
   & Qubit Mapping~\cite{li2020qubit} & \\
 \hline
\end{tabular}
}
\caption{Applicability of Quantum Approaches/Solvers.}
\label{table:2}
\end{table*}


Table \ref{table:1} compares time complexities for various problems in Quantum versus Classical computing. The table consists of two columns: "Classic C." and "Quantum C.", each provides information based on the latest level of our knowledge.

There is an additional column labelled ``Improvement''. This column aims to clarify whether a Quantum solution can be considered as an improvement for a given problem. If a quantum computer solves a problem in time $T$ and a classical computer requires a polynomial function of $T$ (such as $T^2$), it is termed a polynomial speedup. When the time complexity of the best classical algorithm, denoted as $C(n)$, and the time complexity of the quantum algorithm, denoted as $Q(n)$, satisfies $C = 2^{\Omega(Q^\lambda)}$ for some positive constant $\lambda$, the speedup is referred to as superpolynomial. Exponential speedup occurs when a quantum computer requires time $T$, while a classical computer needs an exponentially growing function of $T$ (for example, $2^T$). These categories represent the promising and desired speedups in quantum computers.

In the final column, the "Problem Number(s)" refers to the open problems associated with well-known classical problems revisited in this paper. Further, in Table \ref{table:1}, the notation $|V| = n$ represents the number of vertices, and $|E|$ represents the number of edges in a graph $G(V, E)$.\\

Table~\ref{table:2} shows the potential of quantum computing in addressing NP optimization problems efficiently. It offers a detailed list of classical computational problems (column named "Areas Used") along with the quantum approaches or solvers utilized to address them (column named "Approach/Solver"). 

The table further outlines potential future applications for these approaches and solvers (column named "Problem Number(s)"):

\textbf{QAOA.} Future applications of QAOA involve optimizing combinatorial search problems on graphs (Problem~\ref{ab-QAOA1}), finding approximate solutions for problems in NPO PB (Problem~\ref{p:Scalability:1}), efficiently solving 3-Satisfiability (Problem~\ref{RQAOA}) and Max-3-SAT problems (Problem~\ref{RQAOA}), addressing Combinatorial Search Problems on Graphs (Problem~\ref{RQAOA1}), optimizing the Max-Cut problem (Problem~\ref{p:Scalability:2}), seeking approximate solutions for the Maximum k-Vertex Cover problem (Problem~\ref{p:QAlgorithm:1}), utilizing QAOA in the Theory of Cuts for solution optimization (Problem~\ref{QAOAMaxCut}), and efficiently solving 3-SAT Problems (Problem~\ref{p:Scalability:3}).

\textbf{Quantum Annealing (QA).} Problem~\ref{p:QSoftware:1} suggests utilizing Quantum Annealing (QA) to potentially solve the Tail Assignment Problem efficiently. Problem~\ref{cocchi1} proposes the application of QA to offer approximate solutions for graph partitioning into Hamiltonian subgraphs, particularly on graphs with weighted edges or different constraints on cycle length. Problem~\ref{cocchi2} deals with 3-satisfiability and Max problems. Problem~\ref{p:Error:2} expands the potential use of QA to combinatorial graph search problems. Problem~\ref{MIP1} pertains to the Traveling Salesman Problem (TSP) and its variations. Lastly, Problem~\ref{p:Hybrid:1} centres on the potential advantages of incorporating QA in the context of many-body Rydberg atoms.

\textbf{Quantum Search.} Problem~\ref{p:QAlgorithm:3} discusses the potential to optimize the reduction and selection process in the SAT problem's search. Are there ways to identify all potential areas for a solution to larger problems? Further, can we adapt the techniques in~\cite{varmantchaonala2023quantum} to the XORSAT problem? 

Problem~\ref{p:QAlgorithm:5} discusses the potential advantages of modifying the weights assigned to constraints in evolutionary algorithms (EAs) so that the EA can prioritize certain constraints over others and get better performance. Additionally, the problem suggests integrating the random walk of local search techniques into the construction of GenSAT-based auxiliaries. 

Problem~\ref{p:QAlgorithm:4} discusses the potential for addressing a broader range of problems and extending the application of the quantum cooperative search algorithm to handle other combinatorial optimization challenges beyond its current scope. 

Problem~\ref{p:QSupremacy:0} discusses enhancements to the Eppstein algorithm that have been made to solve the Hamiltonian cycle problem, reducing the time complexity and prompting questions about its potential to improve the performance of current best classical algorithms. 

Problem~\ref{p:QSupremacy:1} pertains to the use of quantum techniques, specifically Grover search and quantum backtracking, to enhance Eppstein's algorithm. The goal is to integrate these quantum methods with classical algorithms to potentially achieve improved performance in solving computational problems.

Problem~\ref{decomposition} deals with subgraph extraction and pruning techniques in a decomposition algorithm for computing the Maximum Cliques of a graph. It suggests tailoring decomposition and simplification strategies to specific types of graphs to optimize algorithm performance and reveal insights into trade-offs between subgraph reduction and runtime efficiency. 

Lastly, Problem~\ref{prob:problemname4} discusses the tradeoff of using the Lovász number. While it reduces subproblems, it increases computational complexity. It also raises the question of finding a more efficient method to compute the Lovász number. Improving its calculation could enhance practicality and applicability in real-world scenarios, leading to more efficient problem-solving strategies. 

\textbf{Quantum Walk.} Problem~\ref{quantumrandomwalk} points out that Laarhoven's algorithm for quantum random walks aims to locate $k$ different marks by running the random walk $\mathcal{O}(k)$ times. However, there are concerns about optimizing the algorithm to improve its complexity and efficiency. Further research is needed to enhance the algorithm's performance and efficiency, leveraging quantum walk properties. 

In Problem~\ref{p:QAlgorithm:6}, quantum random walks are integrated with the Shortest Vector Problem (SVP) to introduce local sensitivity within the random walk graph, aiming to improve SVP performance. The integration of quantum random walks has the potential to advance quantum algorithms in lattice-based cryptography and solve complex computational problems. In problem~\ref{Johnsongraph}, the focus is on incorporating the local sensitivity property into the graph used for the random walk, as opposed to the usage of the Johnson graph. The goal is to investigate whether this modification can yield improved results. There is potential for incorporating the local sensitivity property into the graph structure for the random walk to examine the possible advantages of this alternative approach in addressing the computational problem. 

In Problem~\ref{quantumRandomWalk}, the absence of efficient generic algorithms for identifying multiple marked items prompts inquiry into whether an approach exists within the current framework. This underscores the intricacy of leveraging quantum computing for specific computational tasks and underscores the necessity for additional research to fine-tune quantum algorithms. 

Problem~\ref{p:Error:3} discusses a quantum algorithm~\cite{dalzell2023mind} that shows potential for a super-quadratic speedup compared to classical algorithms. However, it does not quite achieve this speedup. One might question whether the algorithm can be combined with classical techniques, effectively exploiting problem structure. 

Problem~\ref{pf1} deals with the issue of decoherence in quantum computing, which poses a scalability challenge. It proposes investigating alternative quantum computing models, such as topological quantum computing, one-way quantum computing, and quantum walks, to reduce the impact of decoherence and improve quantum computing capabilities. These models present different approaches to quantum information processing that may be less susceptible to decoherence than the conventional architectures.

\textbf{Mean Estimator.} Problem~\ref{problem::ME1} addresses mean estimation in quantum computing. The problem involves computing an estimate, denoted as $\tilde{\mu}$, for the mean $\mu$ of a random variable $X$. One might raise the question of investigating bounds on the diagonal entries of $\Sigma$ for $l_p\text{-norms}$ where $p<2$. This analysis aims to enhance accuracy and efficiency in estimating $\mu$ for informed decision-making in quantum processes. 

Problem~\ref{problem::ME3} aims to compare mean estimators in classical and quantum settings. The precision of the mean estimator in the classical scenario depends on the maximum value of the square root of the variance of each variable $X_j$, multiplied by the logarithm of the dimensionality factor $d$ divided by a small value $\delta$, all divided by the sample size $n$. However, in quantum settings, the precision of the mean estimator is calculated similarly, but the sum of the square root of the variances of each variable $X_j$ is taken instead of the maximum. Both estimators rely on the dimensionality factor $d$. This comparison suggests exploring an optimal approach that can combine the strengths of both classical and quantum methodologies across all settings.

\textbf{Qubit Mapping.} Problem~\ref{li2022} discusses the FIDLS algorithm for qubit mapping. The algorithm aims to minimize the number of additional two-qubit gates needed in the output circuit to increase efficiency. It is suggested that using a more advanced subgraph isomorphism algorithm, such as the approximate subgraph isomorphism algorithm~\cite{siraichi2019qubit}, could potentially enhance the quality of results in qubit mapping. 

In Problem~\ref{p:Simulation:1}, the challenge of directly generating penalties from Satisfiability Modulo Theories (SMT) for large Boolean functions is addressed. The problem prompts the exploration of methods to reconstruct the decomposition function without imposing restrictions, as well as considering the use of additional qubits to expand the scope of an existing penalty function and explore alternative SMT formulations. 

Problem~\ref{p:Simulation:2} explores using 2048 sparsely connected qubit architectures for solving SAT and MaxSAT problems in Quantum Annealing systems as they scale. The problem points out considering the potential for employing more connected topologies to enable larger QA hardware graphs with higher per-qubit connectivity and reduced separation between qubit clusters. New encoding strategies may be needed to address more complex Ising problems. 

Problem~\ref{SATtoIsing} examines the conversion of the Boolean Satisfiability Problem (SAT) into an Ising model using Satisfiability Modulo Theories (SMT). It suggests exploring the potential improvement of existing algorithms by considering the intricacies of embedding onto Quantum Annealing (QA) hardware graphs, encompassing placement and routing. 

Problem~\ref{p:Scalability:6} discusses encoding small Boolean functions using SMT. There is a need for faster and more scalable techniques, focusing on placing variables and computing penalty functions. It is suggested that existing methods involving variable placement are more scalable than those involving simultaneous placement and penalty function computation. 

Problem~\ref{bian} uses the Ising model to solve SAT and MaxSAT problems within the Satisfiability Modulo Theories (SMT) framework. It explores expressing Boolean formulas as quantified Boolean formulas (QBF) or through Shannon expansion. It raises the question of investigating new techniques for efficiently handling quantified formulas within quantum computing to potentially enhance computational performance in tackling complex SAT and MaxSAT problems. 

Problem~\ref{p:Error:4} refers to a quantum algorithm that offers a more efficient approach to solving the subset sum problem using fewer qubits for encoding. In classical computing, the subset sum problem is NP-complete. Still, in quantum computing, this algorithm demonstrates a substantial enhancement in computational efficiency through a quadratic speedup achieved by harnessing quantum mechanics principles. The problem also emphasizes the potential of reducing error rates in quantum devices to improve the quality and reliability of available qubits, involving strategies such as enhancing coherence times and minimizing crosstalk between qubits. 

\textbf{Quantum Particle Swarm Optimisation.} Problem~\ref{p:QAlgorithm:7} deals with the Hybrid KNN Quantum Cuckoo Search Algorithm (KQCSA) designed for the Knapsack problem. This innovative algorithm merges K-nearest neighbour concepts with quantum computing principles to improve solution efficiency. The KQCSA combines classical computational techniques like the cuckoo search algorithm with quantum computing strategies, exploring the potential for cross-pollination of ideas and techniques across different quantum algorithms. This exploration could potentially lead to the discovery of new synergies and optimizations that enhance the performance of quantum algorithms in solving complex computational problems.

Problem~\ref{p:Hybrid:2} refers to a hybrid heuristic approach that combines Quantum Particle Swarm Optimization (QPSO) with a local search method to solve the Multidimensional Knapsack Problem. This approach aims to efficiently explore the solution space and deliver improved solutions for NP-hard and combinatorial optimization problems by leveraging the strengths of both QPSO and local search. The collaboration between quantum and classical techniques in this hybrid approach demonstrates the potential for addressing challenging computational problems more effectively. 

Addressing the multidimensional knapsack problem, Problem~\ref{quantumParticle} refers to the Diversity-Preserving Quantum Particle Swarm Optimization (DQPSO) approach. This approach seeks to potentially enhance the conventional QPSO by integrating a diversity-preserving strategy to prevent premature convergence of the algorithm. The problem also raises the prospect of leveraging different pseudo-utility ratios in combination to possibly further enhance the performance of the repair operator. This inquiry suggests exploring the synergies between different utility ratios to optimize the repair process and improve the overall efficiency of the DQPSO algorithm. 

Problem~\ref{lai2020} examines using a diversity-preserving quantum particle swarm optimization algorithm to solve the classical 0-1 multidimensional knapsack problem with numerous constraints. However, there are concerns about the algorithm's variability in performance across multiple runs, especially in complex instances. Various approaches to enhance the algorithm's robustness can be considered to address this issue, such as examining the impact of different parameters and exploring adaptive parameter tuning or integrating diversity maintenance mechanisms. These efforts aim to improve the algorithm's resilience and reliability in solving complex instances of the 0-1 multidimensional knapsack problem.

\textbf{Quantum Merging Trees.} Problem~\ref{p:QSecurity:1} discusses a framework introduced in~\cite{naya2020optimal}, focusing on Merging Trees. By leveraging this framework, researchers can address K-list problems efficiently and effectively. The problem raises the question of cryptographic applications for quantum k-list algorithms, such as lattice algorithms or decoding random linear codes. K-list algorithms play a crucial role in cryptographic applications due to their ability to efficiently solve complex problems related to cryptography.

Problem~\ref{p:QSecurity:2} discusses the optimization strategies that have been meticulously developed and refined to achieve optimal results for all merging trees. These strategies have been implemented successfully, as indicated by~\cite{naya2020optimal}. However, there is a potential for further exploration by extending this framework to enhance quantum algorithms. Specifically, the issue of the complexity of the r-th encryption merge tree algorithms, surpassing the time limit of $2^{0.3 n}$, prompting the question of whether this bound can be exceeded. 

Problem~\ref{naya} refers to k-list problems that involve finding the k smallest or largest elements in a list. Merging trees are used to solve them efficiently. The problem mentions that despite various techniques being investigated in the context of quantum computing, none have yet demonstrated quantum advantages in this specific area. The lack of success in achieving quantum advantages in this context highlights a gap in quantum algorithm development for addressing classical computational challenges related to k-list problems and merging trees.


\textbf{Quantum Machine Learning.} In Problem~\ref{knapsackML}, a hybrid algorithm is proposed for approaching medium and large instances of the multidimensional knapsack problem. The algorithm in~\cite{garcia2021knn} suggests using dynamic parameters to enhance the optimization process, possibly leading to improved efficiency and effectiveness in addressing complex instances of the problem.

Problem~\ref{li2022-2} refers to a method proposed in reference~\cite{li2020qubit}. The setting is crucial during the searching step and facilitates the execution of two-qubit gates in the logical circuit by employing a beneficial SWAP combination. One might ask if there are machine learning and deep learning algorithms that can efficiently determine the optimal action during the search process, particularly when the search depth increases significantly. It refers to Equation (2) in~\cite{li2020qubit}, which defines a value function that needs to be maximized. 

\subsection{Inferential Analysis}

The k-distinctness (Element Distinctness) problem has received increased attention in recent years, as shown in Table~\ref{WorksOverTimeTable2},~\ref{WorksOverTimeTable3}. On average, there have been two results per year since 2021. However, we have not encountered any open problems mentioned in the works related to this problem. It would be beneficial if someone proves a lower bound on the time complexity of this problem on the quantum side, or suggests possible directions for further improvements. Possibly, by considering the Quantum Walk method as previously proved useful~\cite{belovs2012learning} (Also, see Table~\ref{table:2}). 


\begin{center}
\begin{table*}[!htp]
\centering
\begin{tabular}{|c|c|c|c|c|c|c|c|c|c|c|}
\hline
 & 1998 & 1999 & 2000 & 2002 & 2004 & 2006 & 2007 & 2008 & 2009 & 2010 \\ \hline
k-SAT(k > 2) & \cite{ohya1998np} & \cite{ohya1999quantum} &  & \cite{10.1023/A:1015644508998} &  &  & \cite{CHENG2007123}, \cite{leporati2007three} &  &  &  \\ \hline
Maximum Independent Set &  &  &  &  &  &  & \cite{doern2007quantum} &  &  &  \\ \hline
Hamiltonian Cycle &  &  &  &  &  &  & \cite{dorn2007quantum} & \cite{schott2008nilpotent,staples2008new} &  &  \\ \hline
Eulerian Tour &  &  &  &  &  &  & \cite{dorn2007quantum} &  &  &  \\ \hline
Maximum Matching &  &  &  &  &  &  &  &  & \cite{dorn2009quantum} &  \\ \hline
Graph Coloring &  &  &  &  &  & \cite{cameron2006quantum} &  &  &  &  \\ \hline
Maximum Clique &  &  & \cite{childs2000finding} &  &  &  &  &  &  &  \\ \hline
Hitting-Set Problem &  &  &  &  &  &  &  &  &  & \cite{chang2010quantum} \\ \hline
DiscreteLogarithm &  & \cite{shor1999polynomial} & \cite{hales2000improved} &  & \cite{mosca2004exact} &  &  &  &  &  \\ \hline
\end{tabular}
\caption{Works on Famous Classical Problems from 1998 until 2010}
\label{WorksOverTimeTable1}
\end{table*}
\end{center}

\vspace*{-\baselineskip} 

\begin{center}
\begin{table*}[!htp]
\centering
\begin{tabular}{|c|c|c|c|c|c|c|c|c|}
\hline
 & 2011-2012 & 2013 & 2014 & 2015 & 2016 & 2017 & 2018 & 2019 \\ \hline
k-distinctness (Element Distinctness) & \cite{belovs2012learning} &  &  &  &  &  &  &  \\ \hline
k-SAT(k > 2) &  &  &  &  &  & \cite{benjamin2017measurement} &  &  \\ \hline
Maximum Independent Set &  &  &  &  &  &  & \cite{pichler2018quantum}, \cite{8477865} &  \\ \hline
Hamiltonian Cycle &  &  &  &  &  & \cite{staples2017hamiltonian} & \cite{srinivasan2018efficient} & \cite{mahasinghe2019solving} \\ \hline
Eulerian Tour &  &  &  &  &  &  &  &  \\ \hline
Travelling Salesman Problem &  & \cite{warren2013adapting} &  &  &  & \cite{moylett2017quantum} & \cite{srinivasan2018efficient} & \cite{kieu2019travelling} \\ \hline
Maximum Matching &  & \cite{fowler2013minimum} &  &  &  & \cite{gabow2017weighted} &  &  \\ \hline
Maximum Cut &  &  &  &  &  &  & \cite{crooks2018performance} & \cite{guerreschi2019qaoa} \\ \hline
Graph Isomorphism &  &  & \cite{Gaitan_2014} &  & \cite{dziemba2016adiabatic} &  &  &  \\ \hline
Minimum Vertex Cover &  &  & \cite{chang2014quantum} &  &  &  &  & \cite{pelofske2019solving} \\ \hline
Set Cover in Graphs & \cite{Johnson2011Quantum} &  & \cite{cai2014practical} &  & \cite{Humble_2016} & \cite{hamilton2017identifying} &  &  \\ \hline
Bisection Problem &  &  &  & \cite{younes2015bounded} &  &  &  &  \\ \hline
Maximum Clique &  &  &  &  &  &  &  & \cite{pelofske2019solving} \\ \hline
Protein Folding & \cite{kassal2011simulating} &  &  &  &  &  &  &  \\ \hline
DNA Motif Model Discovery &  &  &  & \cite{cao2015adiabatic} &  &  &  &  \\ \hline
Mixed-Integer Programming &  &  & \cite{cai2014practical} &  &  &  &  &  \\ \hline
Knapsack Problem &  &  &  &  & \cite{haddar2016hybrid} &  &  &  \\ \hline
Subset-Sum & \cite{10.1007/978-3-642-20465-4_21} & \cite{cryptoeprint:2013/199}, \cite{bernstein2013quantum} &  &  &  &  & \cite{helm2018subset}, \cite{grassi2018quantum} &  \\ \hline
Discrete Logarithm &  &  &  &  & \cite{ekeraa2016modifying} & \cite{roetteler2017quantum},~\cite{ekeraa2017quantum} &  & \cite{delaplace2019improved} \\ \hline
\end{tabular}
\caption{Works on Famous Classical Problems from 2011 until 2019}
\label{WorksOverTimeTable2}
\end{table*}
\end{center}

\vspace*{-\baselineskip} 

\begin{figure*}[htp]
    \centering
    \includegraphics[scale=0.3]{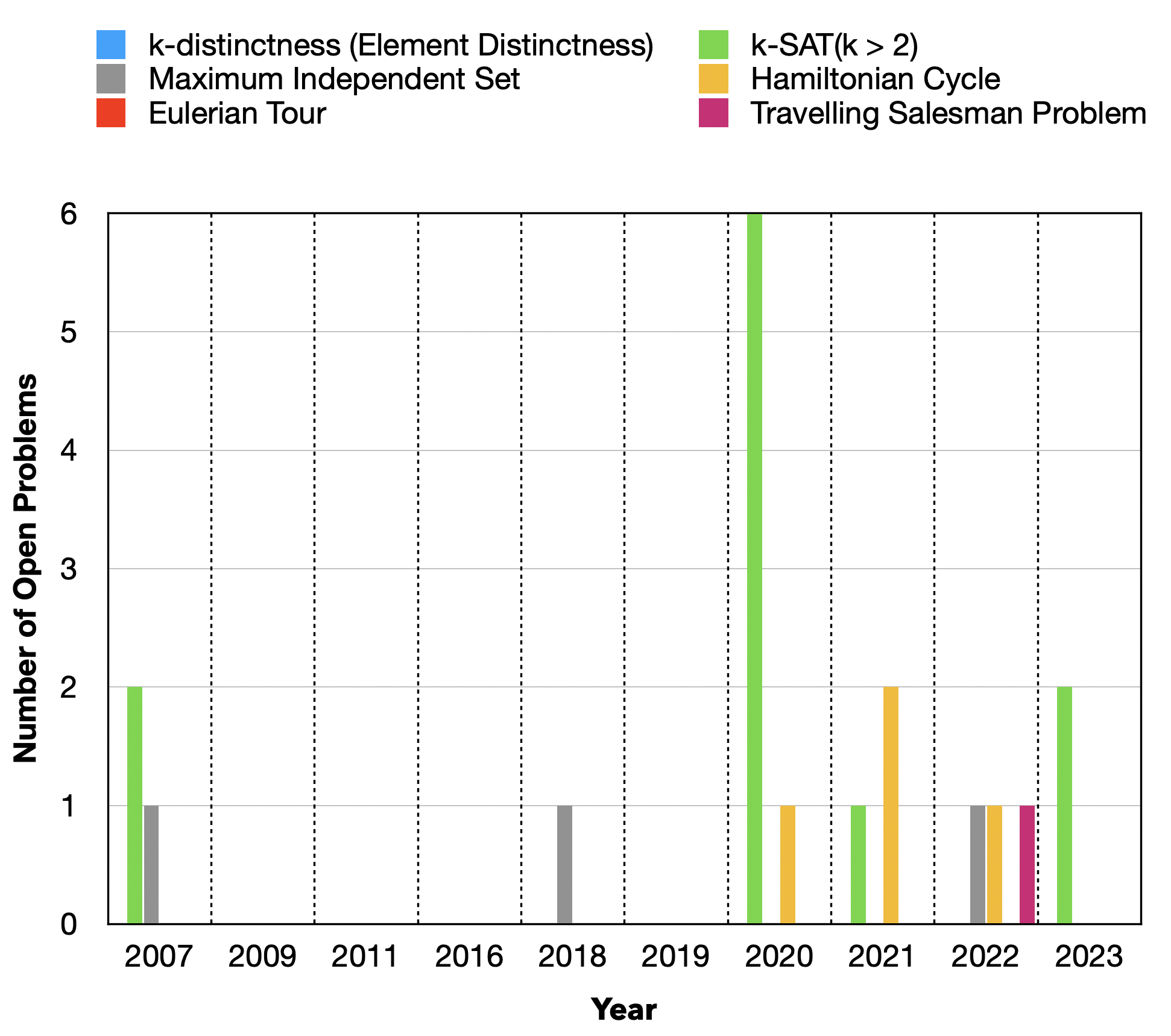} 
    \caption{Number of Open Problems Corresponding to Well-known Problems in Computer Science}
    \label{NumberOfOpenProblems}
\end{figure*}

The k-SAT problem, where $k$ is greater than $2$, has been the subject of extensive research. It attracted significant attention from 1998 to 2007, resulting in five publications. However, there was a ten-year gap from 2007 to 2017 with no new developments. Since 2020, there has been a renewed interest in k-SAT, particularly in 2023 with four new works related to the problem, as outlined in Table~\ref{WorksOverTimeTable1},~\ref{WorksOverTimeTable2},~\ref{WorksOverTimeTable3}. Interestingly, numerous problems regarding k-SAT have accumulated over time, as depicted in Figure~\ref{NumberOfOpenProblems}. Two of which emerged after 2007, noted in~\cite{CHENG2007123}, and surprisingly, six more were added in 2020 despite only two works being published related to k-SAT. Even though there was more effort in 2023 with four works about k-SAT, only two new open problems were introduced to the community. In total, it seems that this problem is a hot topic and there is a good chance for the researchers to obtain more results around it. As a possibility, one may investigate putting QAOA, Quantum Annealing, and Quantum Search in use to get further results. As shown in Table~\ref{table:2}, the above methods were previously utilized for 3-SAT, in particular. 


The Maximum Independent Set problem is just as compelling as the k-SAT problem, and there have been a similar number of results for both over time. However, we observed that results in the literature about the Maximum Independent Set problem only emerged in 2007, which is later than those found for k-SAT dating back to 1998. It is worth noting that the focus on the Maximum Independent Set problem has increased from 2018 to 2023, with an average of two publications each year. Interestingly, the number of open problems accumulated during the years is only three: one in 2007, another in 2018 (not directly related to the Maximum Independent Set problem), and the last one in 2022. It seems that further investigations are needed for this problem to identify new directions for future research or to examine to what extent quantum computation can help improve the known results for the Maximum Independent Set problem. Is there a quantum lower bound, or have we not yet found a way to significantly improve the known time complexity of the Maximum Independent Set problem (See Table~\ref{table:1} where we mentioned the known classical and quantum time complexity for the Maximum Independent Set problem)? Perhaps checking the possibility of using Quantum Annealing is the first step, as it was used before~\cite{8477865} (See Table~\ref{table:2}).

The Hamiltonian Cycle problem has been widely discussed in the literature, with eleven related publications. Similar to the Maximum Independent Set problem, the earliest results for the Hamiltonian Cycle problem also date back to 2007. Notably, there is a gap in research results from 2009 to 2017, and no publications are noted for 2023 (refer to Table~\ref{WorksOverTimeTable1},~\ref{WorksOverTimeTable2},~\ref{WorksOverTimeTable3}). The problem garnered more attention between 2017 and 2022, with an average of more than one publication per year. During this period, the community was presented with four open problems related to the  Hamiltonian Cycle problem. Additional research is necessary to address new avenues for future studies and explore how quantum computation could enhance the established outcomes for the Hamiltonian Cycle problem. Notably, the time complexity for the Hamiltonian Cycle problem is polynomial on the quantum side, prompting whether we can prove optimal bounds. This is especially significant, given that attempting to solve the same problem classically would require exponential running time (See Table~\ref{table:1}).

There has been a lack of interest in the Eulerian Tour problem within the community, given that only one publication from 2007 addresses related results. Furthermore, we have not come across any open problems regarding future directions. This lack of interest is due to the lower bound obtained in~\cite{dorn2007quantum}. As a result, further improvements do not seem particularly compelling.

There seems to be increasing attention given to the Travelling Salesman problem since 2013. The peak period spans from 2017 to 2022, with a publication every year except for 2020. Furthermore, there is a good chance to find future directions, as the only open problem we found was introduced in 2022. We recommend exploring new directions, particularly by investigating the potential of Quantum Annealing and Quantum Search. This is supported by earlier results mentioned in Table~\ref{table:2}, where both methods were utilized to achieve results for the Travelling Salesman problem.

Since 2009, quantum computation techniques have been applied to the Maximum Matching problem, with new results emerging every two or three years. In 2022, two publications appeared, following the results in 2021. Since 2009, three open problems related to the Maximum Matching problem have remained, and two more were added in 2021. This suggests that ongoing research in this area could benefit from a concerted effort to seek new results or explore new directions. Perhaps revisiting Quantum Annealing offers insight as it previously helped obtain some results. Or maybe it leads toward using another method?


\begin{center}
\begin{table*}[!htp]
\centering
\begin{tabular}{|c|c|c|c|c|}
\hline
 & 2020 & 2021 & 2022 & 2023 \\ \hline
k-distinctness (Element Distinctness) &  & \cite{chang2021quantum} & \cite{jeffery2022multidimensional},\cite{doi:10.1137/1.9781611977073.109},\cite{legall_et_al:LIPIcs.ITCS.2022.97} & \cite{jin2023quantum},\cite{jin2023quantum} \\ \hline
k-SAT(k > 2) & \cite{zhang2020procedure},    \cite{bian2020solving} &  &  & \cite{varmantchaonala2023quantum},  \cite{10071022}, \\ 
& ~ & ~ & ~ & \cite{nusslein2023solving}, \cite{yu2023solution} \\ \hline
Maximum Independent Set & \cite{Serret_2020}, \cite{farhi2020quantum}, \cite{farhi2020quantum2} & \cite{chang2021quantum}, \cite{saleem2021quantum} & \cite{ebadi2022quantum}, \cite{kim2022rydberg} & \cite{manyem2023maximum} \\ \hline
Hamiltonian Cycle & \cite{Ge_2020}, \cite{cook2020quantum} & \cite{cocchi2021graph} & \cite{jiang2022quantum}, \cite{cui2022solving} &  \\ \hline
Travelling Salesman Problem &  & \cite{paler2021nisq} & \cite{salehi2022unconstrained} &  \\ \hline
Maximum Matching &  & \cite{kimmel2021query} & \cite{mcleod2022benchmarking}, \cite{witter2022query} &  \\ \hline
Maximum Cut &  & \cite{boulebnane2021predicting}, \cite{basso2021quantum} & \cite{farhi2022quantum} &  \\ \hline
Graph Isomorphism & \cite{li2020qubit} & \cite{9559130}, \cite{gheorghica2021gr3} & \cite{de2022quantum} & \cite{mariella2023quantum}, \cite{mondada2023subgraph} \\ \hline
Minimum Vertex Cover & \cite{cook2020quantum} &  & \cite{zhang2022applying} & \cite{wang2023quantum} \\ \hline
Set Cover in Graphs &  &  &  &  \\ \hline
Graph Coloring &  &  & \cite{bravyi2022hybrid} & \cite{bochniak2023quantum} \\ \hline
Maximum Clique & \cite{miyamoto2020quantum} & \cite{ramanujan2021approximate} &  & \cite{kerger2023asymptotically} \\ \hline
Graph and String Edit Distance &  & \cite{boroujeni2021approximating} & \cite{Incudini_2022} &  \\ \hline
Longest Common Subsequence & \cite{9317938} &  & \cite{legall_et_al:LIPIcs.ITCS.2022.97}, \cite{ doi:10.1137/1.9781611977073.109}, \cite{10.1145/3519935.3520061} & \cite{jin2023quantum} \\ \hline
Mixed-Integer Programming & \cite{bernal2020integer}, \cite{ajagekar2022hybrid}, \cite{chang2020hybrid} &  &  &  \\ \hline
Knapsack Problem & \cite{lai2020diversity} & \cite{van2021quantum}, \cite{garcia2021knn} &  & \cite{awasthi2023quantum} \\ \hline
Subset-Sum & \cite{bonnetain2020improved}, \cite{helm2020power}, \cite{xu2020scalable}, \cite{naya2020optimal} &  & \cite{zheng2022quantum}, \cite{biesner2022solving} &  \\ \hline
DiscreteLogarithm & \cite{ShaiEvra2020},~\cite{ekeraa2020post}, &  &  & \cite{hhan2023quantum}, \cite{aaronson2023certified},  \\
& \cite{haner2020improved} & ~ & ~ & \cite{leverrier2023efficient}, \cite{leverrier2023decoding} \\ \hline

\end{tabular}
\caption{Works on Famous Classical Problems from 2020 until 2023}
\label{WorksOverTimeTable3}
\end{table*}
\end{center}

\vspace*{-\baselineskip} 

\begin{figure*}[htp]
    \centering
    \includegraphics[scale=0.25]{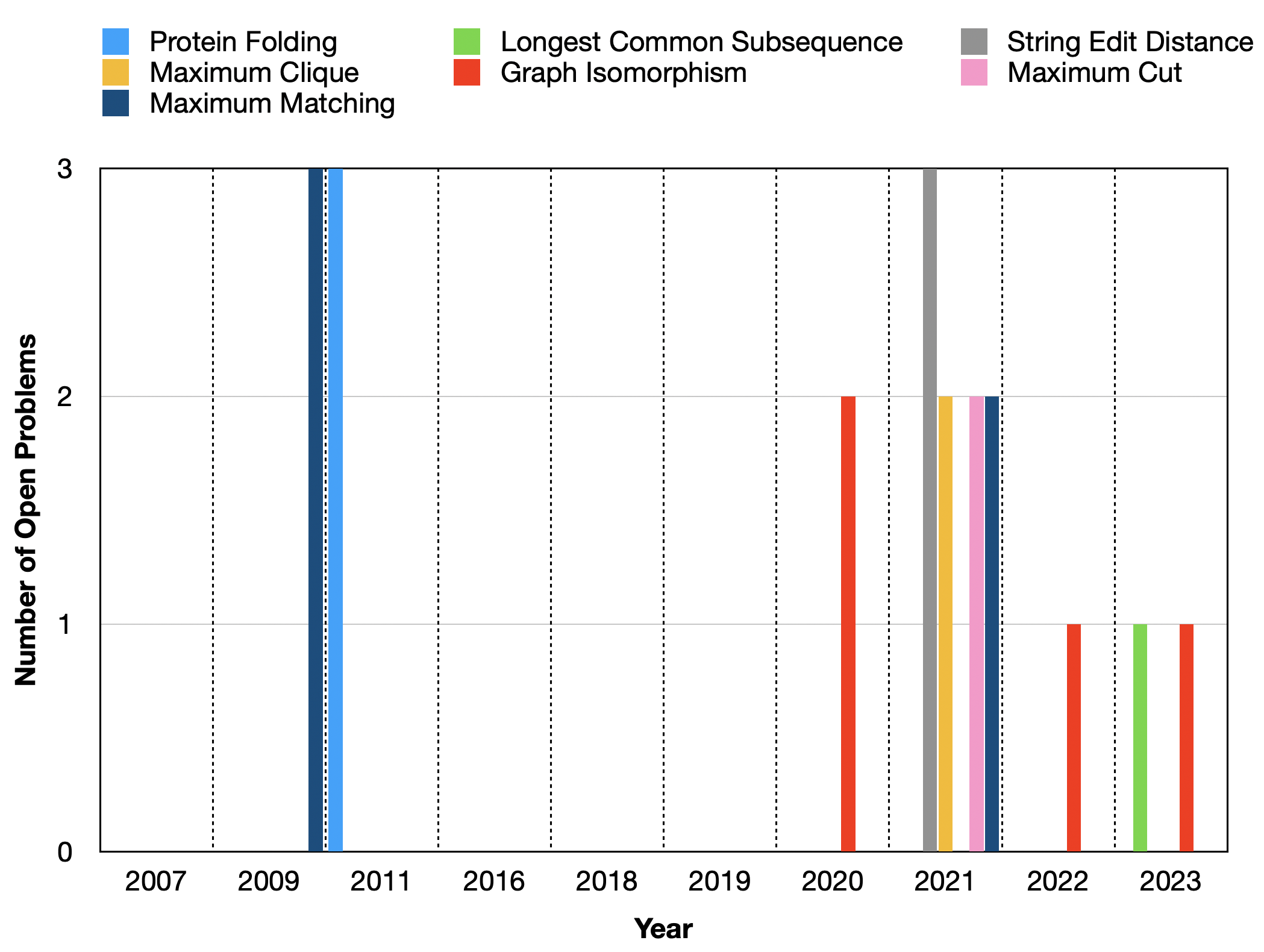}
    \caption{Number of Open Problems Corresponding to Well-known Problems in Computer Science}
    \label{NumberOfOpenProblems3}
\end{figure*}


The Maximum Cut problem has received attention from the research community since 2018 and has maintained its relevance, with an average of one publication per year related to it until 2022. Notably, only two open problems have emerged in the community since 2021. It is worth noticing that four out of five works on the Maximum Cut problem have utilized QAOA. Consequently, there is potential for further exploration of the possible applications of QAOA or the discovery of alternative methods to either open new lines of inquiry or address existing open problems.

The Graph Isomorphism problem has garnered attention from the research community since 2014. After a one-year hiatus, interest in the problem was reignited in 2016, leading to further research. 
Notably, there has been a three-year gap between publications of results. However, the problem has been under regular study since 2020, with an average of at least one paper published yearly. 
Additionally, two open problems were introduced in 2020 followed by another two in 2022 and 2023 according to Figure~\ref{NumberOfOpenProblems3}. 
It may be worthwhile to investigate whether the methods and approaches described in Table~\ref{table:2} could be applied to either open new directions or resolve the existing open problems, as no such attempts have been made thus far.

The Minimum Vertex Cover problem was introduced in 2014 in the field of quantum computing but resumed in 2019 after a gap of four years. After 2019, some studies have been done to solve the problem of maximum vertex coverage by quantum algorithms every year except 2021.
A quantum annealing approach has been used to solve the maximum vertex coverage problem in 2019.

Research on the Set Cover Problems in Graphs using quantum computers commenced in 2011 and persisted in 2014, 2016, and 2017 (See Table~\ref{WorksOverTimeTable2},~\ref{WorksOverTimeTable3}). However, there have been no subsequent results since 2017. An approach based on the Quantum Annealing method was introduced in 2017 for the Set Cover Problems in Graphs, as shown in Table~\ref{table:2}. Therefore, further exploration of Quantum Annealing may yield additional results or prompt the consideration of alternative methods.

In 2006, a study investigated the use of quantum computation to address the Graph Colouring problem. Interestingly, there was a substantial lapse in the publication of results on this problem. It was not until 2022 that the topic was revisited, and further new results were published in 2023 (Refer to Table~\ref{WorksOverTimeTable2},~\ref{WorksOverTimeTable3}). Consequently, exploring the potential application of Quantum methods and approaches to the Graph Colouring problem could be worthwhile, as no such attempts have been made thus far. This exploration could also pave the way for future avenues for addressing this problem.

The Bisection problem has received limited attention from the community, with a single paper published in 2015 (See Table~\ref{WorksOverTimeTable2}). Subsequent work on this problem within the field of quantum algorithms has been scarce. The only related work utilizes an interactive partial negation and partial measurement technique to derive a solution. An interesting question arises as to whether Quantum approaches/methods similar to those outlined in Table~\ref{table:2} could be applied to address this issue.

In 2000, the Maximum Clique problem was studied from the perspective of quantum computation. However, a solution remained elusive until 2019. In 2019, Pelofske et al.~\cite{pelofske2019solving} introduced a solution for large minimum vertex cover problems using a quantum annealer, leading to subsequent publications in 2020, 2021, and 2023 (See Table~\ref{table:2}). In 2021, two open problems concerning the Maximum Clique problem in Graphs were presented to the community (as shown in Fig.~\ref{NumberOfOpenProblems3}), suggesting the potential use of quantum approaches for their resolution. One may further investigate this direction to obtain further results. 

The Graph Edit Distance problem quantifies the similarity between graphs, similar to how the Levenshtein distance measures similarity between strings. This has recently garnered attention from the research community in 2021 and 2022, prompting the exploration of quantum-based solutions. In 2021, a quantum constant approximation algorithm was utilized to address this problem. Later in 2022, a QUBO solution was implemented to run on quantum annealers and gate-based quantum computers~\cite{Incudini_2022}. Three open problems related to the Graph Edit Distance were introduced in 2021. Further investigations may lead to uncovering alternative methods (or modifying those used in 2021 and 2022) in this new line of research, perhaps those mentioned in Table~\ref{table:2}.

Since 2020, the Longest Common Subsequence has been explored in the community, to propose quantum solutions. On average, there has been at least one publication per year related to this problem. The peak was in 2022 with three works. However, there is a gap in 2021 and no more than one paper in 2023. Nonetheless, an open problem related to the Longest Common Subsequence was introduced in 2023. This line of research seems to be emerging, and we recommend re-exploring the quantum techniques to uncover more avenues for addressing this problem.

\begin{figure*}[htp]
    \centering 
    \includegraphics[scale=0.3]{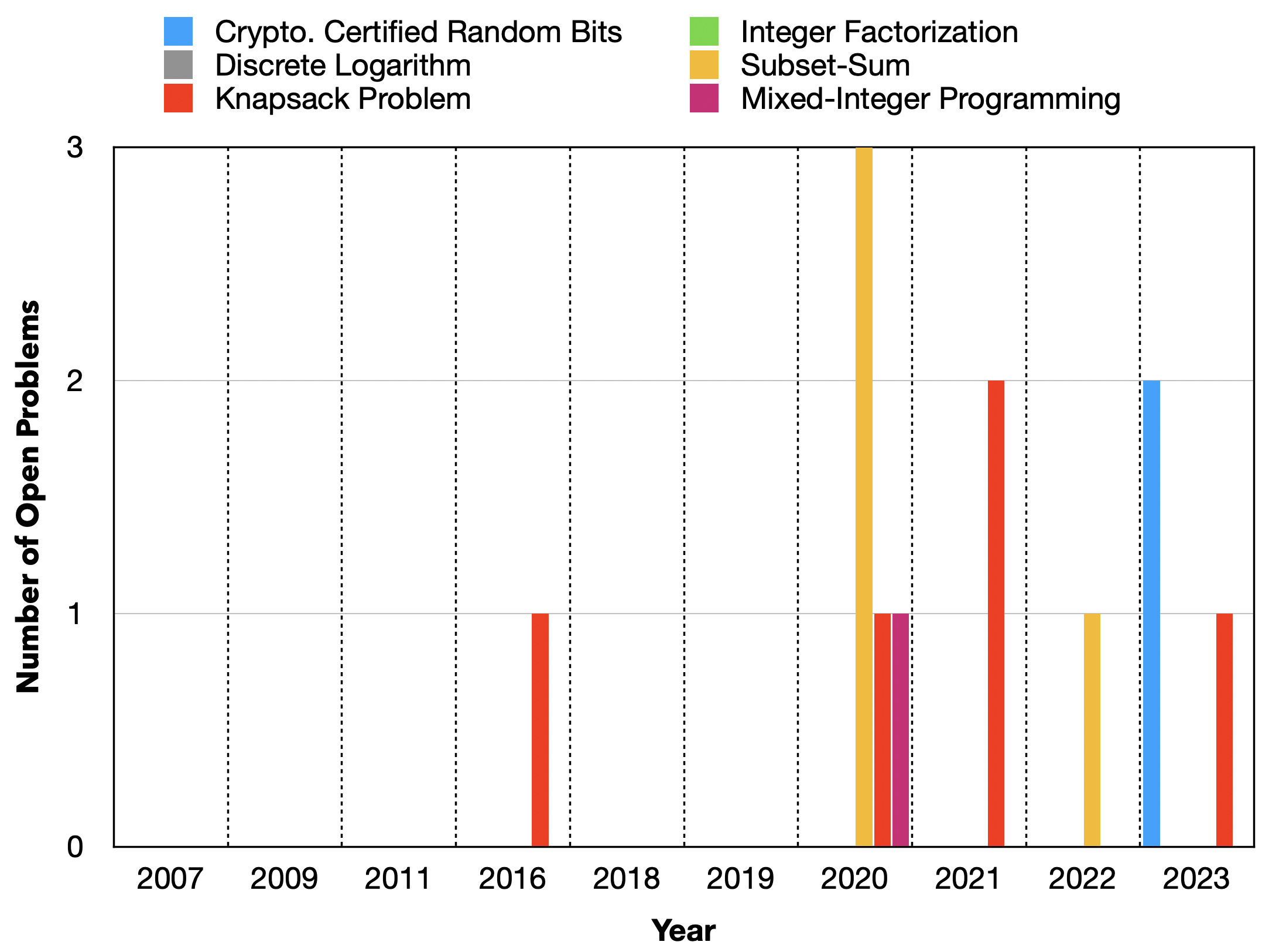}
    \caption{Number of Open Problems Corresponding to Well-known Problems in Computer Science}
    \label{NumberOfOpenProblems2}
\end{figure*}

Protein Folding in quantum algorithms was addressed for potential solutions in the years 2011-2012. Since then, there has been limited research on this problem, despite three open problems being identified at that time. Solutions for this problem have primarily centered on the adiabatic and circuit models, both of which have demonstrated significant advantages. In 2015, researchers also turned their attention to using quantum algorithms to tackle the problem of discovering the DNA motif model in bioinformatics. Furthermore, the Hitting-Set problem, another bioinformatics challenge, was resolved using Grover's algorithm in 2010.

There have been attempts to develop quantum algorithms for mixed-integer programming since 2014, but these attempts were halted until 2020 when a combination of quantum annealing and classical optimization techniques was explored, leading to three more results. However, there have been no new findings since 2020. Notably, an open problem was identified in 2020. Perhaps re-examining the quantum techniques in Table~\ref{table:2} could pave the way for new approaches or propose a solution for the open problem identified that year.

In 2016, an endeavour was undertaken to propose solutions for the Knapsack problem, followed by a three-year hiatus before resuming in 2020 and further progress in 2021 and 2023. Five open problems related to the Knapsack problem were identified in the realm of quantum computing between 2016 and 2022. It is worth highlighting that various quantum techniques such as Quantum Annealing, Quantum Particle Swarm Optimization, and Quantum Machine Learning have been put forward to tackle this classic problem in computer science. Further exploration of these methods and potentially new approaches may shed light on the future of solving this problem.

An early exploration of the Subset-Sum problem in quantum computers was published in 2011, and the research continued until 2022. However, there was a hiatus from 2014 to 2017. Nevertheless, at least one publication per year, on average, was released between 2018 and 2022. The peak occurred in 2020 with four results, while none were introduced in 2019 and 2021. Notably, four open problems were identified in 2020 and 2022 (see Fig.~\ref{NumberOfOpenProblems2}). It is worth mentioning that the Quantum Walk technique is one of the methods used for the Subset-Sum. Revisiting this method or exploring the other approaches/methods listed in Table~\ref{table:2} might offer a path to solving the open problems and/or uncovering new research directions.

The research on the discrete logarithm problem traces back to the revised and extended version of Shor's algorithm (originally appeared in 1994), which was refined in 1999. It continued in 2000 and 2004. In 2017, an algorithm was presented for computing the short discrete logarithm with a polynomial number of executions. Further, a detailed resource estimation for implementing Shor's algorithm and its impact on the security of elliptic curve point operations appeared in the community. Later in 2020, an improvement on quantum circuits for elliptic curve scalar multiplication was suggested. In 2023, a study explored the quantum complexity of Discrete Logarithms and related group-theoretic problems, providing quantum lower bounds and a model for generic hybrid quantum-classical algorithms. Further exploration of the problem and potential new approaches may shed light on finding new research directions.

\section{Quantum Algorithms Complexity}
\label{sec:previousworks}

The complexity theory has traditionally been concerned with algorithms based on classical computers. It is beginning to consider how quantum algorithms are playing a role in the current developments of quantum computing.


Bounded-error Quantum Polynomial-time (BQP) is a class of decision problems in computational complexity theory. This class represents the problems that can be solved by a quantum computer in polynomial time, with an error probability of at most $\frac{1}{3}$ for all instances. It is the quantum analog to the classical complexity class BPP (Bounded-error Probabilistic Polynomial time).
Mohr A.~\cite{mohr2014quantum} considers what is known and what is speculated about the relationship between BQP and the well-known classical classes P and NP. The category of classical counterpart is Bounded-Error Probabilistic Polynomial Time (BPP). A quantum computer is restricted to probational algorithms because of the nondeterministic nature of quantum mechanics.  On the other hand, whether this new paradigm offers anything of real value is unclear, despite quantum computing's astonishing speed increase compared to classic computers. Indeed, many believe that a classical Turing machine can solve any problem solvable by the existing model of quantum computing. The prospect of hypercomputation is intriguing, although the possibility of achieving such a machine remains theoretical~\cite{mohr2014quantum}.

Keeping in mind the struggle to connect classical computation and quantum in mind, consider a random oracle. Yamakawa and Zhandry~\cite{yamakawa2022verifiable} proved unconditionally unless otherwise stated, that relative to the oracle, with probability one, it holds that:
\begin{itemize}
    \item BQP may solve some NP search problems, but the same is not true for BPP machines.
    \item There are functions one may invert them conveniently using quantum computation. Nonetheless, they are one-way and collision-resistant against classical adversaries. Similarly, the authors stated that the latter holds for digital signatures and CPA-secure public key encryption, requiring a classical scheme. Nonetheless, the above separation does not extend to the other cryptographic objects e.g. PRGs, necessarily.
    \item There are proofs of quantumness conditioned to nothing, publicly verifiable, with minimal rounds of interaction. In the case of having uniform adversaries, the proofs are non-interactive. Nonetheless, the proofs are two messages public coin for non-uniform adversaries.
    \item The results of Yamakawa and Zhandry do not contradict the conjecture of Aaronson and Ambanis~\cite{aaronson2014need}. As a conjecture, they mentioned that considering the Aaronson-Ambanis conjecture is true, one may conclude that there is publicly verifiable certifiable randomness with minimal rounds of interaction.
\end{itemize}

Yamakawa and Zhandry~\cite{yamakawa2022verifiable} further observed that one may replace the oracle with a hash function such as SHA2 and obtain Minicrypt instantiations of the above results.

Away from the BQP, yet another property of quantum computation was exploited in the literature. It is known that the structure of optimally entangled states is constrained by quantum theory's constant character~\cite{alber1999entanglement}. Abrams et al.~\cite{abrams1998nonlinear} illustrated that one may exploit nonlinear time evolution to allow a quantum computer to solve NP-complete and \#P problems in polynomial time. The authors showed how to accomplish this exponential speed-up using the Weinberg model of nonlinear quantum mechanics. All known experiments have confirmed the linearity of quantum mechanics to a high degree of accuracy. So, the result in~\cite{abrams1998nonlinear} may be seen as evidence.

Interestingly, an entirely different approach is somewhat helping the quantum side with the classical. Wang~\cite{wang2022classically} leveraged existing classical techniques to improve quantum optimization. In particular, he first pre-processed the input with a classical algorithm, obtaining an approximate solution. Second, a quantum circuit to search the \emph{neighbourhood} of the solution of the classical algorithm, aiming for higher-quality solutions. They introduced the Classically-Boosted Quantum Optimization Algorithm (CBQOA). One may solve a wide range of combinatorial optimization problems with CBQOA. E.g. all unconstrained problems, constrained problems such as Max Bisection, Maximum Independent Set, Minimum Vertex Cover, Portfolio Optimization, Traveling Salesperson, etc. A crucial component of CBQOA is a practice-friendly continuous-time quantum walk (CTQW) on a properly constructed graph, connecting the feasible solutions. CBQOA utilizes CTQW, the output of a classical procedure, constructing a superposition of feasible solutions. The algorithm solves constrained problems without applying changes to their cost functions. It then narrows down evaluating the quantum state to the feasible subspace. The algorithm also relies not on indexing the feasible solutions. Wang illustrates the applications of CBQOA to Max 3SAT, and Max Bisection as well. The author further provides empirical results that CBQOA outperforms previous approaches.

Apart from the above approach and turning to the prover-verifier model, consider the Quantum Merlin Arthur (QMA). That represents the set of languages for which, when a string is in the language, there is a polynomial-size quantum proof that convinces a polynomial time quantum verifier of this fact with high probability. In addition, the verifier with high probability rejects every linear size quantum state where no string is included in the language.

Some methods suggest using the class $\text{QMA}^{+}(2)$, that is, the unentangled quantum proofs with non-negative amplitudes. Jeronimo and Wu~\cite{10.1145/3564246.3585248} showed that under such settings, one may design proof verification protocols for (increasingly) hard problems. This is done via logarithmic size quantum proofs and also by having a constant probability gap that allows distinguishing \emph{yes} from \emph{no} instances. Specifically, the authors designed global protocols for small set expansion, unique games, and PCP verification. Consequently, Jeronimo and Wu had the result of 
$\text{NP} \subseteq \text{QMA}^{+}_{\log}(2)$ 
with a constant gap. Taking advantage of the mentioned constant gap, the authors could \emph{scale up} the aforementioned result to $\text{QMA}^{+}(2)$. So, they were able to obtain  $\text{QMA}^{+}(2) = \text{NEXP}$ by setting up stronger in-detail properties of the PCP for NEXP.


Gottesman and Irani~\cite{gottesman2009quantum} have studied the complexity of problems that involve satisfying constraints remaining the same under translations in at least one spatial direction. Given an input $N$ in binary. The authors proved the hardness of a classical tiling problem on an $N \cross N$ 2-dimensional grid; And a quantum problem involving finding the ground state energy of a 1-dimensional quantum system of $N$ particles. The classical problem is then proved to be NEXP-complete and the quantum problem is ${\QMA}_{\EXP}$ complete. Thus, an algorithm for these problems that runs in time polynomial in $N$ would imply that $\EXP =$ NEXP or BQEXP$ = \QMA_{\EXP}$, respectively. Tiling is known to be NEXP-complete. Nevertheless, all previous reductions require that $(1)$ the set of tiles and their constraints $(2)$ the varying boundary conditions,  be given as part of the input. Gottesman and Irani considered the problem instance encoded solely in the size of the system.


There are even more interesting results about the QMA complexity class. One may generalize the satisfiability problem in classical computation to a constraint satisfaction problem in the quantum model, called Quantum Satisfiability. Consider a quantum k-SAT problem. Each constraint is then determined by a \emph{k-local projector}. Also, a constraint is satisfied by all states in its null space. A  previous work by Bravyi~\cite{bravyi2011efficient} showed that one may solve quantum 2-SAT on a classical computer. The author further proved that quantum k-SAT with $k \geq 4$ is QMA-complete. Previously, Quantum 3-SAT was recognized as being in QMA~\cite{bravyi2011efficient}. Nonetheless, its computational hardness was unknown. Gosset and Nagaj~\cite{gosset2016quantum} then proved that quantum 3-SAT is QMA-hard. So, this problem is complete for this complexity class.\\




Another direction in the literature is exploiting the properties of quantum entanglement to help improve the known classical complexity of certain problems. More specifically, consider the classical matching theory, that is, a mathematical method for finding pairs between elements from two sets. One may optimize it according to certain criteria. Classical matching theory can be defined as matrices that do not contain negative entries. In quantum theory, the positive operator concept is a generalization of matrices that have nonnegative entries. In terms of complete positive operators, or equivalently with bipartite density matrices, Gurvits~\cite{GURVITS2004448} reformulate the Edmonds problem. Namely, the central problem in the field of algebraic complexity, which J. Edmonds presented in 1967. This problem aims to decide if the span of a given tuple of square matrices contains a non-singular matrix. In other words, it investigates whether a given linear subspace of a matrix has a full rank. One of the most important cases when Edmonds’ problem can be solved in polynomial deterministic time, i.e., an intersection of two geometric matroids, corresponds to unentangled (a.k.a. separable) bipartite density matrices.

Denote $M(N)$ as the linear space of complex $N \cross N$ matrices. To solve Edmonds' problem, Gurvits introduced a general class of linear subspaces $M(N)$. In this class, there is a polynomial deterministic time algorithm to solve Edmonds’ problem. Finally, Gurvits proved that the weak membership problem for the convex set of separable normalized bipartite density matrices is NP-hard.



Khadiev and Safina~\cite{khadiev2023quantum} suggested a quantum algorithm for the dynamic programming approach for problems on directed acyclic graphs (DAGs) like $G(V, E)$. Denote $|V| = n$ as the number of vertices of $G(V, E)$. Further, let $\hat{n}$ be the number of vertices with at least one outgoing edge, and $|E| = m$ be the number of edges. Their algorithm's running time is proportional to $\mathcal{O}(\sqrt{\hat{n}m} \log \hat{n})$. Moreover, executing the best-known deterministic algorithm up to the date of the publication of~\cite{khadiev2023quantum} takes $\mathcal{O}(n + m)$. The authors proved that their algorithm may solve problems that involve functions OR, AND, NAND, MAX, and MIN. For example, a Boolean formula represented by Zhegalkin polynomial~\cite{zhegalkin1927technique, gindikin1985algebraic, yablonsky1989introduction}. Also, a Boolean circuit with shared input and non-constant depth evaluation. In addition, the single source longest paths search for weighted DAGs problem and the diameter search problem for unweighted DAGs.\\

Moving on from the above directions, consider the Quantum Annealing algorithms, introduced as a solution to combinatorial optimization problems. They utilise quantum fluctuations to escape local minima in energy landscapes to solve complex NP-hard problems. Cruz-Santos et al.~\cite{cruz2019qubo} aimed to apply quantum annealing to the theory of cuts, which is a crucial field in theoretical computer science. The authors developed a method to convert the Minimum Multicut Problem into the Quadratic Unconstrained Binary Optimization (QUBO) representation and have addressed the technical challenges that arise when submitting a problem to the quantum annealer processor. Two constructions of the quadratic unconstrained binary optimization functions for the Minimum Multicut Problem have been provided and compared the tradeoffs between the two mappings. Also, numerical scaling analysis is conducted on several classical approaches.


\begin{definition}
Given a quantum circuit $C$, consider the quantum computation verification problem (QCV). The aim of QCV is convincing a client that the output of evaluating $C$ reported by a server is the result the server claims~\cite{aharonov2017interactive,gheorghiu2019verification,reichardt2013classical}. 
\end{definition}

A restriction in QCV is that the client has limited computational power. Nonetheless, the client does not require quantum processing power. That is, one might think of it as the classical verification of quantum computation problem (CVQC). The quantum computations on the server side determine the upper bound of the overall time complexity. As an example, refer to a single-server CVQC protocol suggested by Mahadev~\cite{mahadev2018classical}. This protocol offers $\mathcal{O}(\textbf{poly}(\kappa) |C|^3)$ s.t. $|C|$ is the size of the circuit to be verified, and $\kappa$ is the security parameter. Notice that many other protocols, including multiparty quantum computation, zero knowledge, and obfuscation have the same \emph{cubic} blow up in their time complexity~\cite{alagic2020non,bartusek2021secure,bartusek2021candidate,chia2020classical,chung2022constant,vidick2020classical}.

Zhang~\cite{zhang2022classical} improved the above bound to $\mathcal{O}(\textbf{poly}(\kappa) |C|)$. The protocol in~\cite{zhang2022classical} assumes the existence of noisy trapdoor claw-free functions~\cite{brakerski2018cryptographic}. Considering the above assumption, Zhang showed that their protocol is secure under the quantum random oracle model~\cite{dagdelen2010random}. Further, they suggested a classical channel remote state preparation protocol for states in $\{ \ket{+_{\theta}} = \frac{1}{\sqrt{2}} (\ket{0} + e^{i\theta \pi /4}\ket{1}) : \theta \in \{0, 1, \cdots 7\} \}$. The protocol in~\cite{zhang2022classical} allows parallel verifiable preparation of $L$ independently random states, in the latter form. This is up to a constant overall error. Also, on a possibly unbounded server-side simulator. The author in~\cite{zhang2022classical} suggests is $\mathcal{O}(\textbf{poly}(\kappa) L)$ runtime, and constant rounds. Compared to the previous work~\cite{gheorghiu2022quantum,cojocaru2019qfactory,gheorghiu2019computationally,fu2022computational}, the above bound is indeed an improvement.

\subsection{Models of Quantum Complexity Growth}

The change of states of a closed quantum mechanical system over time is caused by a specific class of transformations, mathematically known as unitary transformations. Denote the quantum complexity of a unitary transformation or quantum state as the size of the shortest quantum computation that runs the unitary or prepares the state. One might find it reasonable to expect that the complexity of a quantum state governed by a chaotic many-body Hamiltonian grows linearly with time s.t. the time is exponential in the system size. However, it is uneasy to rule out a shortcut that improves computation efficiency. So, it is hard to derive lower bounds on quantum complexity for particular unitaries, or states without making additional assumptions. 

Brandao et al.~\cite{brandao2021models} establish a growth of the quantum complexity in the time evolution of several models. They prove that with overwhelming probability, an element sampled from an approximate unitary $k$ - design has a strong complexity that scales at least linearly in $k$. Unitary k-designs is a framework that lets one approximate randomization over the unitary group with a finite set of unitaries. Without using less than
$k+1$ copies of unitaries from these distributions, the approximation cannot be distinguished from the random distribution. They also state that using the known relations between the evolution time and the design order $k$ establishes a lower bound on the growth of quantum complexity. For random quantum circuits in particular, the authors make substantial progress on conjectures by Brown and Susskind~\cite{PhysRevD.97.086015}. In addition, by using an established linear relation between the circuit size and design order, they prove a linear growth of quantum complexity.


Aaronson~\cite{aaronson2018pdqp} proved that one may combine two different hypothetical enhancements to quantum computation. Namely, quantum advice and non-collapsing measurements. This allows a quantum computer to compute every decision problem in polynomial time. The latter holds although neither enhancement implies remarkable computational advantage on its own. This assists a related result by Raz~\cite{raz2005quantum}. The proof uses locally decodable codes.


Dalzell et al.~\cite{dalzell2023mind} introduced a quantum algorithm for several families of binary optimization problems, with guarantees on runtime. Problems including QUBO, Ising Spin Glasses ($p\text{-spin model}$), and $k\text{-local}$ Constraint Satisfaction Problems ($k\text{-CSP}$). The authors proved that either $(1)$ their algorithm finds the optimal solution, requiring a time complexity proportional to $\mathcal{O}^*(2^{(0.5-c)n})$ for an $n\text{-independent}$ constant $c$. This implies a $2^{cn}$ advantage over Grover’s algorithm. Or, $(2)$ for arbitrarily small choice of a constant denoted as $\eta$, there exist \emph{sufficiently many} low-cost solutions s.t. classical random guessing produces a $(1-\eta)$ approximation to the optimal cost value, in subexponential time. They further showed that considering a large fraction of random instances from the $k\text{-spin}$ model, and every fully satisfiable or slightly frustrated $k\text{-CSP}$ formula, then case $(1)$ holds.




Unsal and Oruc~\cite{unsal2023faster} suggested quantum algorithms for routing concentration assignments on full-capacity fat-and-slim concentrators, bounded fat-and-slim concentrators, and regular fat-and-slim concentrators. Using the classical approaches, the time complexity of the concentration assignment is proportional to $\mathcal{O}(n)$ on all the above concentrators s.t. $n$ is the number of inputs. Assisted by Grover’s quantum search, the algorithm in~\cite{unsal2023faster} offers the execution time of $\mathcal{O}(\sqrt{nc} \ln c)$ in which c denotes the capacity of the concentrator. So, when $c \ln^2 c \in o(n)$, the runtime of the algorithm in~\cite{unsal2023faster} asymptotically grows slower than the classical variants. Consider a small positive number $0 < \mu < 1$. Then, for $c = n^\mu$, it holds that $c \ln^2 c \in o(n)$. Therefore, the time complexity of the algorithm in~\cite{unsal2023faster} becomes proportional to $\mathcal{O}(n^{0.5(1 + \mu)} \ln n)$.

\subsection{Quantum Randomness and Classical Complexity}


Given a classical proof (or witness) from a computationally unbounded prover. The Quantum Classical Merlin Arthur (QMA) consists of all the decision problems one may verify by a polynomial-time quantum algorithm, using the aforementioned prover~\cite{bittel2023optimizing}. BQP contains some problems that are believed to be hard for classical computers. E.g. factoring large numbers or simulating quantum systems. Kretschmer constructed a quantum oracle relative to which $\text{BQP} = \text{QMA}$. Nonetheless, the cryptographic pseudorandom quantum states and pseudorandom unitary transformations exist for the above oracle. This is a counterintuitive result because pseudorandom states can be \emph{broken} by quantum Merlin-Arthur adversaries. As a result of the distinction between algorithms operating on quantum and classical inputs, the author explained how this nuance arises. Provided that there are no pseudorandom states when $\text{BQP} = \text{PP}$. The author then proved that some computational complexity assumption is needed to create a pseudorandom state. The authors in~\cite{https://doi.org/10.4230/lipics.tqc.2021.2} discussed the implications of these results for cryptography, complexity theory, and shadow tomography.


Arora et al.~\cite{arora2023quantum} suggested a comprehensive characterization of the computational power of shallow quantum circuits. Such circuits are combined with classical computation. In particular, for classes of search problems, the authors proved that relative to a \emph{random oracle}, several properties hold. First, they proved that the conjecture of Jozsa~\cite{jozsa2006introduction} has to be refused. Becuase $\text{BPP}^{\text{QNC}^{\text{BPP}}} \neq \text{BQP}$. Consequently, an \emph{instantiatable} separation between the classes was obtained in~\cite{arora2023quantum}. This was done by replacing the oracle with a cryptographic hash function. The latter also implied a resolution to one of Aaronson’s ten semi-grand challenges~\cite{aaronson2005ten} in quantum computing. Second, they showed that $\text{BPP}^{\text{QNC}} \not\subseteq \text{QNC}^{\text{BPP}}$. Also, they proved $\text{QNC}^{\text{BPP}} \not\subseteq \text{BPP}^{\text{QNC}}$ holds. So, one may observe that an interaction between the classical and the shallow quantum computation exists. In particular, for some problems, performing adaptive measurements in a single shallow quantum circuit sounds more favourable than performing polynomially many shallow quantum circuits, having no adaptive measurements. Third, they stated that there is a 2-message proof of quantum depth protocol. One may instantiate the proof of quantum depth using the one for quantumness construction previously discussed in the work of Yamakawa and Zhandry~\cite{yamakawa2022verifiable}. One may observe that the above protocol lets a classical verifier certify that a prover has to perform a computation of some minimum quantum depth.

Clifford and Clifford~\cite{clifford2018classical} studied the classical complexity of the exact Boson Sampling problem where the objective is to produce provably correct random samples from a particular quantum mechanical distribution. Aaronson and Arkhipov proposed the computational framework in STOC ’11 in 2011 as an attainable demonstration of ‘quantum supremacy’, that is a practical quantum computing experiment able to produce output at a speed beyond the reach of classical (that is non-quantum) computer hardware. Aaronson and Arkhipov demonstrated that exact Boson Sampling is not efficiently solvable by a classical computer unless $P^{\#P} = BPP^{NP}$ and the polynomial hierarchy collapses to the third level~\cite{aaronson2010computational}. Clifford and Clifford give an algorithm running in $\mathcal{O}(n2^n + poly(m, n))$ time and $\mathcal{O}(m)$ space. The algorithm is simple to implement and has low constant factor overheads which able it to solve the exact Boson Sampling problem for system sizes far beyond current photonic quantum computing experimentation~\cite{clifford2018classical}.

\subsection{Approximation for Quantum Problems}


As mentioned in the previous subsection, Quantum Classical Merlin Arthur (QCMA) is a quantum generalization of the classical complexity class NP. Given a classical proof (or witness) from a computationally unbounded prover. QCMA consists of all the decision problems that can be verified by a polynomial-time quantum algorithm, using the above prover. With classical proof, QCMA is also known as QMA~\cite{bittel2023optimizing}.

Gharibian and Kempe~\cite{gharibian2012hardness} defined a quantum generalization of the polynomial hierarchy and initiated its study. They showed that there are complete problems for the second level of this quantum hierarchy. Further, these problems are hard to approximate. They further obtained the hardness of approximation for the class QMA using these techniques. Inspired by the classical results of Umans on the hardness of approximation for the second level of the classical polynomial hierarchy,  the authors' approach is based on the use of dispersers. They proved the hardness of the approximation of the problems, including a quantum version of the Succinct Set Cover problem. Further, for a variant of the local Hamiltonian problem with hybrid classical-quantum ground states~\cite{gharibian2012hardness}.


A crucial element in computation is by including quantumness or quantum correlations. These are lasting in entanglements yet often separable states. Consequently, quantifying the quantumness of a state in a quantum system indeed brings useful insights. Identifying the nearest classical-classical state helps in understanding the degree of quantum correlation~\cite{lu2023approximation}. Nevertheless, the exact quantification of quantumness is an NP-hard problem. 

Lu et al.~\cite{lu2023approximation} considered approximating the exact solution. They took the Frobenius norm, constructed an objective function, and offered a gradient-driven descent flow on Stiefel manifolds. This could determine the quantity. The authors proved that the objective value decreases along the flow. They also provided supporting experimental results. Further, the ability to decompose quantum states into tensor products of certain structures is guaranteed by their method. The method also maintains primitive quantum assumptions. Moreover, the provided numerical results verify that their method is practice-friendly as well.


Consider the Max-Cut (MC) problem and its quantum analog, Quantum Max-Cut (QMC).

Kallaugher and Parekh~\cite{kallaugher2022quantum} explored the space complexity of the above graph streaming problems. It is trivial to suggest a $2\text{-approximation}$ for MC using $\mathcal{O}(\log n)$ space. Nonetheless, the classical space complexity of MC was proven by Kapralov and Krachun~\cite{kapralov2019optimal} to be $\Omega(n)$ for all $(2-\varepsilon)\text{-approximation}$. In~\cite{kallaugher2022quantum}, the authors generalized the above cases and proved an $\Omega(n)$ space lower bound for $(2-\varepsilon)\text{-approximation}$ for MC and QMC. Notice that the lower bound also holds when the algorithm may maintain a quantum state. 

Further, it is known that the QMC has a straight-forward $4\text{-approximation}$. Kallaugher and Parekh~\cite{kallaugher2022quantum} proved the tightness with an algorithm returning a $(2 + \varepsilon)\text{-approximation}$, for the quantum max-cut value of a graph, in $\mathcal{O}(\log n)$ space. The authors resolved the approximability of MC and QMC requiring $o(n)$ space.

Moving on from QMC, let $y$ be a real random variable. Consider $n$ \emph{independent} samples $y_1, y_2, \cdots, y_n$. One might estimate the mean of $y$ i.e. $\mu = E[y]$ from the samples. A straightforward approach is to output the mean of the samples: $\hat{\mu} = (y_1 + \cdots + y_n)/n$. This is an unbiased estimator with standard deviation $\sigma / \sqrt{n}$ s.t. $\sigma = \textbf{stddev}[y] = E[(y - \mu)^2]$. One may observe that Chebyshev’s inequality implies: $$\text{Pr}[|\hat{\mu} - \mu| \geq 10 \sigma / \sqrt{n}] \leq 1 \%$$ 

Kothari and O’Donnell~\cite{kothari2023mean} assumed \emph{the code} that generates $y$ is given. For instance, a randomized or quantum circuit whose output is $y$. The authors provided a quantum procedure that runs the code $\mathcal{O}(n)$ times to generate $y$, outputting $\hat{\mu}$ that satisfies $| \hat{\mu} - \mu | \leq \sigma/n$, w.h.p. The dependence on $n$ is optimal for quantum algorithms. A crucial point about the results in~\cite{kothari2023mean} is obtaining results better than the classical ones. Namely, classical algorithms can achieve the quadratically worse $| \hat{\mu} - \mu | \leq \sigma/\sqrt{n}$. The previous works before~\cite{kothari2023mean} made further assumptions about $y$, and/or assumed the algorithm knew an a priori bound on $\sigma$, and/or used additional logarithmic factors beyond $\mathcal{O}(n)$~\cite{hamoudi2021quantum,montanaro2015aceleracion,heinrich2003problem,terhal1999quantum,brassard1998quantum,grover1996fast,beaver1996correlated}. The heart of the work in~\cite{kothari2023mean} is Grover’s algorithm, yet with complex phases.

Another interesting result is obtained by Cornelissen et al.~\cite{cornelissen2022near}. They proposed a near-optimal quantum algorithm for estimating the mean of a vector-valued random variable with finite mean and covariance, in Euclidean norm. The authors considered the \emph{binary oracle model}. This generalizes in the classical sample complexity. It also is a setting used frequently in previous works, e.g.~\cite{abrams1999fast,brassard2011optimal,brassard2002quantum,hamoudi:LIPIcs.ESA.2021.50,HEINRICH20021,montanaro2015quantum}. Cornelissen et al. extended the theory of multivariate sub-Gaussian estimators~\cite{lugosi2019mean} to the quantum setting. The authors pointed out that under the quantum settings, one may not prove that there is a univariate estimator that can be utilized to become a multivariate estimator, with at most a logarithmic overhead in the dimension. If the sample complexity is considered smaller than the dimension, there exists a quantum advantage for the mean estimation~\cite{heinrich2004power}. Denote $d$ as the dimensionality factor. The lower bound proved in~\cite{heinrich2004power} implies that a $\log(d)$ overhead for the quantum multivariate mean estimation problem is possible. The crucial point shown in~\cite{cornelissen2022near} is that apart from the above settings, there is a quantum estimator, outperforming all classical ones. In particular, the quantum estimator reduces the error by a factor close to the square root of the ratio between the dimensionality factor, and the complexity of the sample.



Cornelissen and Hamoudi~\cite{cornelissen2023sublinear} introduced a quantum algorithm, estimating Gibbs partition functions in sublinear time w.r.t. the logarithm of the size of the state space. The authors obtained a speed-up over the work of {\v{S}}tefankovi{\v{c}} et al.~\cite{vstefankovivc2009adaptive}. They utilized the properties of quantum Markov chains to conserve the quadratic speed-up in precision and spectral gap, obtained in previous works. The authors offered polynomial improvements to compute the partition function of the Ising model, counting the number of k-colourings, matching, or independent sets of a graph, and estimating the volume of a convex body. The approach in~\cite{cornelissen2023sublinear} included developing variants of the quantum phase and amplitude estimation algorithms. To reduce the variance quadratically faster compared to the classical empirical mean, the authors extended the latter subroutines into a nearly unbiased quantum mean estimator.




\begin{definition}
    The Tail Assignment problem is the task of assigning individual aircraft to a given set of flights, minimizing the overall cost. 
\end{definition}

\begin{definition}
    The Quantum Approximate Optimization Algorithm (QAOA) is a hybrid quantum-classical algorithm, which helps solve combinatorial optimization problems.  
\end{definition}

Vikst{\aa}l et al.~\cite{vikstaal2020applying} simulated the QAOA applied to instances of this problem derived from real-world data of the Tail Assignment problem. These instances were reduced to fit on quantum devices with 8, 15, and 25 qubits.  For instance, the reduction process leaves only one viable solution that allows us to map out an Exact Cover problem on Tail Assignment~\cite{vikstaal2020applying}.

\subsection{Quantum Query Complexity}
 

It is known that quantum computers can exponentially improve the running time of certain computational tasks, compared to their classical counterparts. The setting of \emph{black box} or \emph{query model} helps one to show such exponential speedups. In this model, there are \emph{black-box accesses} to the input. While trying to minimize the number of queries, an algorithm under this assumption aims to compute a function of the input.

Aaronson and Ambainis~\cite{aaronson2015forrelation} used a property-testing problem called \emph{Forrelation}. In this problem, one decides whether a Boolean function is highly correlated with the Fourier transform of a second function. Consider all \emph{partial functions}\footnote{Recall that a partial function is a function not defined for some inputs of the right type. Namely for some of the domains. E.g. division is a partial function since division by $0$ is undefined on the Reals.} on $N$ bits that can be computed with an advantage $\delta$ over a random guess by making $q$ quantum queries. Aaronson and Ambainis showed that such functions can be computed classically as well, with an advantage $\delta/2$ by a randomized decision tree. In particular, it requires $\mathcal{O}_q(N^{1-\frac{1}{2q}} \delta^{-2})$ many queries. Let the \emph{k-Forrelation problem} be a partial function taking $q = \lceil k/2 \rceil$ quantum queries. The authors in~\cite{aaronson2015forrelation} conjectured that k-Forrelation is a suitable candidate for exhibiting such an extremal separation.

Further, Bansal and Sinha~\cite{bansal2021k} considered the most basic access model in their work. In particular, each query returns a bit of the input. They proved the conjecture of Aaronson and Ambainis by showing a tight lower bound of $\Tilde{\Omega}(N^{1-\frac{1}{k}})$ for the randomized query complexity of k-Forrelation s.t. $\delta = 2^{-\mathcal{O}(k)}$.


Apart from the works related to the Forrelation, consider a classical query algorithm (decision tree) and a guessing algorithm that tries to predict the query answers. One may then suggest a quantum algorithm with query complexity $\mathcal{O}(\sqrt{GT})$ s.t. $T$ denotes the query complexity of the classical algorithm (i.e. depth of the decision tree). Also, $G$ represents the maximum number of guessing algorithm's wrong answers~\cite{Beigi2020quantumspeedupbased, v012a018}. Beigi et al.~\cite{beigi2022time} proved that, provided some constraints on the classical algorithms, one may implement this quantum algorithm in time $\Tilde{\mathcal{O}}(\sqrt{GT})$. The algorithm in~\cite{beigi2022time} is based on non-binary span programs and their efficient implementation. Let $n$ be the number of vertices of a given graph. The authors pointed out that various graph-theoretic problems, including bipartiteness, cycle detection, and topological sort can be solved in time $\mathcal{O}(n^{3/2} \log^2 n)$ and with $\mathcal{O}(n^{3/2})$ quantum queries. Further, one may find a maximal matching with $\mathcal{O}(n^{3/2})$ quantum queries in time $\mathcal{O}(n^{3/2} \log^2 n)$, and maximum bipartite matching in time $\mathcal{O}(n^{2} \log^2 n)$.

\subsection{Quantum Random Walk}

\begin{definition}
Quantum random walk is the counterpart of classical random walks for particles that cannot be precisely localized due to quantum uncertainties. Quantum random walks are described in terms of probability amplitudes~\cite{aharonov1993quantum}. 
\end{definition}

Apers et al.~\cite{apers2019unified} introduced a quantum walk search framework, unifying and strengthening quantum walk frameworks. 
Their framework detects and finds marked elements in the electric network setting. 
It also permits interpolating between the hitting time framework. 
This minimizes the number of walk steps, and the MNRS framework~\cite{magniez2007search}, minimizes the number of checking the elements for being marked. 
The previous frameworks relied on quantum walks and phase estimation only. 
Nonetheless, the algorithm used in~\cite{apers2019unified} benefits from a technique named quantum fast-forwarding as well. Further, the authors showed how, in certain cases, one may perform simplifications by applying the quantum walk operator several times.

Another interesting application of quantum random walk is in the shortest vector problem (SVP). Which, is an important problem for the cryptanalysis of lattice-based cryptography. Chailloux and Loyer \cite{chailloux2021lattice} improved the results of Laarhoven \cite{laarhoven2016search} for SVP.
They introduced an algorithm with a (heuristic) running time of $2^{0.2570d+o(d)}$ s.t. $d$ is the lattice dimension. 
The authors also suggested time-memory trade-offs. Namely, one may quantify the amount of quantum memory and quantum random access memory of the algorithm. 
From a broader perspective, their idea is to replace Grover’s algorithm used in~\cite{laarhoven2016search} in a key part of the sieving algorithm. 
This is done by a quantum random walk in which one adds a layer of local sensitive filtering.


Marsh and Wang~\cite{marsh2019quantum} aimed to find approximate solutions for computational problems contained in NPO PB\footnote{polynomially bounded NP optimization complexity} class efficiently. The authors then apply QAOA to the minimum vertex cover problem. They studied a generalization of the QAOA state evolution to alternating quantum walks and solution-quality-dependent phase shifts. They also use quantum walks to integrate the problem limitations of NPO problems. Marsh and Wang~\cite{marsh2019quantum} have shown that a hybrid quantum-classical variational algorithm suits finding the value of a certain expectation value. It is also efficient for NP optimization problems that have polynomially bounded measures. The results in~\cite{marsh2019quantum} illustrated that for solving the minimum vertex problem, a hybrid quantum-classical variational scheme shows promising results. It only requires hiring a fixed and small number of optimization parameters.


Belovs~\cite{belovs2012learning} showed that the quantum query complexity of $k\text{-distinctness}$ is $\mathcal{O}(n^{\frac{3}{4} - \frac{1}{4} \frac{1}{2^k - 1}})$ s.t. $k$ is a constant and $k \geq 4$ holds. Also, before the work of Jeffery and Zur~\cite{jeffery2022multidimensional}, the time complexity of the latter was $\Tilde{\mathcal{O}}(n^{1 - \frac{1}{k}})$. Nevertheless, Jeffery and Zur~\cite{jeffery2022multidimensional} proved a tighter upper bound of $\Tilde{\mathcal{O}}(n^{\frac{3}{4} - \frac{1}{4} \frac{1}{2^k - 1}})$ for the time complexity. This matches the query complexity, yet up to polylogarithmic factors. They provided a technique to design quantum walk search algorithms. This is specifically an extension of the electric network framework. The technique helped to obtain the tighter upper bound for the time complexity. 

Consider the welded trees problem. The authors in~\cite{jeffery2022multidimensional} suggested a solution requiring $\mathcal{O}(n)$ queries and $\mathcal{O}(n^2)$ time. The result is a direct application of the technique mentioned earlier. Namely, the quantum walk framework suggested in~\cite{jeffery2022multidimensional} achieves exponential improvement in running time.

\section{Quantum Solutions to Classical Problems}
\label{sec:quantumSolToClassProb}

Quantum algorithms aim to reduce the computational time of combinatorial problems, which usually introduce a large search space~\cite{varmantchaonala2023quantum}. There can be improvements to the running time by bounding an area and keeping the focus on it in the search space.

The Grover search algorithm, one of the most popular quantum algorithms, is a good solution for solving the NP problem but requires a large number of quantum bits (qubits) to function.

Further, Quantum Annealers (QAs) physically exploit quantum effects. They are specialized quantum computers that minimize objective functions over discrete variables. Many QA platforms allow the optimization of quadratic objectives defined over binary variables (i.e. qubits) a.k.a \emph{Ising Problems}. The QA systems implemented by D-Wave followed a Moore-like growth in terms of scaling. There are architectures with $2048$ sparsely-connected qubits. Continued exponential growth and increased connectivity are anticipated.

\subsection{Satisfiability Problem}

\begin{definition}[SAT problem]
The SAT problem determines whether a given propositional Boolean expression, $\Psi(x_1, x_2, \cdots )$, of Boolean variables, $x_1, x_2, \cdots $ is satisfiable. Namely, checking if there exists a set of boolean values for the variables satisfying the formula. A special case is called 3-SAT. Its formula consists of multiple clauses, i.e. a conjunction of $N_C$ clauses $\Psi(x_1, x_2,\cdots, x_n) = \wedge^{N_C}_{j=1} C_j$. Each clause $Cj = L_{j,1} \vee l_{j,2}$ or ~$ l_{j,1} \vee l_{j,2} \vee l_{j,3}$ is a disjunction of at most three literals, ~$l_{j,1}, l_{j,2}, l_{j,3}\in\{x_k,\Bar{x}_k|k = 1,\cdots, n\}$. 
\end{definition}


Classical algorithms were previously applied to solve the satisfiability problem. Due to the insufficient computational power of classical computers, effective methods are still being explored. Quantum algorithms are among the options, as they can parallelism. Nonetheless, many quantum algorithms require a large number of qubits to solve a computationally simpler problem. 

Zhang et al.~\cite{zhang2020procedure} suggested an optimized data structure to solve the Boolean satisfiability problem. They utilized Grover's algorithm first. Second, they proposed a formula, balancing variables in consideration of complexity. Via simplification, the authors built quantum circuits to decrease the number of qubits required in Grover's algorithm.


Cheng et al. \cite{CHENG2007123} proposed a new quantum cooperative search algorithm to make Grover's search algorithm work with a small number of quantum bits in solving the 3-SAT problem. The proposed algorithm, in the use of traditional 3-SAT algorithms such as evolutionary algorithms and heuristics local search algorithms, replaces some quantum bits with classical bits to identify assignments for these classical bits. Moreover, a mathematical analysis suggests the optimal configuration of the proposed algorithm. In most cases, the experimental results show that a quantum cooperative search algorithm composed of Grover's search and heuristic local search performs better than other pure traditional 3SAT algorithms. Furthermore, the experiments confirm a mathematical analysis of an appropriate number of quantum bits~\cite{CHENG2007123}.


Varmantchaonala et al.~\cite{varmantchaonala2023quantum} studied a particular parameter. They suggested an approach based on the Grover algorithm. Their approach considers the aforementioned parameter, reducing the search space for the SAT problem. Let $l$ be the number of \emph{good} solutions amongst the initial solutions. Then the algorithm in~\cite{varmantchaonala2023quantum} requires the time complexity of $\mathcal{O}(\sqrt{N/lS})$ s.t. $N$ is the number of potential solutions. Also, $S$ denotes the reduction or subdivision factor of the space. In case there are no such $l$ solutions, the time complexity becomes proportional to $\mathcal{O}(\sqrt{N/S})$.


Leporati and Felloni~\cite{leporati2007three} proposed three quantum algorithms that solve the 3-SAT decision problem in polynomial time, in the semi-uniform setting. Their construction assumes an external observer may measure a null vector and a non-null vector. Consider every instance $\phi_n$ of 3-SAT. The first algorithm in~\cite{leporati2007three} constructs a quantum Fredkin circuit. This circuit computes a superposition of all classical evaluations of $\phi_n$ in the first output line. Now, suppose a register of a quantum register machine is given. Similar to the first algorithm, the second and third algorithms compute the same superposition on the register, and as the energy of a given membrane in a quantum $P$ system, respectively. Provided a non-unitary operator. Assume that the operator can be realized as a quantum gate. That is, an instruction of the quantum register machine, and as a rule of the quantum $P$ system. The operator can be described using creation and annihilation operators. As stated in~\cite{leporati2007three}, one may apply the operator to the result of the computation performed by the algorithms. So, the solution of 3-SAT for the instance $\phi_n$ can be extracted.

Moving to another method, consider Conflict-Driven Clause Learning (CDCL), that is, an algorithm for solving SAT. In the case of a Boolean formula, the SAT problem requires the assignment of variables so that the entire formula can be evaluated as true. Tan et al.~\cite{10071022} proposed HyQSAT, a hybrid approach that integrates QA with Conflict-Driven Clause Learning. It enables end-to-end acceleration for solving SAT problems. Instead of all embedded clauses for QA hardware, the authors used a quantitative estimation of conflict frequency and breadth-first traversal to determine their embedded order. HyQSAT takes advantage of both quantum parallelism and classical algorithms. Further, the experiments in~\cite{10071022} showed that, compared to the state of classical physics, HyQSAT can achieve a real quantum speed of 12.62X on DWave 2000s.

Yet, another method is called the Ohya-Masuda Quantum algorithm, which uses a quantum computer and a classical amplifier for solving the 3-SAT problem~\cite{ohya1999quantum}. The 3-SAT may be solved in polynomial time by quantum computer if the superposition of two orthogonal vectors $\ket{0}$ and $\ket{1}$ can physically be detected~\cite{ohya1998np}. Nonetheless, the issue is that it is inconvenient to distinguish the items with a low measurement probability. The algorithm in~\cite{ohya1999quantum} consists of three steps. The first one performs a unitary transformation on the input state. The second one carries out a nonunitary transformation that gives a classical mixture of states. Lastly, the third step represents a logistic map amplification of the result of the second step. However, Dugic in~\cite{10.1023/A:1015644508998} described that the crucial second step is not possible to be achieved in general. The heart of the Ohya-Masuda algorithm is producing a classical mixture out of pure states. Then, amplifying classical information on the last qubit state. In the second step of the algorithm, a non-unitary transformation is used. In a realistic situation, it is unlikely that such a transformation will occur. And, in most cases, an error related to an imperfect measurement will change the final state unavoidably. So, this error and its amplification give the wrong result in the end~\cite{10.1023/A:1015644508998}.


Apart from the above methods, Benjamin et al.~\cite{benjamin2017measurement} investigated how a quantum algorithm performs when solving the 3-SAT problem. The authors point out that a cycle of post-selected measurements drives \emph{monotonically} the computer’s register toward a steady state. This implies a correlation to the classical solution(s). Let $\theta$ be an internal parameter that determines the degree of correlation and the probability of success. So, $\theta$ controls the algorithm’s runtime. $\theta$ can optionally be evolved gradually during the algorithm’s execution. Evolving the parameter then creates a Zeno-like effect. One may view this as an adiabatic evolution of a Hamiltonian s.t. it remains frustration-free. It lower bounds the corresponding gap as well. Consider using up to thirty-four qubits when performing the exact numerical simulations of small systems. The approach in~\cite{benjamin2017measurement} offers a competitive running time compared to a classical 3-SAT solver. A classical 3-SAT solver itself outperforms a brute-force approach i.e. applying Grover’s search.


Another interesting result is obtained by N{\"u}{\ss}lein et al.~\cite{nusslein2023solving}, where they introduced an approach to translate arbitrary 3-SAT instances to QUBO. Such 3-SAT instances are used by quantum annealing or the quantum approximate optimization algorithm. The approach in~\cite{nusslein2023solving} requires a smaller number of couplings and fewer physical qubits than its previous methods. This advantage results in higher solution quality. In addition, the authors verified the benefits of their approach in practice by testing it on a D-Wave quantum annealer.


It is known that when one changes the clause density in 3-SAT and Max-3-SAT problems, using QAOA for solving the problem, the quantum cost shows an easy-hard-easy or easy-hard pattern, respectively~\cite{yu2023solution}. 

Using numerical simulations, Yu et al.~\cite{yu2023solution} showed with up to fourteen variables and analytical arguments that the adaptive-bias QAOA (ab-QAOA) improves performance remarkably in the hard region of the 3-SAT and Max-3-SAT problems. ab-QAOA needs three levels for 10-variable 3-SAT problems as compared to twenty-two for QAOA. The comparison is under the setting of having a similar accuracy and is reported for the \emph{on average} case. In addition, for 10-variable Max-3-SAT problems, there are seven and sixty-two levels, respectively. The above improvement is obtained from an entanglement of a more targeted and limited generation during the evolution. The authors in~\cite{yu2023solution} show that it is not strictly necessary for the classical optimization to be in the ab-QAOA. This is because the local fields are used to guide the evolution. The authors thus proposed an optimization-free ab-QAOA that uses significantly fewer quantum gates for solving the hard-region 3-SAT and Max-3-SAT problems.


Biana et al.~\cite{bian2020solving} considered the feasibility of QA architectures implemented by D-Wave, as they scale, for solving SAT and MaxSAT problems. They developed technical ways for encoding SAT and MaxSAT (with some limitations though) into Ising problems s.t. is compatible with \emph{sparse} QA architectures. The authors provided first showed how the mapping can be done, theoretically. Second, they offered encoding techniques, having on-the-fly placement and routing. They combined offline Satisfiability and Optimization Modulo Theories. They also conducted preliminary empirical tests on a current generation 2048-qubit D-Wave system. The results provide supporting insights on the feasibility of their approach for certain SAT and MaxSAT problems.

\subsection{Maximum Independent Set} 

\begin{definition}[Internal Set Problem]
An internal set problem is a graph optimization problem that involves determining subsets of vertices on the graph, in which there are no two vertices connected. 
\end{definition}

\begin{definition}[Maximum Independent Set (MIS) Problem]
\label{MIS}
The Maximum Independent Set (MIS) Problem tries to find the largest set of vertices that don't have two adjacent vertices. In other words, the aim is to obtain the widest set of vertices possible so that there are no edges connecting any two sets of vertices~\cite{manyem2023maximum}. 
\end{definition}

\begin{definition}
The Unit-Disk MIS (UD-MIS) problem is the MIS problem restricted to unit-disk graphs. 
\end{definition}

\begin{figure*}[t]
    \centering
    \includegraphics[width=\textwidth]{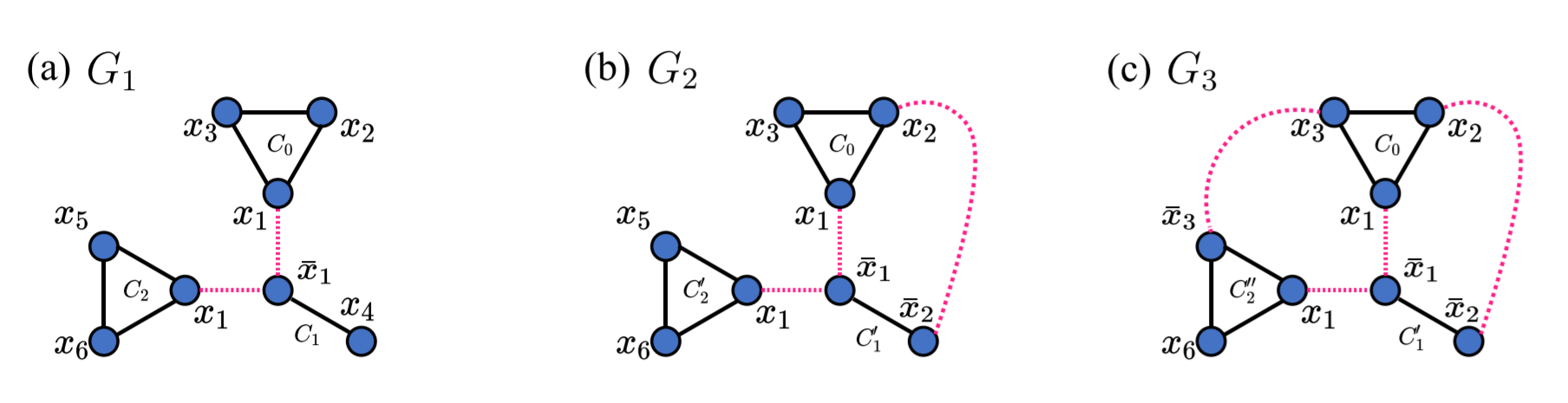}
    \caption{The figure is taken from~\cite{jeong2023quantum}. The MIS graphs reduced from the 3-SAT instance, where vertices represent literals $(x_1, \cdots, x_6  \text{ and negations})$, solid edges intra-clause logics, and dashed edges the inter-clause logics (between literals and their negations).}
    \label{fig:3-SAT graph}
\end{figure*}

Pichler et al.~\cite{pichler2018quantum} pointed out that one may convert the 3-SAT problem into the Maximal Independent Set (MIS) problem for a given graph $G(V, E)$ as follows:
\begin{align*}
    V &= \{ \ (j, k) \ | \ l_{j,k} \in C_j\} \\
    E &= E_1 \cup E_2\\
    E_1 &= \{ \ [(j, k_1), (j, k_2)]|k_1 \neq k_2\}\\
    E_2 &= \{ \ [(j_1, k_1), (j_2, k_2)] \ | \ k_1 \neq k_2, \ l_{j_1,k_1}\neq l_{k_2,k_2}\}
\end{align*}

\noindent such that $V$ is the set of vertices of which a component $(j, k)$ corresponds to the literal $l_{j,k}$ in the $j$-th clause $C_j$. Also, $E$ is the set of all edges, namely the union of two edge sets $E_1$ and $E_2$. In particular, $E_1$ connects vertices in the same clause and $E_2$ connects vertices in different clauses negating each other. Figure \ref{fig:3-SAT graph} is an instance of the MIS graph.
The authors in~\cite{pichler2018quantum} described an architecture for quantum optimization. They further analyzed it to deal with the MIS problem using neutral atom arrays trapped within optical tweezers. As shown in~\cite{pichler2018quantum}, the solutions to MIS problems can be efficiently encoded in the ground state of interacting atoms in 2D arrays by utilizing the Rydberg blockade mechanism. The Rydberg blockade mechanism prevents the simultaneous excitation of nearby atoms to the state of Rydberg (excited state) by the same laser pulse. This happens when two atoms, the target, and the control, are in proximity. 
One may describe the terms of energy of the Rydberg system with its Hamiltonian, which is the energy operator of every physical system. The energy terms in the Hamiltonian of the system have some intrinsic properties that make it possible to simulate MIS on them. 

Assume the state of a physical system is described by $\psi(x, t)$. So, the total energy (kinetic plus potential) of system $\mathbf{E}$, is in the form of: $$ \mathbf{E} \psi (x,t) = \hat{\mathbf{H}}(x, t) \psi(x, t)$$
$\mathbf{H}(x, t)$ is the operator of total energy, called Hamiltonian~\cite{griffiths_schroeter_2018}. The spectrum of a system is the different eigenvalues that can show how much energy the design has.

As can be seen in~\cite{pichler2018quantum}, the equation below describes the Hamiltonian of many body Rydberg atoms: $$\hat{H}(t) = U \sum_{(j, k)\in E}\hat{n}_j\hat{n}_k - \frac{\hbar}{2} \sum_{j \in V} \left( \delta(t)\hat{\sigma}_j^z - \Omega(t) \hat{\sigma}_j^x \right)$$

In the above equation, $U$ is the van der Walls interaction between nearest-neighbour in fixed distance $d$, $\delta$ is the time-dependent laser detuning,  and $\Omega$ is the time-dependent Rabi frequency~\cite{kim2022rydberg}. For short, laser detuning is adjusting the laser to a slightly different frequency than the frequency of the quantum system. Also, Rabi frequency is the frequency at which the energy level of an atom fluctuates in an electromagnetic field.


Dörn~\cite{doern2007quantum} presented a reduced and upper limit of quantum complexity in terms of independent set problems for graphs. Specifically, the author provided quantum algorithms to compute a graph's maximal and maximum independent sets.  The results have improved the best classical bounds of complexity for some graph problems~\cite{doern2007quantum}.


Ebadi et al.~\cite{ebadi2022quantum} focused on the maximum independent set problem. The authors used a quantum device based on coherent, programmable arrays of neutral atoms trapped in optical tweezers. They aimed to investigate the quantum optimization algorithms for systems ranging from $39$ to $289$ qubits, in two spatial dimensions. Also, with effective depths, sufficient for the quantum correlations so that they may spread across the graph. Accessible by quantum evaluation, the useful depth is bounded by Rydberg state lifetime and intermediate-state laser scattering. One may subdue this by increasing the control laser intensity and intermediate-state detuning~\cite{ebadi2022quantum}.

Serret et al.~\cite{Serret_2020} calculated the quantified requirements on system size and noise levels that platforms must fulfil to achieve quantum benefits in solving the problem. The authors computed the average approximation ratio, using noisy simulations of Rydberg platforms of up to twenty-six atoms. The atoms interacted through realistic van der Waals interactions. The above can be attained with a simple quantum annealing-based heuristic within a fixed temporal computational budget. For near-future noise levels, the authors found that ratios of at least $0.84$ can be reached for approximation. Consequently, quantum advantages can be obtained by two possible hardware-related conditions: $(i)$ An increase in the size of the quantum system to a few thousand atoms; $(ii)$ An improvement in the quantum system's coherence properties. The authors emphasized that one might achieve a higher quantitative approximation ratio by pure algorithmic improvements, namely, beyond the simple quantum annealing method used in~\cite{Serret_2020}.


Kim et al.~\cite{kim2022rydberg} introduced a programmable quantum-wired Rydberg simulator for Kuratowski subgraphs and a six-degree graph. These are instances of non-planar and high-degree graphs, respectively. Using the above simulator, the authors stated that one may find solutions for the maximum independent set (MIS) problem. MIS aims to choose the maximum number of vertices s.t. no edges connect them directly to each other. The authors in~\cite{kim2022rydberg} showed that the ground state (the state with minimum energy) of the Rydberg atom system is equivalent to the solution to the MIS. 

\begin{figure*}[t]
    \centering
    \includegraphics[width=300pt]{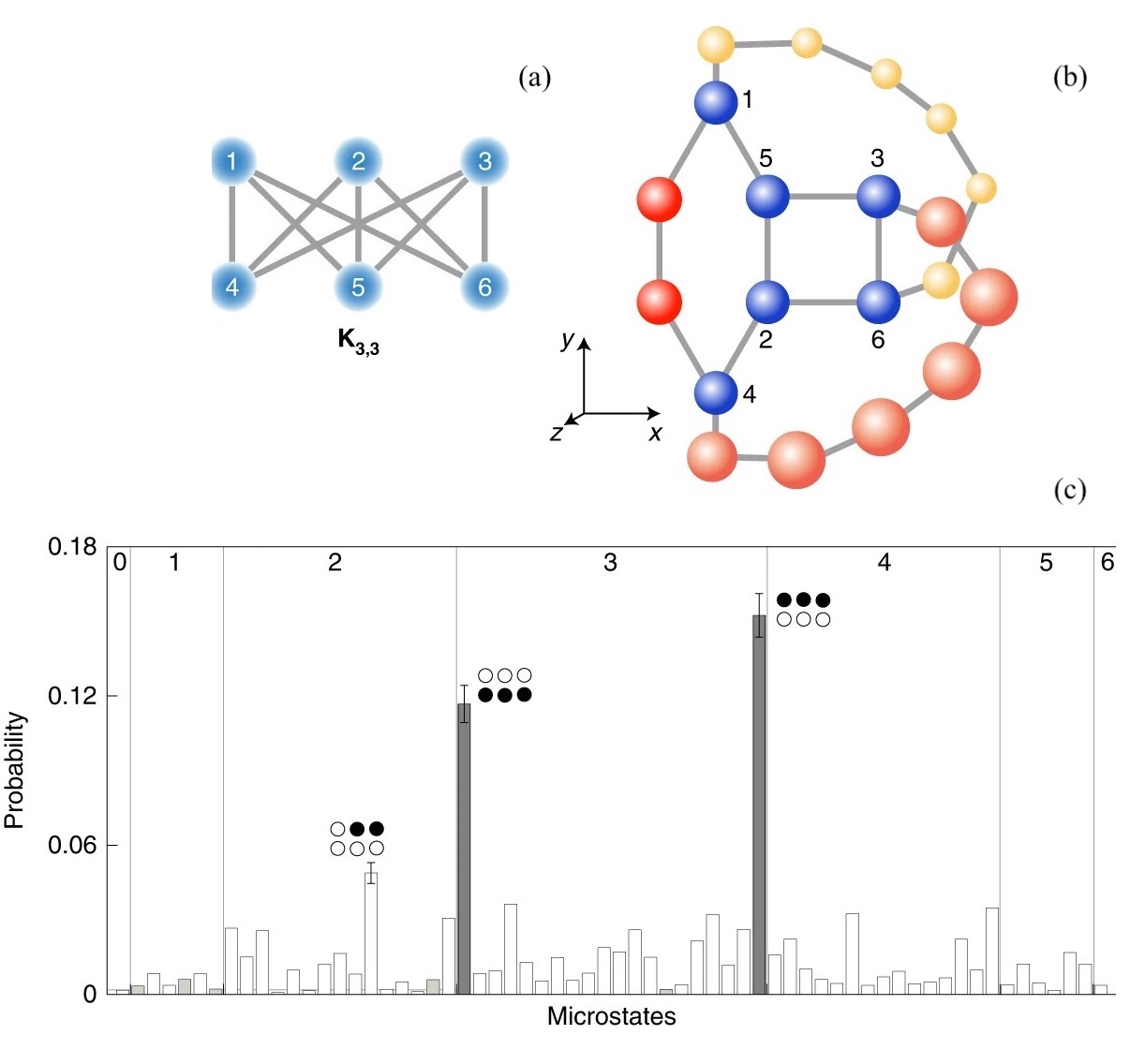}
    \caption{Taken from ~\cite{kim2022rydberg}\\
    a) $K_{3, 3}$ Graph, b) experimental $K_{3,3}$ in 3D (blue circles) making by quantum wires. c) The experimental probability distribution of $K_{3,3}$ be the solution of MIS}
    \label{fig: Rydberg atom}
\end{figure*}

There are two limitations to the Rydberg atoms system as a quantum simulator. First, one may not simulate non-planar graphs with 2D Rydberg atoms. Second, graphs with high-degree vertices cannot be encoded. This is because the blockade radius sets the size of Rydberg atom interactions. Consider the quantum wires as chains of auxiliary wire atoms that mediate strong interactions between distant atoms. Using quantum wires, one can tackle MIS~\cite{kim2022rydberg} (Figure \ref{fig: Rydberg atom}). 


Finding an independent set with the largest possible size for a given graph is required to solve the maximum independent set problem. It is known that this problem is NP-hard, which means that an efficient algorithm is unlikely to find the maximum independent set of graphs. The authors provided the first results of experiments involving arbitrarily structured MIS inputs using a D-Wave QPU (Commercial quantum processing unit), which is increasingly used to solve complex combinatorial optimization problems. The approach of Yarkoni et al.~\cite{8477865} shows how one may map MIS to QUBO problems. In their work, the authors had a limiting factor, the number of QPU (Quantum Processing Unit) parameters to be tuned to optimize performance~\cite{8477865}.


Chang et al.~\cite{chang2021quantum} propose a bio-molecular algorithm for solving the independent-set problem for every graph $G(V, E)$ s.t. $|V| = n$ and $|E| = m$. The algorithm requires $\mathcal{O}(n^2 + m)$ biological operations, $\mathcal{O}(2^n)$ DNA strands, $\mathcal{O}(n)$ tubes and the longest DNA strand, $\mathcal{O}(n)$. The authors demonstrate a new kind of Boolean circuit inferred from the biomolecular solutions with $m$ NAND gates, $(m + n(n +1))$ AND gates, and $\frac{n(n+1)}{2}$ NOT gates. The circuit can find the maximal independent-set(s) to the independent-set problem for every graph $G(V, E)$. Chang et al.~\cite{chang2021quantum} illustrated a new kind of quantum-molecular algorithm. The algorithm can find the maximal independent set(s) with the lower bound $\Omega(2^{\frac{n}{2}})$ queries and the upper bound $\mathcal{O}(2^{\frac{n}{2}})$ queries.

The quantum lower bound of solving the element distinctness problem was previously $\Omega(2^{\frac{2n}{3}})$ queries, in case of using a quantum walk algorithm. The element distinctness problem is determining the distinctness of every element of a list. The problem with the input of $n$ bits can be determined by $2^n$ comparisons. The authors in~\cite{chang2021quantum} demonstrate that their quantum-molecular algorithm improves the quantum lower bound to $\mathcal{O}(2^{\frac{n-1}{2}})$ queries.


One may apply QAOA to combinatorial search problems on graphs. The quantum circuit has $p$ applications of a unitary operator. This respects the locality of the graph. Consider a graph with a bounded degree. Having $p$ small enough, measurements of distant qubits give uncorrelated results in the state output by the QAOA. Farhi et al.~\cite{farhi2020quantum} considered random graphs with $dn/2$ edges. They kept $d$ fixed and let $n$ grow large. Specifically, they focused on finding large independent sets in such graphs. Suppose that $p$ is less than a d-dependent constant time $\log n$. The authors prove that the QAOA performs within a factor of $.854$ of the optimal answer, in terms of size, for $d$ large. They use the Overlap Gap Property of almost optimal independent sets in random graphs. Also the locality of the QAOA. The logarithm is slowly growing. So, even at one million qubits, one may only prove that the algorithm is blocked if $p$ is in single digits. At higher $p$, the algorithm \emph{sees} the whole graph. Thus, one cannot indicate that implies a limitation of performance.


Farhi et al.~\cite{farhi2020quantum2} use a Quantum Approximate Optimization Algorithm, as a local algorithm, for searching problems on graphs with the cost function of sums corresponding to edges. When defining an edge term, the QAOA unitary at depth $p$ provides an operator that depends only on a subgraph consisting of edges that are at most $p$ away from the edge in question.

Consider $d$ a fixed number and $p$ as a small constant time $\log{n}$. On random $d\text{-regular}$ graphs, these neighbourhoods are almost all trees. So, one may determine the performance of the QAOA by its actions taken on an edge in the middle of the tree. 

Both bipartite and general random $d\text{-regular}$ graphs are trees, locally. Thus, the QAOA’s performance remains the same in both cases. Consequently, one may show that QAOA with $(d - 1)^{2p} < n^A$ for all $A < 1$ and a large $d$, approximates Max-Cut on bipartite random $d\text{-regular}$ graphs with the ratio of $\frac{1}{2}$. Maintaining the same setting, for the Maximum Independent Set, the best approximation ratio is a $d\text{-dependent}$ constant. It goes to $0$ as $d$ gets larger~\cite{farhi2020quantum2}.

Consider finding approximate solutions to cases of the Maximum Independent Set problem with nearly twice as many nodes as the number of qubits available on a single quantum processor. Saleem et al.~\cite{saleem2021quantum} simulated the implementation of the quantum divide and conquer algorithm for the above. They introduced a quantum divide and conquer algorithm that enables the use of distributed computing for constrained combinatorial optimization problems. Three major components of the algorithm are as follows: $(1)$ classical partitioning of a target graph into multiple subgraphs, $(2)$ variational optimization over these subgraphs $(3)$ a quantum circuit cutting procedure that allows the optimization to take place independently on separate quantum processors~\cite{saleem2021quantum}.

\subsection{Graph Traversal Problems}

\begin{definition}[Hamiltonian Cycle Problem]
A Hamiltonian cycle is a cycle in a graph $G(V, E)$ that passes through every vertex $v \in V$. $G$ is said to be Hamiltonian if it contains at least one Hamiltonian cycle. 
\end{definition}


For graph traversal problems on quantum computers, Dörn in~\cite{dorn2007quantum} studied the complexity of such algorithms. In particular, he explored Eulerian tours, Hamiltonian tours, traveling salesman problems, and project scheduling. The author showed quantum algorithms and quantum lower bounds for these problems. In particular, the author proved that the quantum algorithms for the Eulerian tour and the project scheduling problem are optimal in the query model.


Ge and Dunjko~\cite{Ge_2020} find an approach that can be used to improve Eppstein’s algorithm for the cubic Hamiltonian cycle problem and reach a polynomial speedup for any fraction of the number of qubits to the size of the graph. The Eppstein algorithm is a graph theory algorithm that finds the number of nearest paths that allow cycles to be carried out between n given pair of vertices on a direct graph. The Cubic Hamiltonian Cycle is in a cube graph, where each vertex has the 3rd degree.
The authors also introduce the divide-and-conquer hybrid approach. They outline the general criteria for the kind of classical algorithms to which their framework is applicable. The results in~\cite{Ge_2020} also imply that one may achieve a trade-off between the speedup the authors obtain, and the size of the Hamiltonian cycle problem. 
Two key issues currently prevent our algorithms from being practical, asymptotic speedups and the assumption of ideal noiseless settings~\cite{Ge_2020}.


Mahasinghe et al.~\cite{mahasinghe2019solving} review the quantum computational methods, before their work, for solving the Hamiltonian cycle problem in different computational frameworks. For instance, quantum circuits, quantum walks, and adiabatic quantum computation. The authors then present a QUBO formulation, suitable for adiabatic quantum computation, on a D-Wave architecture. Further, they derive a physical Hamiltonian from the QUBO formulation and discuss its adequateness in the adiabatic framework. Finally, the authors discuss the complexity of running the Hamiltonian cycle QUBO on a D-Wave quantum computer and compare it to existing quantum computational methods.


For realizing quantum algorithms, Jiang~\cite{jiang2022quantum} suggested the notion of explicit, as well as implicit oracle. The author further provides how to construct the quantum circuit of the Grover algorithm with the explicit oracle to solve the Hamiltonian Cycle Problem for the 4-clique and the 5-clique complete graphs. The results suggested quadratic speed-up over the classical unstructured search algorithm. Their implementation and execution were on an IBM quantum computer simulator.


QUBO can be seen as a generic language for optimization problems. QUBOs can be solved with quantum hardware. Like quantum annealers or quantum gate computers running QAOA. The authors in~\cite{cook2020quantum} introduced two QUBO formulations for $k-\text{SAT}$ and Hamiltonian Cycles. Their formulation scales significantly better. For $k-\text{SAT}$, the authors improved the bound on of the QUBO matrix from $\mathcal{O}(k)$ to $\mathcal{O}(\log k)$. or Hamiltonian Cycles, the matrix grows linearly in the number of edges and logarithmically in the number of nodes. The authors showed these two formulations as meta-algorithms. The meta-algorithms facilitate the design of more complex QUBO formulations. They also allow convenient reuse in larger and more complex QUBO formulations.


In terms of $\mathbf{Z}_2$ lattice gauge theory (LGT), Cui and Shi~\cite{cui2022solving} introduced an approach defined on the lattice with the graph as its dual. Let $g$ be the coupling parameter. The ground state becomes a superposition of all configurations with closed strings of spins in the same single spin state, only if $g$ is smaller than $g_c$ (the critical value). It can be obtained by using an adiabatic quantum algorithm with time complexity $\mathcal{O}(\frac{1}{g^{2}_{c}} \sqrt{\frac{1}{\varepsilon} N^{3/2}_{e}(N^{3}_{v} + \frac{N_e}{g_c})})$ s.t. $N_v$ and $N_e$ are the numbers of vertices and edges of the graph, respectively. One may perform a subsequent search among the closed strings to solve the HC problem. The authors show that for some random samples of small graphs, the dependence of the average value of $g_c$ on $\sqrt{N_{hc}}$ is linear s.t. $N_{hc}$ is the number of HCs. The same dependency holds for the average value of $\frac{1}{g_c}$ on $N_e$. So, the authors suggest that the HC problem may be solved in polynomial time for some graphs. They also discuss a possible quantum algorithm using $g_c$ to infer $N_{hc}$.

Consider the action of adjacency matrices, also combinatorial Laplacians on the vector spaces spanned by the graph’s vertices. Operators are accordingly induced on fermion and zeon algebras. Properties of the algebras naturally provide information on the graph’s spanning trees. Further, it gives knowledge about vertex coverings by cycles \& matching. A fermion-zeon convolution results from combining the properties of operators induced on fermions-zeons. For an arbitrary graph, it recovers the number of Hamiltonian cycles. The results obtained by Staples~\cite{staples2017hamiltonian} offer a deeper physical interpretation of the results in~\cite{schott2008nilpotent,staples2008new}, developed using nilpotent adjacency matrices.


Consider the problem of graph partitioning into subgraphs containing Hamiltonian cycles of constrained length. Cocchi et al.~\cite{cocchi2021graph} illustrate in their work that one may use a quantum annealer to solve the above NP-complete problem. They introduce a method called vertex 3-cycle cover. It aims to find a partition of a provided directed graph into Hamiltonian subgraphs with three vertices or more. The authors structured the problem in terms of a quadratic unconstrained binary optimization. Next, they executed it on a D-Wave Advantage quantum annealer. Their tests were on synthetic graphs s.t. each one was constructed by adding several random edges to a set of disjoint cycles. Cocchi et al. proved that the cycle length does not affect the probability of solution. Also, for graphs up to $4000$ vertices and $5200$ edges, they illustrated that a solution was found, close to the number of physical working qubits available on the quantum annealer.




One of the most famous problems in graph theory is the travelling salesman problem. Which closely relates to the Hamiltonian Cycle problem. However, the extent to which quantum computers can speed up algorithms to resolve this problem is not yet known. Moylett et al.~\cite{moylett2017quantum} use a quantum backtracking algorithm to a classical algorithm by Xiao and Nagamochi to demonstrate a quadratic quantum speedup when the degree of each vertex is at least 3. Then, to speed up the classical algorithm for when the degree of each vertex is at least 4, authors use similar techniques to speed up the higher degree graphs by reducing the instances~\cite{moylett2017quantum}.

Salehi et al.~\cite{salehi2022unconstrained} analyze the travelling salesman problem with time windows (TSPTW) in the context of quantum computation. They introduced quadratic unconstrained and higher-order binary optimization formulations as well as the advantages of edge-based and node-based formulations of TSPTW. In addition, the authors conducted experimental realizations on their methods on a quantum annealing device.


 Warren~\cite{warren2013adapting} shows how to guide a quantum computer in selecting an optimal tour for the traveling salesman.  Adiabatic quantum computers are excellent machines for solving TSPs. They solve discrete optimization problems and have computational capacity and speed attributed to their quick selection of a near-optimal outcome from an exponential number of possibilities.  There are currently limited numbers of binary variables that can be handled by adiabatic quantum computers. This is because, in the Ising model, binary variables are mixed with physical quantum bits. As the number of quantum bits grows, it is no longer possible to build a connection between them through existing practical development methods. This means that the size of TSPs that can be implemented is limited~\cite{warren2013adapting}.


Using the phase estimation technique, Srinivasan et al.~\cite{srinivasan2018efficient} proposed a quantum algorithm to solve TSP. They encoded the given distances between the cities (in the context of TSP) as phases. The eigenvectors of the unitary operators that the authors constructed are computational basis states. Also, the eigenvalues of the operators are various combinations of these phases. They applied the phase estimation algorithm to certain eigenstates. This provides all of the total distances possible for every route. Having obtained the distances, one may search through this information to find the least possible distance, also the route taken. This is acquired by the quantum search algorithm for finding the minimum~\cite{durr1996quantum}. For a large number of cities in TSP, the above technique offers a quadratic speedup over the classical brute force method. In their work, Srinivasan et al. illustrated an example of TSP by taking four cities. They presented their results by simulating the codes in IBM’s quantum simulator.

Kieu~\cite{kieu2019travelling} introduced an algorithm for the traveling salesman problem, in the framework of Adiabatic Quantum Computation (AQC). In their work, the initial Hamiltonian for the AQC process recognized canonical coherent states as the ground state. Also, the target Hamiltonian has the shortest tour as the desirable ground state. Further, the author provides estimates/bounds for the computational complexity of the algorithm. Other than the time and space complexity, they emphasized the required energy resources for the physical process of (quantum) computation.


One may expect the noisy intermediate-scale quantum computers (NISQ) to run quantum circuits, introducing up to a few hundred qubits. In terms of constraints related to allocating qubits and running multi-qubit gates, the circuits need to comply with the architecture of NISQ. 

The Quantum circuit compilation (QCC) outputs a compatible circuit by taking a nonconforming circuit. Two types of operations can solve Compilation as a known combinatorial problem. The first one is \emph{qubit allocation} and the second one is \emph{gate scheduling}. Paler et al.~\cite{paler2021nisq} informally proved that the two operations form a discrete ring. Consider the search landscape of QCC. One may think of it as a two-dimensional discrete torus. In this setting, how circuit qubits are allocated to NISQ registers are represented by vertices. Further, Torus edges are weighted by the cost of scheduling circuit gates. The fact that a circuit’s gate list is circular, is incorporated, which suggests the novelty of the work of Paler et al.~\cite{paler2021nisq}. Particularly, compilation may start from an arbitrary gate as long as all the gates are processed, and the compiled circuit has the correct gate order.

\subsection{Maximum Matching on Graphs}

\begin{definition}[Matching]
A matching is a set of non-adjacent edges in an undirected graph. 
\end{definition}

\begin{definition}[Maximum Matching (MM)]
The Maximum Matching (MM) problem aims at finding the matching with the largest number of edges. 
\end{definition}


For the MM, there is a setting called the query setting: Given a graph $G$, as an adjacency matrix or adjacency list, find an MM with the minimum number of queries. A query in the matrix model is: \emph{Does vertex x share an edge with vertex y?} Alos, a query in the list model is: \emph{What is the ith vertex adjacent to vertex x?}

Let $G(V, E)$ be a graph s.t. $|V| = n$ and $|E| = m$. Kimmel et al.~\cite{kimmel2021query} used a classical maximum matching algorithm and the guessing tree method to give an $\mathcal{O}(n^{7/4})$ query bound in the matrix model. Also, they provide $\mathcal{O}(n^{3/4} \sqrt{m + n})$ query bound in the list model for maximum matching s.t. the former and the latter are both obtained for quantum computers and general graphs. The authors' result narrows the gap between the previous trivial upper bounds of $\mathcal{O}(n^2)$ and $\mathcal{O}(m)$ and the quantum query complexity lower bound of $\mathcal{O}(n^{3/2})$.


The work in~\cite{dorn2009quantum} introduces quantum algorithms for the matching problems. The author considers unweighted and weighted graphs with $n$ vertices and $m$ edges. In general graphs, the results in~\cite{dorn2009quantum} shows that finding a maximal matching takes $\mathcal{O}(\sqrt{nm} \log^{2} n)$. In addition, a maximum matching under the same setting takes $\mathcal{O}(n\sqrt{m} \log^{2} n)$. Considering a different setting, the author proves that finding a maximum weight matching in bipartite graphs is proportional to $\mathcal{O}(n \sqrt{m} N \log^{2} n)$ s.t. $N$ is the largest edge weight. Whilst a minimum weight perfect matching introduces the time complexity of $\mathcal{O}(n \sqrt{nm} \log^{3} n)$.


QA suits problems that can be formulated as QUBO problems. E.g. SAT, graph colouring, and traveling salesman. McLeod and Sasdelli~\cite{mcleod2022benchmarking} investigate the performance of D-Wave’s 2000Q (2048 qubits) and Advantage (5640 qubits) quantum annealers empirically. In particular, they benchmark the solvers based on solving a specific instance of the maximum cardinality matching problem. They used the results of a previous work that measured the performance of earlier QA hardware from D-Wave. The authors illustrated that Advantage produces optimal solutions to problems of a larger size, compared to 2000Q. Further, the authors investigated the scaling of the diabatic transition probability, by utilizing the Landau-Zener formula. In particular, they considered the structure of the problem and its implications for suitability to QA. Moreover, they introduced a method, exploring the behavior of minimum energy gaps for scalable problems deployed to QAs. The authors showed for their targeted QA problem that the minimum energy gap does not scale as desired. One might ask if this problem is suitable for benchmarking QA hardware. It might lack the necessary nuance to recognize significant performance improvements between generations.


Given an $L \cross L$ array of qubits. Fowler~\cite{fowler2013minimum} considers the minimum weight perfect matching problem. Provided $(1)$ a modular architecture s.t. there is a finite maximum number of non-functioning qubits in any given connected defective patch $(2)$ gate, leakage, and loss error rates below some set of nonzero values $(3)$ a 2-D array of processors with finite processing power, finite local memory, and a communication line with their nearest neighbour at a finite data rate $(4)$ external memory to store all detection events and matching data. This should last for the duration of a temporally finite quantum algorithm.

Fowler then proves that one may find the global optimum of the problem with $\mathcal{O}(1)$ average cost per round of error detection independent of $L$.

Witter~\cite{witter2022query} created a query-efficient quantum maximum matching algorithm by combining results in classical maximum matching and quantum algorithm design. The author provided improvement on the previous best quantum upper bound for maximum matching in general graphs i.e. $\mathcal{O}(\sqrt{m}n)$~\cite{dorn2009quantum}. And designed a quantum algorithm with query complexity $\mathcal{O}(\sqrt{m}n^{3/4})$. To present edges in Gabow’s classical maximum matching algorithm~\cite{gabow2017weighted}, the approach in~\cite{witter2022query} bounds the queries and applies the decision tree guessing theorem of Beigi and Taghavi~\cite{beigi2020quantum}.

\subsection{Maximum-Cut Problem}

\begin{definition}[Maximum Cut Problem (Max-Cut)]
    The maximum cut problem (max-cut) is a graph partitioning problem. The objective of max-cut is to first partition the set of vertices of a graph into two subsets.  Second, the aim is to maximize the sum of the weights of the edges whose endpoint(s) is in each of the subsets~\cite{Commander2009}. 
\end{definition}


Crooks in~\cite{crooks2018performance} studied the performance of QAOA on the MaxCut combinatorial optimization problem. He aimed to optimize the quantum circuits on a classical computer, using automatic differentiation and stochastic gradient descent. He also hired QuantumFlow, a quantum circuit simulator implemented with TensorFlow. The author suggested that one may amortize the training cost by optimizing batches of instances of problems. In particular, QAOA exceeds the classical polynomial time Goemans-Williamson algorithm in terms of performance, with modest circuit depth. Also, he illustrates that performance with fixed circuit depth is insensitive to problem size. Furthermore, one may implement MaxCut QAOA on a gate-based quantum computer with limited qubit connectivity. This can be done by using a qubit swap network.


Guerreschi and Matsuura~\cite{guerreschi2019qaoa} lower bounded the size of quantum computers with practical utility. They performed simulations of QAOA and concluded that: At least for a representative combinatorial problem, quantum speedup will not be attainable. This will last until several hundreds of qubits are available.


Boulebnane and Montanaro~\cite{boulebnane2021predicting} generalized techniques appeared in~\cite{farhi2022quantum}. Applied to different families of sparse random graphs, the authors evaluated the performance of $\QAOA$ in the infinite-size limit for the MAX-CUT problem. Their analysis rigorously holds only when $n \rightarrow \infty$ and large degree limit. Still, the analysis provides sufficient initial variational parameters to optimize small-scale, very sparse instances. In addition, the authors introduced a numerical algorithm. It evaluates the performance of $\QAOA$ on an average finite-size instance of the SK model. This is where the analytic methods exposed in~\cite{farhi2022quantum} fail. Nevertheless, the algorithm supports low depth only. Further, their work cannot address the D-spin model for $D \geq 3$ and $p > 1$.

The performance of $\QAOA$ improves monotonically with its depth, denoted as $p$. Basso et al.~\cite{basso2021quantum} applied the $\QAOA$ to MaxCut on large-girth $D\text{-regular}$ graphs.
The authors suggested an iterative formula, evaluating the performance for every $D$ at every depth $p$. They illustrated that the $p = 11$ $\QAOA$ outperforms other classical algorithms free of unproven conjectures, also known by the authors. They obtained the above by investigating random $D\text{-regular}$ graphs, at optimal parameters and as $D \rightarrow \infty$.

The iterative formula for $D\text{-regular}$ graphs
is derived by analyzing a single tree subgraph. Nonetheless, the authors proved that it also provides the ensemble-averaged performance of the $\QAOA$ on the Sherrington-Kirkpatrick (SK) model defined on the complete graph. 

The authors generalized their formula to Max-q-XORSAT on large-girth regular hypergraphs. The complexity of their iteration nevertheless grows proportional to $\mathcal{O}(p^2 4^p)$. This outperforms the previous procedure for analyzing $\QAOA$ and allows setting $p = 20$.

Lastly, the authors conjectured that the $\QAOA$, as $p \rightarrow \infty$, achieves the Parisi value. They analyzed the performance of their algorithm in this case. But still, one needs to put the theoretical results into practice using a quantum computer, producing a string with guaranteed performance.

\subsection{Graph Isomorphism}


\begin{definition}[Graph Isomorphism (GI)]
Two $N$-vertex graphs, $G$ and $G_0$, are presented in the Graph Isomorphism problem, and the task is to determine whether there is a permutation of the vertices of $G$ that preserves adjacency and transforms $G_0$. If yes, then $G$ and $G_0$ are considered isomorphic; otherwise, they're non-isomorphic. 
\end{definition}

For solving the subgraph isomorphism problem on a quantum computer based on a gate, Mariella and  Simonetto~\cite{mariella2023quantum} propose a novel variational method. The method is based on a new representation of the adjacency matrices of the underlying graphs, which requires several qubits, which are logarithmically scaled to the number of vertices of the graphs, and a new Ansatz, which can investigate the permutation space efficiently.  Simulations of the approach to graphs up to 16 vertices are then presented, while the approach could be applied to realistic subgraph isomorphism problems in the medium term, given the logarithmic scale~\cite{mariella2023quantum}.


 Cruz et al.~\cite{de2022quantum} introduced a new way of encoding non-directed graphs with $n$ nodes. They used quantum gates on $n$ qubits. One may set up a relation between the adjacency characteristics of the graph nodes, and the angles of each amplitude of the system's quantum state. This relation corresponds to the nodes identified by \emph{one} in the bit sequence that defines each amplitude. Applying the phase estimation algorithm provides insight into all subgraphs of the original graph. The subgraphs are classified by their number of edges. One should note that the number of subgraphs is exponential. So, the algorithm suggested in~\cite{de2022quantum} deals with an exponential number of graphs within polynomial time. This is apart from the task of measuring the system. This also provides insight into invariant(s) on graph isomorphism. The authors provided examples, aiming to illustrate the distinguishing power of their suggested algorithm. Nevertheless, it does not discriminate all graphs as they showed by counterexamples.


GI is a major computer science problem that has been recognized as similar to the problems of integer factorization. Gaitan and Clark~\cite{Gaitan_2014} presented a quantum algorithm to solve arbitrary GI instances and also presented a novel method for determining all automorphisms on that graph. It is demonstrated in~\cite{Gaitan_2014} how the GI problem can be transformed into a combinatorial optimization problem, in which adiabatic quantum evolution makes it possible to solve this problem. The authors then discuss the GI quantum algorithm’s experimental implementation, and close by showing how it can be leveraged to give a quantum algorithm that solves arbitrary instances of the NP-Complete Sub-Graph Isomorphism problem. The computational complexity of an adiabatic quantum algorithm is largely determined by the minimum energy gap $\Delta(N)$ between the ground state and first excited state in the limit of large problem size
$N\gg1$~\cite{Gaitan_2014}.AQC is a form of quantum computing that relies on the adiabatic theorem to do calculations and is closely related to quantum annealing. By the adiabatic theorem, the system remains in the ground state, so in the end, the state of the system describes the solution to the problem~\cite{dziemba2016adiabatic}.


In~\cite{9559130}, the authors presented an algorithm called GR2, as part of an initial contribution that studied and highlighted the flexibility and importance of parallel programming for classic computing and quantum computing. GR2 consists of Pruning methods and a Producer-Consumer pattern to find subgraph isomorphism. This algorithm has been designed and implemented to work with directed and undirected graphs. It may be declared that it is possible to obtain execution times for the size of query graphs used~\cite{9559130}.

Gheorghica~\cite{gheorghica2021gr3} introduced the GR3 Algorithm to find the best execution times for searching subgraph isomorphisms. The GR3's design consists of a combination of Parallel Programming and Quantum Computing. Execution times obtained are extremely low.

Li et al.~\cite{li2020qubit} proposed a new algorithm for qubit mapping, based on subgraph isomorphism and filtered depth-limited search. Their algorithm (FIDLS), can reduce the necessary extra two-qubit gates in the output circuit. Experimental results illustrated that when the circuit had less than $1000$ two-qubit gates, their subgraph isomorphism-induced initial mapping was much better than empty mappings. It also performed better than the naive mappings i.e. assigning the $i^{\text{th}}$ qubit in the logical circuit to the $i^{\text{th}}$ qubit in the quantum device. Furthermore, on a large set of $131$ benchmark circuits, their results showed that FIDLS performs significantly better than SOTA algorithms on IBM Q Tokyo i.e. the architectural graph of which has a relatively large average node degree and a smaller diameter~\cite{li2020qubit}.

Mondada and Andrés-Martínez~\cite{mondada2023subgraph} suggested an algorithm for the subgraph isomorphism problem on port graphs. Their algorithm performs pattern matching in quantum circuits for many patterns simultaneously. This is independent of the number of patterns. The running time of the algorithm is linear for the size of the input quantum circuit, after pre-computation, in which the patterns are compiled into a decision tree. Consider the connected port graphs. Such graphs introduce a property i.e. every edge $e$ incident to $v$ has a label $L_v(e)$ unique in $v$. The algorithm suggested by Mondada and Andrés-Martínez~\cite{mondada2023subgraph} enumerates the subgraph isomorphism problem $H\subseteq G$ for all $m$ pattern matches in time $\mathcal{O}(P)^{P + \frac{3}{2}} \cdot |V(G)|+\mathcal{O}(m)$ s.t. $P$ is the number of vertices of the largest pattern. Also, they can express the bound obtained with the quantum circuit in terms of the maximum number of qubits $N$ and depth $\delta$ of the patterns: 
$\mathcal{O}(N)^{N+\frac{1}{2}} \cdot \delta |V(G)| \log (\delta) + \mathcal{O}(m)$.

\subsection{Minimum Vertex Cover Problem}

\begin{definition}[Minimum Vertex Cover (MVC)]
Given an undirected graph $G=(V, E)$ with vertex set $V$ and edge set $E \subseteq V \cross V$. A subset $V^{'} \subseteq V$ is called a \emph{vertex cover} if every edge in $E$ has at least one endpoint in $V$. Namely, if for every $e = (u, v) \in E$ s.t. $u, v \in V$, it holds true that $u \in V^{'}$ or $v \in V^{'}$. A minimum vertex cover is a vertex cover of minimum size. 
\end{definition}


Pelofske et al.~\cite{pelofske2019solving} proposed a novel decomposition algorithm for the Minimum Vertex Cover (MVC) problem. Their algorithm splits a given MVC instance recursively into smaller subproblems. There is a level of recursion in which each subproblem can be solved with an arbitrary method of choice. E.g. a quantum annealer. The algorithm is exact. This implies that the optimal solution of MVC is guaranteed. It is given that all subproblems can be solved with no error. If one solves subproblems probabilistically (s.t. the error probability is at most $\varepsilon$), then the total error rate for the solution becomes at most $\varepsilon$.\\


Chang et al.~\cite{chang2014quantum} suggested a quantum algorithm for implementing Boolean circuits. Their idea is generated from their DNA-based algorithm for solving the vertex-cover problem. This algorithm is proved to be the optimal quantum algorithm for every graph $G$ with $m$ edges and $n$ vertices. The space complexity in terms of \emph{quantum bits} for their algorithm is as follows: For the worst, average, and best case, one needs 
\begin{center}
    $(2m + 3) + \frac{n^3 + 15n^2 + 26n}{6}$
\end{center}
quantum bits.\\


A generalization of QAOA is called Quantum Alternating Operator Ansatz, designed to compute approximate solutions for combinatorial optimization problems with hard constraints. Cook et al.~\cite{cook2020quantum} studied Maximum k-Vertex Cover under this ansatz due to its modest complexity. This problem is more complex than Max-Cut and Max E3-LIN2.

The approach of Cook et al.~\cite{cook2020quantum} includes $(1)$ a comparison between classical states and Dicke states as starting states, in terms of performance $(2)$ a comparison between the two XY-Hamiltonian mixing operators i.e. ring mixer and the complete graph mixer, in terms of performance $(3)$ analyzing the distribution of solutions via Monte Carlo sampling $(4)$ investigating efficient angle selection strategies.

The authors then showed that Dicke states improve performance compared to classical states. They could also provide an upper bound on the simulation of the complete graph mixer. Further, the authors found that the complete graph mixer improves performance relative to the ring mixer. In addition, their numerical results indicated the standard deviation of the distribution of solutions decreases exponentially in $p$ s.t. $p$ is the number of rounds in the algorithm. This implies an exponential number of random samples when aiming to find a better solution in the next round. Lastly, the authors showed a correlation between angle parameters. This presents high-quality solutions, behaving similarly to a discretized version of the Quantum Adiabatic Algorithm.

Zhang et al. in~\cite{zhang2022applying} provided a quantum circuit solution scheme. They used an algorithm of quantum approximate optimization. A quantum Ising equation and Hamiltonians are obtained based on the Ising model quantized by rotation operators and Pauli operators. Also, a parametric unitary transformation is acquired between the original Hamiltonian and the problem Hamiltonian, as a generator. The final quantum state and the problem of Hamilton's expectation are generated by a continuous evolution of two parametric unitary transformations. The results of simulations in~\cite{zhang2022applying} indicate that the model may solve problems with large probability in linear time. It can further achieve a quintuplet acceleration as well as some feasibility, and efficiency.\\

Wang et al.~\cite{wang2023quantum} presented a quantum solution based on the Grover search algorithm. They also concluded that the quantum algorithm has a square root order of magnitude improvement in efficiency compared to classical computational methods. In this process, Oracle's task is to identify a target solution whereas Grover's job is to increase the probability that it will be found~\cite{wang2023quantum}.

 \subsection{Set Cover in Graphs}
Optimization problems in computer science require finding the best solution from a set of possible changes. To solve an optimization problem using quantum annealing, the logic graph must be embedded into a known hardware graph. Minor embedding is a technique used in adiabatic quantum computation that involves finding a subgraph of a quantum hardware graph, such that a given graph can be obtained from it by contracting edges. To reduce the complexity of the minor embedding problem, the authors introduced a minor set cover (MSC) of a known graph $\mathcal{G}$: a subset of graph minors containing any remaining minor of the graph as a subgraph. All graphs that can be embedded in $\mathcal{G}$ will be embedded in a member of the MSC. Hamilton and Humble~\cite{hamilton2017identifying} demonstrate that the complete bipartite graph $K_{N, N}$ has an MSC of $N$ minors, from which $K_{N+1}$ is identified as the largest clique minor of $K_{N, N}$.

Whereas the register elements are represented by quantum physical subsystems that can store qubits of information, the continuous-time evolution depends explicitly on a Hamiltonian that defines the interactions between register elements~\cite{Kaminsky2004}. An ideal Hamiltonian may allow for arbitrary interactions between elements, but physical and technological limitations often prevent the fabrication of
arbitrary interactions or forms of connections in actual devices. A prominent example is found in the quantum annealer developed by D-Wave Systems, Inc.~\cite{Johnson2011Quantum}, which uses a well-defined hardware connectivity graph called the Chimera lattice and implements problems that can be described using the Ising Hamiltonian with two-body interactions
Cai, MacCready, and Roy (CMR) have presented a randomized algorithm for generating an embedding~\cite{cai2014practical}. To find the shortest path between randomly mapped logical vertices, their approach is based on Dijkstra's algorithm. The method is not guaranteed to work and has the worst case complexity, which scales as $\mathcal{O}(n^9)$ with the input graph order n, although the average case behaviour appears to be $\mathcal{O}(n^3)$.
The CMR embedding algorithm represents a significant portion of the time needed for
a quantum annealing workflow, and for even modest problem sizes it can far exceed the time required for executing a quantum annealing schedule~\cite{Humble_2016}.

\subsection{Graph Coloring}



\begin{definition}[Quantum Chromatic Number (QN)]
    Consider two separated proves, an interrogator, and a given graph $G(V, E)$. The quantum chromatic number (QN) is the minimal number of colours necessary in a protocol with which the provers may assure the interrogator that they have a colouring of $G$. 
\end{definition}

Cameron et al.~\cite{cameron2006quantum} investigated the quantum chromatic number of a graph. The authors then focused on establishing relations between the clique number (CN) and orthogonal representations of $G$. They showed several general facts about this graph parameter as well. A large separation was also found between CN and the QN by considering random graphs. They also proved that there can be no separation between classical and quantum chromatic numbers if the latter is $2$, nor if it is $3$ in a restricted quantum model. In addition, they exhibited a particular case s.t. $|V| = 18$, $|E| = 44$, $CN = 5$, and $QN = 4$.


Bravyi et al.~\cite{bravyi2022hybrid} proved a way to apply the Recursive Quantum Approximate Optimization algorithm (RQAOA) to MAX-k-CUT. Namely, the problem of finding an approximate vertex k-coloring of a graph. The authors performed a comparison between the best-known classical and hybrid classical-quantum algorithms. They initially showed how $\QAOA$ fails to solve the above optimization problem for most of the regular bipartite graphs at every constant level $p$.The approximation ratio one may obtain by using $\QAOA$ hardly outperforms a random assignment of colours to the vertices. 
As their next step, the authors constructed a classical simulation algorithm to simulate level-1 $\QAOA$ and level-1 RQAOA for arbitrary graphs. They had up to $300$ qutrits which were applied to randomly generated 3-colorable constant-degree graphs. The simulation showed that level-1 RQAOA is competitive. That is, for the considered groups of graphs, the approximation ratios were \emph{usually} higher than the generic classical algorithm based on rounding an SDP relaxation. The authors conjecture that higher-level RQAOA may be a beneficial algorithm for NISQ devices.

Using classical Mycielski transformation, one may preserve the size of the largest clique in a new graph $G^{'}$ constructed from a given one $G$, but allow an arbitrary chromatic number for $G^{'}$. Bochniak and Pawe{\l}~\cite{bochniak2023quantum} introduced an analog of the above transformations for quantum graphs. The authors explored how the transformations affect the (quantum) chromatic number. They also investigated the clique numbers associated with them.

\subsection{Bisection Problem}
\begin{definition}[Max-Bisection Problem]
Max-bisection problem aims to divide the vertices of a given graph with an even number of vertices into two subsets of the same size, maximizing the number of edges between the two subsets. 
\end{definition}

\begin{definition}[Min-Bisection Problem]
The min-bisection problem aims to divide the vertices of a given graph with an even number of vertices into two subsets of the same size, minimizing the number of edges between the two subsets. 
\end{definition} 

Younes~\cite{younes2015bounded} introduced a bounded-error quantum polynomial time (BQP) algorithm for the max-bisection and the min-bisection problems. This algorithm runs in $\mathcal{O}(m^2)$ for a graph with $m$ edges and in the worst case runs in $\mathcal{O}(n^4)$ for a dense graph with $n$ vertices. The author represents both problems as Boolean constraint satisfaction problems s.t. the set of satisfied constraints is simultaneously maximized/minimized. An iterative partial negation and partial measurement technique do this. The algorithm targets a general graph. It also achieved a probability of success $1 - \varepsilon$ for $\varepsilon > 0$, using polynomial space resources.

\subsection{Finding Cliques in Graphs}



\begin{definition}[Maximum Clique Problem (MCP)]
The maximum clique problem determines in a graph a clique (i.e.,
a complete subgraph) of maximum cardinality~\cite{wu2015review}. 
\end{definition}

Pelofske et al.~\cite{pelofske2019solving} explored methods to decompose larger instances of MCP into smaller ones. This may be solved subsequently on D-Wave. While decomposing, the authors targeted pruning subproblems that do not contribute to the solution. The authors suggested reduction methods, containing upper and lower-bound heuristics in conjunction with graph decomposition, vertex and edge extraction, and persistency analysis.

Consider the Connected $\mathcal{F}\text{-Deletion}$ problem in which $\mathcal{F}$ is a fixed finite family of graphs. The aim is to compute a minimum set of vertices (alternatively, a vertex set of size at most $k$ for a given $k$) s.t. the set provides a connected subgraph of the given graph. Also, deleting this set excludes every $F \in \mathcal{F}$ as a minor. This problem is known for excluding a polynomial kernel subject to standard complexity hypotheses, even in special cases e.g. $F = \{\mathcal{K}_2\}$, namely Connected Vertex Cover.
Ramanujan in~\cite{ramanujan2021approximate} introduces a $(2 + \varepsilon)\text{-approximate}$ polynomial time compression for the Connected $\mathcal{F}-\text{Deletion}$ problem. In their settings, $F$ contains at least one planar graph. As a corollary, the author suggests an approximate polynomial compression result for the special case of Connected $\eta\text{-Treewidth Deletion}$.


Miyamoto et al.~\cite{miyamoto2020quantum} showed how to apply the combination of quantum search with classical dynamic programming to the minimum Steiner tree problem. The complexity of their quantum algorithm is $\mathcal{O}((1.812)^{k} poly(n))$ s.t. $n$ is the number of vertices in the graph, and $k$ shows the number of terminals.


Kerger et al.~\cite{kerger2023asymptotically} initially addressed the following question: \emph{What problems can be solved efficiently in Congested and Congested Clique quantum models than in the classical ones?} First, they presented two algorithms in the Quantum Congested Clique model of distributed computation. Their algorithms offer a high probability of success. One of the algorithms approximates the optimal Steiner Tree. The other one computes an exact minimum spanning tree for the directed setting. Both algorithms require $\Tilde{\mathcal{O}}(n^{1/4})$ communication rounds,  and $\Tilde{\mathcal{O}}(n^{9/4})$ messages s.t. $n$ is the number of nodes in the network. The authors provide a hybrid setting with classical algorithmic frameworks and quantum subroutines. Namely, the heart of the asymptotic speedup the authors obtain is by hiring an existing framework for using a distributed version of Grover’s search algorithm. 
In addition, the authors characterized the constants and logarithmic factors in their algorithms. Also, for the related algorithms. Such factors were otherwise obscured by $\Tilde{\mathcal{O}}$ notation.

Consider a random graph $G$.  Each edge in an $n\text{-vertex}$ random graph appears with probability $\frac{1}{2}$. Also, a clique is a completely connected subgraph of $G$.
Childs et al.~\cite{childs2000finding} illustrated the results of a numerical study of the problem of finding the largest clique in a random graph. The authors investigated small graphs i.e.  $(n \leq 18)$. The quantum algorithm in~\cite{childs2000finding} appeared to require only a quadratic runtime.


\subsection{Graph and String Edit Distance}

\begin{definition}[Graph Edit Distance (GED)]
GED measures the degree of (dis)similarity between two graphs in terms of the operations needed to make them identical. 
\end{definition}

\begin{definition}[String Edit Distance Problem]
The edit distance between two strings is the smallest number of insertions, deletions, and substitutions necessary to convert one string to another. 
\end{definition}


An important notion of distance for graph-shaped data is the Graph Edit Distance (GED).
As the complexity of computing GED is the same as NP-hard problems, it is reasonable to consider approximate solutions. Incudini et al.~\cite{Incudini_2022} presented a QUBO solution to the GED problem. This gives us two possibilities to implement quantum annealing and variation algorithms running on both types of quantum hardware that exist today: quantum annealer and gate-based quantum computers. Quantum annealing is a process using quantum fluctuations to optimize for finding the general minimum of an objective function over some set of possible solutions.


Boroujeni et al.~\cite{boroujeni2021approximating} introduced a quantum constant approximation algorithm to compute the edit distance in \emph{truly} subquadratic time. The algorithm's runtime is proportional to $\mathcal{O}(n^{1.858})$ in terms of quantum computation. It approximates the edit distance by a factor of seven. Also, the authors extended
the above result to an $\mathcal{O}(n^{1.781})$ quantum algorithm, approximating the edit distance within a larger constant factor.

\subsection{String-Related Problems}


\begin{definition}[Longest Common Extension(LCE) (a.k.a. Longest Common Subsequence (LCS))]
    Given a string $T \in \Sigma^{n}$. Preprocess $T$ into a data structure $D$. So, one may answer queries like $\text{LCE}(i, j)$ s.t. for every given $i, j \in [n]$, $\text{LCE}(i, j)$ is the length of the longest common prefix between $T[i \cdot  \cdot n]$ and $T[j \cdot \cdot n]$(See also~\cite{10.1145/3313276.3316368}). 
\end{definition}

A variant of LCE is provided with a threshold of $d$. It asks whether two length-n input strings have a common substring of length $d$. The two corner cases of $d=1$ and $d=n$, correspond to Element Distinctness and Unstructured Search, respectively. Jin and Nogler~\cite{jin2023quantum} proved that the complexity of LCS with threshold $d$ smoothly hovers between the two extreme cases up to $n^{o(1)}$ factors. The LCS with threshold $d$ has a quantum algorithm with time and query complexity of $\frac{n^{2/3+o(1)}}{d^{1/6}}$. Also, a lower bound of $\Omega(\frac{n^{2/3}}{d^{1/6}})$ holds for the quantum query complexity~\cite{jin2023quantum}. The above results suggest improvement over the previous upper bounds of $\Tilde{\mathcal{O}}(\min \{\frac{n}{d^{1/2}}, n^{2/3} \})$ obtained in~\cite{legall_et_al:LIPIcs.ITCS.2022.97, doi:10.1137/1.9781611977073.109}. The main contribution in terms of technicality in~\cite{jin2023quantum} is the quantum speed-up of the \emph{Synchronizing Set} appeared in~\cite{10.1145/3519935.3520061}.  This technique implies a near-optimal LCE data structure under the quantum setting as well. An application of the quantum string synchronizing set algorithm suggested in~\cite{jin2023quantum} appears in the k-mismatch matching problem. That is, given a pattern, the problem asks if it has an occurrence in the text with at most $k$ hamming mismatches. Jin and Nogler used a structural result obtained previously in~\cite{9317938}. They showed that there is a quantum algorithm for the $k\text{-mismatch}$ matching with $k^{3/4} n^{1/2 + o(1)}$ query complexity. The algorithm also suggests a time complexity proportional to $\Tilde{\mathcal{O}}(k n^{1/2})$. Further, the authors provided an observation on the non-matching quantum query lower bound. Namely, the bound of $\Omega(\sqrt{kn})$.

\subsection{Chemistry and Bio-informatics problems}


Kassal et al.~\cite{kassal2011simulating} addressed how a quantum computer can be hired for the simulation of chemical systems and their properties. They discussed several algorithms with significant advantages for protein folding, known to be NP-hard~\cite{dill1985theory}, and several other problems. Kassal et al. focused on the adiabatic and circuit models. Nevertheless, they point out that these are not the only universal models of quantum computation. One might make further algorithmic progress with other models; Such as topological quantum computing~\cite{kitaev2003fault,nayak2008non}, one-way quantum computing~\cite{raussendorf2003measurement,raussendorf2001one}, and quantum walks~\cite{kempe2003quantum,childs2009universal,lovett2010universal}.


Either experimental or computational methods can be used to study Site-specific DNA-protein interactions. In modern molecular biology, comprehending the relation between the above two approaches, also finding ways to improve them is challenging. To use quantum computers in DNA motif model discovery, a relevant quantum algorithm is necessary. Based on the quantum adiabatic theorem, Cao et al.~\cite{cao2015adiabatic} obtained an adiabatic quantum algorithm and its application to DNA motif model discovery. Their results are in terms of the MISCORE-based motif score. One may use the method to discover DNA motif models with big data by a quantum computer, e.g. D-Wave.


Consider the problem of computing the number of elements in each hitting-set in an instance of the hitting-set problem. Chang et al.~\cite{chang2010quantum} showed that the quantum variant of bio-molecular solutions may play the oracle's role in Grover’s algorithm. Namely, the target state labelling, getting executed before Grover’s searching steps. The authors also performed a three-qubit nuclear magnetic resonance (NMR) experiment to solve the simplest variant of the hitting-set problem.

\subsection{Mixed-Integer Programming}

\begin{definition}[Mixed-Integer Programming (MIP)]
A linear mixed integer program is an optimization problem in which a nonempty subset of integer variables (unknowns) and a subset of real-valued (continuous) variables exist, the constraints are all linear equations or inequalities, and the objective is a linear function to be minimized (or maximized)~\cite{wolsey2007mixed}. 
\end{definition}


Graph minor-embedding problems enjoy the presence of Integer programming (IP) approaches for solving them. Attained from the polynomial equations of Dridi et al.~\cite{dridi2018novel}, the authors in~\cite{bernal2020integer} developed a monolithic IP and a decomposition approach. Both of the above can identify infeasible instances and provide bounds on the quality of the solution. They are agnostic of the source and target graphs. Both approaches were implemented by the authors in~\cite{bernal2020integer} and tested using a range of different source graphs. The graphs had various sizes, densities, and structures. The graphs used by the authors followed the architecture of the chips in D-Wave annealers. The method in~\cite{bernal2020integer} is nevertheless slower overall than another existing heuristic method~\cite{cai2014practical}. Still, the method has proved to be a better approach for highly structured source graphs. Namely, when the heuristic fails with a higher probability.


Ajagekar et al.~\cite{ajagekar2022hybrid} developed QC-based solution strategies, exploiting quantum annealing and classical optimization techniques. This helps solve large-scale scheduling problems in manufacturing systems. They illustrated the applications of their algorithms in two case studies in production scheduling: First, a hybrid QC-based solution for the job-shop scheduling problem is suggested. Second, the authors introduce a hybrid QC-based parametric method for the multipurpose batch scheduling problem with a fractional objective. The algorithms tackle optimization problems formulated in terms of mixed-integer linear and mixed-integer fractional programs, respectively. They also provide feasibility guarantees. Ajagekar et al. conducted a comparison between the performance of the SOTA solvers (exact and heuristic) and their proposed QC-based hybrid solution. The comparison took place on both job-shop and batch-scheduling problems. The hybrid frameworks suggested by the authors use quantum annealing to support their established deterministic optimization algorithms.


Many approaches to quantum optimization solve unconstrained binary programming problems. Nonetheless, mixed-integer linear programming provides more insight into practice. Chang et al.~\cite{chang2020hybrid} developed a new approach for mixed-integer programming. Their approach applies Benders decomposition. This helps decompose the mixed-integer programming into binary programming and linear programming sub-problems. The above sub-problems are solvable by a noisy intermediate-scale quantum processor and a conventional processor, respectively. The algorithm suggested by the authors provably reaches the optimal solution of the original mixed-integer programming problem. It is also tested in practice. In particular, on a D-Wave 2000Q quantum processing unit. The algorithm turned out to be practical for small-scaled test cases. Inspired by power system applications, the authors tested the algorithm on mixed-integer programming as well.

\subsection{Knapsack Problem}

\begin{definition}[Knapsack Problem]
Given a set of weighted items, each with a value, the knapsack problem aims to determine which items to include in the collection so that the total weight is less than or equal to a given limit and the total value is as large as possible~\cite{mathews1896partition}. 
\end{definition}

The decision version of the knapsack problem is NP-complete~\cite{gary1979computers}. There are definitions for the problem that help obtain further results. In particular: 

\begin{definition}[Low Density Knapsack Problem]
Given $n$ integers $a=(a_1, \cdots, a_n)$ and a target integer $S$, find an n-bit vector $e = (e_1 \cdots, e_n) \in \{0, 1\}^{n}$ s.t. $e \cdot a = \sum_{i} e_i a_i = S$. The density of the knapsack instance is defined as $d = n / (\log_2 \max_i a_i)$. Given a random instance $a$, the density is related to the number of solutions. 
\end{definition}


Certain densities allow efficient algorithms, related to lattice reduction~\cite{lagarias1985solving,lyubashevsky2005parity}. But, the best algorithms known for the problem when $d$ is close to $1$ are exponential-time.

Haddar et al.~\cite{haddar2016hybrid} introduced a hybrid heuristic approach, combining the Quantum Particle Swarm Optimization technique with a local search method. The authors then solved the Multidimensional Knapsack Problem. Their approach includes a heuristic repair operator as well. This operator uses problem-specific knowledge rather than the penalty function technique, hired for constrained problems. Experiments in~\cite{haddar2016hybrid} were conducted on a set of benchmark problems. Their results illustrated the competitiveness of the authors' method in contrast to the SOTA heuristic methods.


Dam et al.~\cite{van2021quantum} introduced two techniques, making QAOA more satisfactory for constrained optimization problems. Their first technique shows how one may use the result of a previous greedy (classical) algorithm to define an initial quantum state and mixing operation. This allows adjusting the quantum optimization algorithm and exploring the answers around the initial greedy solution. One may use their second technique to avoid the local minima close to the greedy solutions. The authors used unit-depth quantum circuits and executed the quantum algorithm on known hard instances of the Knapsack Problem. The experimental results indicate that the adjusted quantum optimization heuristics perform better than many classical heuristics in most cases.


Garc{\'\i}a  and Maureira~\cite{garcia2021knn} introduced a hybrid algorithm, using the k-nearest neighbor technique, improving over a quantum cuckoo search algorithm applied in resource allocation. The authors performed numerical experiments, obtaining insights from the involvement of the k-nearest neighbour technique in their final result. The multidimensional knapsack problem was explored in~\cite{garcia2021knn} as well, for the validation of their procedure. Further, the authors made a comparison to SOTA algorithms. They illustrated that their hybrid algorithm offers better results in most cases.

Lai et al.~\cite{lai2020diversity} have explored a quantum particle swarm optimization algorithm, integrating a diversity-preserving strategy for population management. Also, they included a local optimization method based on variable neighbourhood descent, improving solutions. The method in~\cite{lai2020diversity} was evaluated on the classic NP-hard 0–1 multidimensional knapsack problem (MKP). The experimental results, obtained by the authors, on $270$ instances showed that their algorithm is efficient compared to several SOTA MKP algorithms. This supremacy holds in two terms: the solution quality and the computational efficiency on the instances with a small or medium-sized number $(m\leq 10)$ of constraints.

Awasthi et al.~\cite{awasthi2023quantum} studied quantum computing approaches for multi-knapsack problems. They explored some of the SOTA quantum algorithms, using different quantum software and hardware tools. The Variational Quantum Eigensolver (VQE) is a hybrid quantum-classical algorithm that finds the minimum eigenvalue of a given Hamiltonian operator s.t. the operator uses a variational technique. In~\cite{awasthi2023quantum}, several gate-based quantum algorithms are considered. E.g. QAOA and VQE, as well as quantum annealing. Further, an exhaustive study of the solutions and the estimation of runtimes is provided. The results of Awasthi et al. illustrated that adapting a standard algorithm such as QAOA may considerably improve the quality of the delivered results. 

\subsection{Subset-Sum Problem}

\begin{definition}[Subset-Sum]
\label{KnapsackProblem}
Given natural numbers $w_1 \cdots w_n$, and a target number $W$, the subset-sum problem asks if there is a subset of $\{ w_1 \cdots w_n \}$ that adds up to precisely $W$~\cite{kleinberg2006algorithm}? Note that the subset-sum problem is a special case of knapsack problem. 
\end{definition}

\begin{definition}[Johnson Scheme]
Named after Selmer M. Johnson, the Johnson Scheme is the association scheme s.t. its elements are the m-subsets of an n-set. 
\end{definition}

\begin{definition}[Johnson Distance]
Consider half of the size of the symmetric difference between two m-sets in a Johnson Scheme. Denote this relation as Johnson Distance. 
\end{definition}

\begin{definition}[Johnson Graph]
The Johnson Graph $J(n,m)$ is the graph that describes the distance-1 relation in the Johnson scheme. The graph is distance-regular of diameter $\min(m, n - m)$. 
\end{definition}


Quantum computing has unique advantages for problems like subset-sum (SSP) due to quantum parallelism and superposition. On this basis, Zheng et al.~\cite{zheng2022quantum} constructed a relationship between the SSP and quantum circuit model. They proposed a feasible quantum algorithm to solve SSP and obtain quadratic speedup through applied quantum amplitude amplification iteration.


In 2013, Bernstein, Jeffery, Lange, and Meurer constructed a quantum subset sum algorithm with heuristic time complexity $2^{0.241 n}$. They enhanced the classical subset sum algorithm of Howgrave Graham and Joux with a quantum random walk technique~\cite{cryptoeprint:2013/199}. Helm and May~\cite{helm2018subset} improve on this by defining a quantum random walk for the classical subset sum algorithm of Becker, Coron, and Joux~\cite{10.1007/978-3-642-20465-4_21}. The new algorithm only needs heuristic running time and memory $2^{0.226 n}$, for almost all random subset sum instances.


Bernstein et al.~\cite{bernstein2013quantum} introduced a subset-sum algorithm. They used a heuristic asymptotic cost exponent below $0.25$. Their algorithm combines the results of Howgrave-Graham-Joux with a streamlined data structure for quantum walks on \emph{Johnson graphs}.


Bonnetain et al.~\cite{bonnetain2020improved} improved classical and quantum heuristic algorithms for subset-sum, building upon several new ideas. First, they used extended representations $(\{-1,0,1,2\})$ to improve the best classical and quantum algorithms before their work. In the quantum setting, they proved how to use a quantum search so that one can speed up the process of filtering representations. The authors built an \emph{asymmetric HGJ} algorithm, leading to the first quantum speedup on subset-sum in the model of classical memory with quantum random access. They also obtained a quantum walk algorithm for subset-sum in the MNRS framework. Although its complexity still relies on a Heuristic, the authors showed how to partially overcome it. And, they obtained the first quantum walk that requires only the classical subset-sum heuristic.


Helm and May~\cite{helm2020power} proposed quantum subset sum algorithms, not using QRAM, and breaking the square root Grover bound using linear many qubits. In particular, they obtain a running time of $2^{0.48n}$. This result is built upon the representation technique and the quantum collision finding algorithm, suggested by Chailloux, Naya-Plasencia, and Schrottenloher. Helm and May, nonetheless, improved the above result to $2^{0.43n}$. The critical point about the trade-off they introduce requires using classical memory only. Namely, not QRAM. There is a time-memory-qubit trade-off. Their result is obtained by using the Schroeppel-Shamir list construction technique. The second algorithm needs to be compared to purely classical time-memory subset sum trade-offs. For instance, those of Howgrave-Graham and Joux. In particular, the algorithms of Helm and May improve on the above completely classical algorithms for all memory complexities less than $2^{0.2n}$.


Biesner et al.~\cite{biesner2022solving} investigated how the subset sum problem has an important role in automating the financial auditing process; And how this provably NP-hard problem can be restated as a well-known problem architecture. This leads to the possibility of solving the subset-sum by applying the gradient descent technique on the energy landscape of Hopfield networks. The authors showed their algorithm finds correct sum structures for synthetic and real-world data. They provided an overview of what adiabatic quantum computers are capable of, for a particular task and its current limitations. Also, they evaluated the capability of quantum annealers for the subset sum problem. The authors discovered that for problems with a small range of values, their algorithm finds the correct solutions.


Photons have high propagation speed, robustness, and low detectable energy levels. So, photons can be alternatives to conventional computers when dealing with NP-complete problems. Xu et al.~\cite{xu2020scalable} presented a scalable chip built-in photonic computer. The computer efficiently solves the Subset Sum Problem (SSP). In particular, through a femtosecond laser direct writing technique, the authors mapped SSP into a three-dimensional waveguide network. They proved that the photons \emph{dissipate} into the networks, and search for SSP solutions in parallel. The approach in~\cite{xu2020scalable}, exhibits a significant improvement in running time even compared to supercomputers. The results verify that light may realize computations intractable for conventional computers. The authors in~\cite{xu2020scalable} show that SSP can be a good choice of benchmarking platform for performing comparisons between photonic and conventional computers.

Given $k$ lists of bit-strings, the k-xor or Generalized Birthday Problem requires finding a k-tuple among the bit-strings whose XOR becomes $0$. Naya-Plasencia and Schrottenloher~\cite{naya2020optimal} suggested algorithms for the k-xor problem, improving the complexities previously appeared in~\cite{grassi2018quantum}. The algorithms improved most values of $k$ under both settings of QACM (quantum-accessible classical memory) and low-qubits. The authors considered limited input size. Their work resulted in k-encryption running exponentially faster than double-encryption. It also allowed reaching the quantum time–memory product known, at the time of publishing their work, for solving the subset-sum problem. Every algorithm in~\cite{naya2020optimal} may be used by replacing XORs with sums modulo $2^n$. 

Naya-Plasencia and Schrottenloher~\cite{naya2020optimal} defined the framework of merging trees, permitting writing strategies for solving k-list problems (under both classical and quantum settings) abstractly and systematically. They obtained their optimization results using Mixed Integer Linear Programming.

\subsection{Coding and Cryptography}


\begin{definition}[Discrete Logarithm]
For every prime number $p$, the multiplicative group $(\mod p)$ forms a cyclic group. Namely, there exist generators $g$ such that $1, \ g, \ g^2, \ \cdots, \ g^{p-2}$ cover all the non-zero residues~\cite{hardy1979introduction,knuth1981seminumerical} $(\mod p)$. Given a prime number $p$ and such a generator $g$, the discrete logarithm of a number $x$ with respect to $p$ and $g$ is the integer $r$ such that $0 \leq r \leq p-1$ and $g^r \equiv x (\mod p)$~\cite{shor1999polynomial}.
\end{definition}

Classically, there are algorithms that can solve the Discrete Logarithm problem in polynomial time using quantum computation techniques~\cite{shor1999polynomial,hales2000improved,mosca2004exact}. 

Eker{\aa} and H{\aa}stad~\cite{ekeraa2017quantum} extended on Eker{\aa}'s work~\cite{ekeraa2016modifying} to allow for different trade-offs between the number of times the algorithm needs to be executed, its complexity, and the requirements it imposes on the quantum computer. Their algorithm computes the short discrete logarithm with only a polynomial number of executions. By using lattice-based techniques in a classical post-processing stage, the algorithm provides options for balancing execution frequency and computational complexity. And, it does not require explicit knowledge of the group's order. Further, the algorithm can solve factoring RSA integers with lower quantum resource requirements.

The work of Eker{\aa} and H{\aa}stad~\cite{ekeraa2017quantum} has been revisited by Eker{\aa}~\cite{ekeraa2020post}, revealing a higher success probability than previously reported. Building on this improved understanding, an enhanced post-processing algorithm has been proposed. This new algorithm allows for better tradeoffs and requires fewer runs. To support these claims, Eker{\aa} developed a classical simulator to sample the probability distribution induced by the quantum algorithm for given logarithms. This simulator represents a significant contribution and is instrumental in demonstrating that the work of Eker{\aa} and H{\aa}stad offers an advantage over Shor, not only in individual runs but also overall, particularly when targeting cryptographically relevant instances of RSA and Diffie–Hellman with short exponents.

Roetteler et al.~\cite{roetteler2017quantum} provided detailed resource estimates for implementing Shor's algorithm, including the number of qubits and Toffoli gates required for elliptic curve point operations (ECC). They developed reversible algorithms for modular quantum arithmetic, such as addition, subtraction, and multiplication, specifically tailored for ECC. The authors utilized the Weierstrass model to refer to the elliptic curves and elaborated on the group law for point addition. Additionally, they delved into the computational complexity of the Elliptic Curve Discrete Logarithm problem and explored the impact of quantum algorithms on the security of ECC.

The research by Roetteler et al.~\cite{roetteler2017quantum} was carried forward by H{\"a}ner et al.~\cite{haner2020improved}. They introduced enhanced quantum circuits designed for elliptic curve scalar multiplication in elliptic curve groups. Their approach involved optimizing reversible integer and modular arithmetic components using windowing techniques and refining the placement of computing steps. Additionally, they enhanced previous quantum circuits for modular inversion by reconfiguring the binary Euclidean algorithm. They developed an affine Weierstrass point addition circuit with reduced depth and fewer $T$ gates compared to earlier circuits. While previous studies focused mainly on minimizing the total number of qubits, this research presents different trade-offs across various cost metrics, including the number of qubits, circuit depth, and T-gate count.

Hhan et al.~\cite{hhan2023quantum} studied the quantum complexity of discrete logarithms and related group-theoretic problems. The authors considered their settings under generic algorithms that do not exploit the properties of the group encoding. They set up a generic model of quantum computation named the quantum generic group model. This model is the quantum variant of its classical counterpart. The authors pointed out that one may describe Shor’s algorithm for the discrete logarithm problem and related algorithms in their suggested model. Under their model, Hhan et al. showed quantum lower bounds, also (almost) matching algorithms of the discrete logarithm and related problems In particular, consider a cyclic group $\mathcal{G}$ of prime order. The authors proved that the depth of group operation queries made by every generic quantum discrete logarithm algorithm is lower bounded by $\Omega(\log \mathcal{G})$. This result implies that Shor’s algorithm is asymptotically optimal among the generic quantum algorithms, even under parallel settings. The authors further observed that some known variations of Shor’s algorithm may benefit from classical computations to reduce the number and depth of quantum group operations. They suggested a model for generic hybrid quantum-classical algorithms. It shows that these algorithms are almost optimal in this model. All generic hybrid quantum-classical algorithms for the discrete logarithm problem with a total number of (classical or quantum) group operations $Q$ must make $\Omega(\log |G|/ \log Q)$ quantum group operations of depth $\Omega(\log \log |G| - \log \log Q)$. Specifically, if $Q = \text{polylog} |G|$, classical group operations reduce the number of quantum queries by a factor of $\mathcal{O}(\log \log |G|)$. And, the quantum depth remains as $\Omega(\log \log |G|)$.


Post-quantum cryptanalytic experiments, if one desires to run them on a large scale, require low-memory algorithms. In their work, Delaplace et al.~\cite{delaplace2019improved} prove that combining techniques from representations, multiple collision finding, and the Schroeppel-Shamir algorithm~\cite{schroeppel1981t} leads to improved low-memory algorithms.
Consider a random subset sum instances $(a_1, \cdots, a_n, t)$ for which modulo $2^n$ is defined. Denote the amount of memory an algorithm consumes by $M$. Delaplace et al. improve the Dissection technique for small memory $M < 2^{0.02n}$. For the mid-memory regime i.e. when $2^{0.13n} < M < 2^{0.2n}$, they obtain better results as well.


Aaronson and Hung~\cite{aaronson2023certified} suggested an application for near-term quantum devices. That is, constructing cryptographically certified random bits so that one may use them in proof-of-stake cryptocurrencies, as well as in other areas. The protocol adapts the currently available \emph{quantum supremacy} experiments. This would be based on random circuit sampling. Let $n$ be the number of qubits. The authors particularly showed that under plausible hardness assumptions, every time the outputs of the latter experiments pass the now-standard Linear Cross-Entropy Benchmark, they necessarily contain $\Omega(n)$ min-entropy. Aaronson and Hung~\cite{aaronson2023certified} obtained a net gain in randomness using a small seed to generate pseudorandom challenge circuits. To have a response in return for the challenge, the quantum computer has to generate output strings. The strings have to be verified first and then one may provide them as inputs into a randomness extractor. This will produce certified nearly-uniform bits. Hence \emph{bootstrapping} from pseudorandomness to genuine randomness. The authors showed that their protocol reasonably works. This holds from two points of view. First, under a hardness assumption named Long List Quantum Supremacy Verification. In particular, they gave ground for it in the random oracle model. Second, they considered an eavesdropper who could share arbitrary entanglement with the device. The authors' proof was unconditionally in the random oracle model against the aforementioned eavesdropper. One should note that the output of the protocol is unpredictable. Even to an unbounded adversary in terms of computational power, who may see the random oracle. A drawback of the above protocol is the exponential cost of verification. Specifically, there cannot be more than $n \sim 60$ qubits. Namely, a setting in which attacks are costly, yet not impractical.


Leverrier and Z{\'e}mor~\cite{leverrier2023efficient} introduced and analyzed a decoder for quantum tanner codes. These codes may correct adversarial errors of \emph{linear} weight. Notice that the previous decoders for quantum low-density parity-check codes supported weights of $\mathcal{O}(\sqrt{n \log n})$~\cite{ShaiEvra2020}. Further, Leverrier and Z{\'e}mor contributed to the link between quantum tanner codes and the Lifted Product codes of Panteleev and Kalachev~\cite{panteleev2022asymptotically}. They proved that one may adapt their decoder to the latter. The decoding algorithm in~\cite{leverrier2023efficient} alternates between sequential and parallel procedures. It also offers convergence in linear time. 

Leverrier and Z{'e}mor~\cite{leverrier2023decoding}, in their previous work~\cite{leverrier2023efficient}, introduced sequential and parallel decoders for quantum tanner codes. The authors highlighted the potential of the new code introduced in~\cite{leverrier2023efficient}, improving the performance of quantum computers capable of executing long-range gates between arbitrary qubits. To utilize such capabilities completely, it is necessary to develop efficient decoding algorithms to manage errors during the execution of a quantum algorithm. This requires highly parallelizable decoders that operate in logarithmic time, as new errors continue to accumulate while a classical decoder tries to identify earlier errors. The authors thus presented such an algorithm for the family of Quantum Tanner codes.

\section{Open Problems in Applying Quantum Techniqes}
\label{sec:openproblems}
 Quantum computing differs from classic computing, and we mentioned more than a hundred previous works to see how quantum computing can help solve fundamental classical challenging problems. In the following, we consider problems individually, what we have, and what remains open.

\subsection{QAOA}
\label{subsec: QAOA}

    \begin{problem}
    \label{ab-QAOA1}
        Adaptive-bias QAOA (ab-QAOA) is used to improve performance in the hard region of the 3-SAT and the Max-3-SAT problems~\cite{yu2023solution}. Initialization of bias-field parameters $\Vec{h}$, which can be generated randomly, is a key issue. Consider the way to obtain the same infidelity using an extensive computational resource. This is a matter of understanding how accurate the initial guess on $\Vec{h}$ scales with $n$. Is it then possible to take a non-random approach, using a classical heuristic, to get the initial value of $\Vec{h}$?
    \end{problem}

    \begin{problem}
    \label{p:Scalability:1}
        It is known that the gap between the QAOA and the abQAOA (adaptive-bias QAOA) is greatest when the plateau barrens~\footnote{The barren plateau phenomenon explains that the gradient of parameters of the parameterized quantum circuits will vanish exponentially in terms of the system size~\cite{Zhao_2021}} are most visible. Equally, numerical evidence suggests that abQAOA has a better scaling behavior than QAOA in terms of system size accuracy (See~\cite{yu2023solution} for more information). Is it possible to study a barren plateau phenomenon with abQAOA? 
    \end{problem}
    
    \begin{problem}
    \label{RQAOA}
        Recursive quantum approximate optimization algorithm (RQAOA) could be a useful application for NISQ devices in combination with effective classical computation that it may apply to large problems of global relevance (See~\cite{bravyi2022hybrid}). The potential of RQAOA at higher levels $P$ has already been shown, which might not be simulant classical. Is it then possible to develop performance guarantees for $\text{RQAOA}_p$, comparable to the robust results achieved by current classical algorithms? 
        For instance, it was previously shown that for the \emph{ring of disagrees}, RQAOA achieves the optimal approximation ratio. Are there more promising classes of graphs that could yield achievability results?
    \end{problem}

    \begin{problem}
    \label{RQAOA1}
        It seems beneficial to apply NISQ devices combined with classical computation for recursive quantum approximate optimization algorithm (RQAOA)~\cite{bravyi2022hybrid}. Is it then possible to explore establishing performance guarantees for $\text{RQAOA}_p$? Can the next step be finding more classes of graphs s.t. achievability results can be established for them?
    \end{problem}

    \begin{problem}
    \label{p:Scalability:2}
        The Sherrington-Kirkpatrick (SK) model is a Hamiltonian with randomly selected interactions between all pairs of sites. Under the settings of the infinite-size limit of the SK model, the performance of QAOA and its optimal parameters can be analyzed. However, the runtime becomes exponential in the number of algorithm layers, not the number of qubits. There is a numerical algorithm to evaluate the performance of $\QAOA$ on an average finite-size instance of the SK model. Nonetheless, the algorithm remains limited to low depth. Are there further cases in which the performance of $\QAOA$ can be evaluated classically on large problem instances?
    \end{problem}

    \begin{problem}
    \label{p:QAlgorithm:1}
    \label{p:Error:1}
        It is known that $\QAOA$ can be applied to MaxCut on large-girth $D\text{-regular}$ graphs (See~\cite{basso2021quantum}). Is it then possible to find an iterative formula for the $\QAOA$’s performance that is more efficient than the existing work? If so, probing the $\text{large-}p$ behaviour of the QAOA becomes more convenient.
    \end{problem}

    \begin{problem}
    \label{QAOAMaxCut}
        One may apply $\QAOA$ to MaxCut on large-girth $D\text{-regular}$ graphs. There is an iterative formula to evaluate performance for every $D$ at every depth $p$ (See~\cite{basso2021quantum}). Nevertheless, the existing work only investigated the cases in which $p$ is relatively tiny. Can such results be recast to support the $p \rightarrow \infty$ limit, in terms of continuous functions corresponding to $\gamma$ and $\beta$ s.t. $\gamma = (\gamma_1, \cdots, \gamma_p)$ and $\beta = (\beta_1, \cdots, \beta_p)$ are defined vectors.
    \end{problem}

    \begin{problem}
    \label{p:Scalability:3}
        The work in~\cite{boulebnane2021predicting} evaluated the performance of QAOA in the infinite-size limit for the MAX-CUT problem. However, they cannot address the D-spin model for $D \geq 3$ and $p > 1$. Is there a different analysis for the above case?
    \end{problem}

\subsection{Quantum Anealing}
\label{subsec: QA}

    \begin{problem}
    \label{p:QSoftware:1}
    \label{p:QA:1}
        It is known how to solve the Set Packing Problem on D-wave quantum annealers using a general formulation. Is it possible then to make a library to solve optimization problems with quantum annealers, using the results in~\cite{venere2023design}? The library could be a novel abstraction layer that completely hides the complexity of quantum annealer use, making it easier to understand. 
        How can this characterization framework on different quantum annealers be evaluated? Also, is it possible to work towards a comparison with quantum computers based on the gate paradigm, while finding a methodology to map the annealing problem on such architectures? 
    \end{problem}
    
    \begin{problem}
    \label{cocchi1}
        One may find a partition of a given directed graph into Hamiltonian subgraphs, with three or more vertices~\cite{cocchi2021graph}. The above problem is denoted as vertex 3-cycle cover. Can the above be extended, considering weighted graphs, or where constraints on the cycle length exist?
    \end{problem}

    \begin{problem}
    \label{cocchi2}
        The work in~\cite{cocchi2021graph} showed it is possible to solve the graph partitioning into subgraphs containing Hamiltonian cycles of constrained length, by using a quantum annealer. For graphs in which weights are assigned to edges or when different restrictions on cycle length exist, is it possible to extend Graph Partitioning into Hamiltonian Subgraphs?
    \end{problem}

    \begin{problem}
    \label{p:Error:2}
        The optimization problem of maximum cardinality matching can be applied to the latest QA hardware from D-Wave, the 2000Q, and Advantage quantum annealers~\cite{mcleod2022benchmarking}. Is performance likely to be improved by further considering other annealing variables? For instance, by optimizing custom annealing schedules, chain strengths, chain break resolution methods, and annealing offsets. Can the benefits of parameter tuning be assessed as well?
    \end{problem}

    \begin{problem}
    \label{MIP1}
        Integer programming techniques can be used for minor embedding in quantum annealers~\cite{bernal2020integer}. Can symmetries and the invariant formulation be hired as well? 
    \end{problem}


    \begin{problem}
    \label{p:Hybrid:1}
        Is it possible to extend the formulations in the work discussed in~\cite{salehi2022unconstrained} to problems based on TSP like the vehicle routing problem and its variants? The results of the aforementioned work suggest that hybrid algorithms may be better for solving large instances of such problems.
    \end{problem}

\subsection{Quantum Search}
\label{subsec: QS}
    \begin{problem}
    \label{p:QAlgorithm:3}
        There can be improvements to the running time of the SAT problem by bounding an area and keeping the focus on it in the search space. Making the selected area smaller results in a better running time. Nonetheless, no solution might be found because of a not-right-selected area. 
        Are there approaches to finding an optimal reduction and selection? Even more interestingly, for the larger problems, several areas may lead to an answer. Can all such areas be found to guarantee the algorithm returns a solution if exists? Moreover, is it possible to adapt the techniques discussed in~\cite{varmantchaonala2023quantum} to the XORSAT problem?

    \end{problem}

    \begin{problem}
    \label{p:QAlgorithm:4}
        In developing an evolutionary algorithm (EA) auxiliaries, the adaptation of constraint weights may improve the performance of EA~\cite{CHENG2007123}. Is it possible to adopt the random work of the local search when constructing the GenSAT-based auxiliaries? 
    \end{problem}

    \begin{problem}
    \label{p:QAlgorithm:5}
        The work in~\cite{CHENG2007123} proposed a quantum cooperative search that makes Grover’s search algorithm work on 3-SAT problems with a small number of qubits. Can the quantum cooperative search (QCS) algorithm be conducted to handle combinatorial optimisation problems other than 3-SAT? 
    \end{problem}

    \begin{problem}
    \label{p:QSupremacy:0}
        The Eppstein algorithm is a graph theory algorithm that finds the number of nearest paths that allow cycles to be carried out between a given pair of vertices on a direct graph. The work in~\cite{Ge_2020} subsequently improved Eppstein’s algorithm for solving the Hamiltonian cycle, from $\mathcal{O}(1.2599^n poly(n))$ to $\mathcal{O}(1.2509^n poly(n))$ and $\mathcal{O}(1.2312^n poly(n))$. Can this framework be used to accelerate the actual best classical algorithms? This would give an asymptotic acceleration over the best classical algorithms for a quantum computer the size of any constant fraction of the problem size.
    \end{problem}

    \begin{problem}
    \label{p:QSupremacy:1}
        The work in~\cite{Ge_2020} improved Eppstein’s algorithm subsequently using Grover search. Is it possible to use fitting methods like the quantum backtracking techniques and subsequent improvements in the hybrid approach of~\cite{Ge_2020}?
    \end{problem}

    \begin{problem}
    \label{decomposition}
        The work in~\cite{kerger2023asymptotically} investigated subgraph extraction and pruning techniques for a decomposition algorithm designed to compute the Maximum Cliques of a graph. It seems that the Lov\'asz bound is most effective in pruning subgraphs but also most computationally intensive. A heuristic upper bound leads to a good trade-off between subgraph reduction and runtime for this bound. Is it possible to use a different decomposition and simplification strategy for each problem type of graph, so that the viability of tradeoff algorithms could be explored further?
    \end{problem}
    
    \begin{problem}
    \label{prob:problemname4}
        The experimental results of the work discussed in~\cite{kerger2023asymptotically} implied that the bounds of the Lov\'asz number offer the highest decrease in subproblems. Nonetheless, the tradeoff was paying a high computational cost. Is there a way to find a faster way to compute the Lov\'asz number? This would make the above bounds more practical and usable in real-world scenarios as a future direction.
    \end{problem}

\subsection{Quantum Walk}
\label{subsec: QW}

    \begin{problem}
    \label{quantumrandomwalk}
        Laarhoven~\cite{laarhoven2016search} proposed an algorithm for quantum random walks to locate $k$ different marks. The algorithm suggests running the random walk, including the setup, $\mathcal{O}(k)$ times, to find the marks. However, it remains unclear whether this approach can be optimized to improve the overall complexity of the algorithm.
    \end{problem}

    \begin{problem}
    \label{p:QAlgorithm:6}
        Can an enhancement to the Shortest Vector Problem (SVP), as outlined in the work by Chailloux et al.~\cite{chailloux2021lattice} in their paper \emph{Lattice-Based Cryptography}, involve incorporating a layer of \emph{local sensitive filtering} through a quantum random walk, potentially by introducing a \emph{local sensitivity value} within the random walk graph, as an alternative to using the Johnson graph?
    \end{problem}
    
    \begin{problem}
    \label{Johnsongraph}
        Is it possible to improve the results of the work in~\cite{apers2019unified} by embedding the local sensitivity property in the graph, on which the random walk was performed in the existing work, rather than working on the Johnson graph?
    \end{problem}

    \begin{problem}
    \label{quantumRandomWalk}
        The work in~\cite{apers2019unified} pointed out that there are no better generic algorithms, if one desires to find $k$ different marked, in contrast to running the entire random walk (including the setup) $\mathcal{O}(k)$ times. Is there a better way to do this under the existing settings? 
    \end{problem}
    
    \begin{problem}
    \label{p:Error:3}
        The algorithm suggested in~\cite{dalzell2023mind} offers a mechanism with no trivial classical analog. The quantumness of its technique might provide a chance for achieving a super-quadratic speedup. The algorithm is unable to achieve the latter in comparison to classical algorithms with improvements obtained previously. Is it then possible to illustrate if the algorithm and the speedup mechanism of~\cite{dalzell2023mind} may be combined with a classical technique to exploit problem structure e.g. backtracking and branch-and-bound?
    \end{problem}

    \begin{problem}
    \label{pf1}
        The main obstacle to expanding the existing quantum computers is decoherence~\cite{kassal2011simulating}. Consider other models such as topological quantum computing, one-way quantum computing, and quantum walks. Can further algorithmic progress be made, perhaps using the above models?
    \end{problem}

\subsection{Mean Estimator}
\label{subsec: MeanEstimator}

    \begin{problem}
    \label{problem::ME1}
        Consider the mean estimation problem~\cite{cornelissen2022near}. The problem aims to compute an estimate $\Tilde{\mu}$ to the mean $\mu = \mathbb{E}[X]$ s.t. $X$ is a random variable, representing the output of some black-box process. Suppose that the covariance matrix $\Sigma$ of $X$ exists. It is not assumed in~\cite{cornelissen2022near} that $\Sigma$ is known beforehand. For some $l_{p}\text{-norms}$, especially for $p  < 2$, is it possible to investigate the bounds on the individual diagonal entries of $\Sigma$?
    \end{problem}


    \begin{problem}
    \label{problem::ME3}
        Using the one-dimensional result and a union bound, one may obtain a mean estimator under the classical settings~\cite{cornelissen2022near}. Denote $d$ as the dimensionality factor. The aforementioned estimator offers a precision of $$\max_{j \in [d]} \sqrt{\text{Var}[X_j] \log(d/\delta) / n}$$ Nonetheless, the quantum settings in~\cite{cornelissen2022near} yield $$\sqrt{\sum_{j \in [d]} \text{Var}[X_j]} \log(d/\delta) / n$$ As a future direction, one can investigate whether some combination of the above approaches may be proved optimal under all settings.
    \end{problem}

\subsection{Qubit Mapping}
\label{subsec: Qubit Mapping}

    \begin{problem}
    \label{li2022}
        An algorithm called FIDLS~\cite{li2020qubit} exists in the literature for qubit mapping based on subgraph isomorphism and filtered depth-limited search. The extra two-qubit gates required in the output circuit could be reduced by this algorithm, significantly. Is it possible to enhance the quality of the results with a better, possibly customized, subgraph isomorphism algorithm (e.g. see the approximate subgraph isomorphism algorithm proposed in~\cite{siraichi2019qubit})?
    \end{problem}

    \begin{problem}
    \label{p:Simulation:1}
        Generating penalties directly from Satisfiability Modulo Theories (SMT)~\cite{bian2020solving} is very difficult for large Boolean functions. Because the number of constraints grows much faster than the number of available parameters.  Function decomposition and chains can help solve this problem, but chains limit the resulting gaps. Other methods that do not impose such restrictions can reconstitute the decomposition function. Is it possible to use additional qubits to increase the gap of an existing penalty function? Are there other SMT formulations for these problems?
    \end{problem}

    \begin{problem}
    \label{p:Simulation:2}
        The feasibility of 2048 sparsely connected qubits architectures has already been explored in the literature, for solving SAT and MaxSAT problems as QA systems scale~\cite{bian2020solving}. The existing techniques allow encoding SAT and MaxSAT, with certain limitations, in an Ising problem compatible with sparse QA architecture. Is it possible to use more connected topologies? One may have larger QA hardware graphs with higher per-qubit connectivity and less separation between clusters (tiles) of qubits. New encoding strategies might be required as well because these changes will result in the ability to solve more difficult Ising problems. The encoding process will significantly improve by taking advantage of additional connectivity using new methods for problem decomposition, placing small Boolean functions, penalty modelling, etc.
    \end{problem}

    \begin{problem}
    \label{SATtoIsing}
        The formulation of SAT to Ising using Satisfiability Modulo Theories (SMT) has already been discussed in the literature~\cite{bian2020solving}. By choosing k-feasible cuts with small $k$, proper mappings provide decompositions into small Boolean functions. Boolean function decomposition and minimization are mature classical subjects. Can the existing algorithms be improved by taking into consideration the specifics of the embedding (placement and routing onto a QA hardware graph) that follow them?
    \end{problem}

    \begin{problem}
    \label{p:Scalability:6}
        Formulating SATtoIsing as a problem in Satisfiability or Optimization Modulo Theories (SMT/OMT) is possible. Are there faster and more scalable methods of encoding small Boolean functions using SMT? Currently, methods for simultaneously placing variables and computing penalty functions are not as scalable or well-studied as methods that keep variables in one place~\cite{bian2020solving}.
    \end{problem}

    \begin{problem}
    \label{bian}
        SAT was formulated in the literature to the Ising model for solving SAT and MaxSAT problems by Satisfiability Modulo Theories (SMT)~\cite{bian2020solving}.
        A Boolean formula is a quantified Boolean formula (QBF) or Shannon expansion if $F(\underline{\mathbf{x}})$ is a QBF, then $\forall x_iF(\underline{\mathbf{x}})$ and $\exists x_iF(\underline{\mathbf{x}})$ are QBFs. $\forall x_iF(\underline{\mathbf{x}})$ is equivalent to $(F(\underline{\mathbf{x}})_{x_i=\top}\land~F(\underline{\mathbf{x}})_{x_i=\bot})$ and $\exists x_iF(\underline{\mathbf{x}})$ is equivalent to $(F(\underline{\mathbf{x}})_{x_i=\top} \lor F(\underline{\mathbf{x}})_{x_i=\bot})$.
        The problem was also reduced to SMT by applying Shannon’s expansion to the Penalty function.
        Is it possible to investigate emerging novel techniques for solving directly the quantified formulas for computing penalty functions via SMT/OMT, avoiding the expensive Shannon expansion?
    \end{problem}

    \begin{problem}
    \label{p:Error:4}
        A feasible quantum algorithm exists in the literature, utilizing fewer qubits to encode and achieving a quadratic speedup for solving the subset sum problem~\cite{zheng2022quantum}. Are there further ways to reduce the error rate of real quantum devices and improve the available qubits?
    \end{problem}

\subsection{Quantum Particle Swarm Optimization}
\label{subsec: QPSO}

    \begin{problem}
    \label{p:QAlgorithm:7}
        A hybrid KNN quantum cuckoo search algorithm (KQCSA) exists in the literature for solving the Knapsack problem~\cite{garcia2021knn}. The concept of $k\text{-nearest}$ neighbour is incorporated in this hybrid algorithm, in the movement of the solutions and its updating. Can the concept of $k\text{-nearest}$ neighbours from KQCSA be adapted to additional quantum metaheuristics such as quantum particle swarm optimization (QPSO)?
    \end{problem}

    \begin{problem}
    \label{p:Hybrid:2}
        A hybrid heuristic to solve the Multidimensional Knapsack Problem already exists that combines Quantum Particle Swarm Optimization with a local search method~\cite{haddar2016hybrid}. How can the above hybrid approach solve other NP-hard and combinatorial optimization problems?
    \end{problem}

    \begin{problem}
    \label{quantumParticle}
        A diversity-preserving quantum particle swarm optimization (DQPSO) approach already exists to solve the multidimensional knapsack problem, enhancing the conventional Quantum particle swarm optimization (QPSO) method. To avoid premature convergence of the QPSO algorithm, a diversity-preserving strategy has been used to maintain the population's diversity. The DQPSO algorithm consists of six components, including the repair operator to ensure the feasibility of generated solutions. Can different pseudo-utility ratios be used in a combined way to enhance the effectiveness of the repair operator?
    \end{problem}

    \begin{problem}
    \label{lai2020}
        A diversity-preserving quantum particle swarm optimization algorithm already exists for solving the classic 0–1 multidimensional knapsack problem~\cite{lai2020diversity}. In instances with many constraints, the algorithm's performance can vary across several runs. Can the robustness of the algorithm be improved?
    \end{problem}

\subsection{Quantum String Matching Algorithm}
\label{subsec: QSMA}
    \begin{problem}
    \label{p:QComplexity:1}
        The work in~\cite{jin2023quantum} samples $n / \tau^{1-o(1)}$ synchronizing positions consistently, in the string depending on their $\text{length-}\Theta(\tau)$ contexts. Each synchronizing position can be reported by a quantum algorithm in $\Tilde{\mathcal{O}}(\tau^{1/2 + o(1)})$ time. Is there a way to improve the $\tau^{o(1)}$ factors in the sparsity and the time complexity of the string synchronizing set of the existing result to poly-logarithmic?
    \end{problem}

    \begin{problem}
    \label{p:QComplexity:2}
        It is known that \emph{String Synchronizing Sets} has applications in classical string algorithms. Are there more applications in quantum string algorithms? E.g. given a maximum of $k$, what is the quantum query complexity to compute the \emph{edit distance} of two strings s.t. their distance is no larger than $k$?
    \end{problem}

\subsection{Quantum Merging Trees}
\label{subsec: QMT}

    \begin{problem}
    \label{p:QSecurity:1}
        Consider the framework introduced in~\cite{naya2020optimal} for merging trees. It permits writing strategies for solving k-list problems (in terms of classical and quantum computation) abstractly and systematically. What are cryptographic applications for quantum k-list algorithms e.g. lattice algorithms or decoding random linear codes~\cite{both2018decoding}?
    \end{problem}

    \begin{problem}
    \label{p:QSecurity:2}
        
        The existing results achieved optimal results for all merging trees by implementing a set of optimization strategies that have been meticulously developed and refined~\cite{naya2020optimal}. However, it is worth exploring the possibility of extending this framework to improve quantum algorithms further. Notably, the complexity of the $r$-th encryption merge tree algorithms exceeds the time limit of $2^{0.3n}$, raising the question of whether this bound can be surpassed. As the team continues to investigate this topic, it is crucial to consider how this framework can be optimized to meet the evolving demands of the field.
        
    \end{problem}

    \begin{problem}
    \label{naya}
        Is it feasible to expand upon the merging trees~\cite{naya2020optimal}, which draw inspiration from the classical literature on k-list problems? Although various techniques have been explored, none have yielded quantum advantages.
    \end{problem}

\subsection{Quantum Congested Clique Model}
\label{subsec: QCC}
            
    \begin{problem}    
    \label{prob:problemname6}
        Algorithms already exist in the quantum Congested Clique model (QCCM) for computing approximately optimal Steiner trees and exact directed minimum spanning trees. Many constants and $\log$ factors are tracked for the algorithms suggested and the related ones, in the work in~\cite{kerger2023asymptotically}. It seems that the standard minimum spanning tree problem (MST) can be solved in a relatively small constant number of rounds in the classical Congested Clique. Is there a way to improve the existing results?
    \end{problem}

    \begin{problem} 
    \label{prob:problemname5}
        What algorithms solving the problems mentioned in the work in~\cite{kerger2023asymptotically}, offer the highest efficiency w.r.t. rounds needed in the $\text{Congested Clique}$ under a setting related to $n$, in which the algorithms are not suitable for practice?
    \end{problem}

    \begin{problem}
    \label{p:QAlgorithm:9}
        Consider the bounded-degree and minimum-degree spanning tree problems. It is known that even the bounded-degree decision problem on an unweighted graph is NP-complete. Can other techniques, apart from the ones that appeared in the work in~\cite{kerger2023asymptotically} provide insight into the problem?
    \end{problem}

\subsection{Verification and Certification}
\label{subsec: ver&cer}
    \begin{problem}
    \label{p:QSupremacy:2}
       Classical algorithms may obtain depth-independent, $\Omega(1)$ LXEB advantage for BosonSampling. The open question is \emph{how much}? Answering the above question seems to be a prerequisite to designing a certified randomness protocol. Also, it puts a lower bound on how efficient a BosonSampling experiment has to be before it can be used for such a protocol~\cite{aaronson2023certified}.
    \end{problem}
    
    \begin{problem}
        Consider the class $\text{QMA}^{+}(2)$ as the unentangled quantum proofs with non-negative amplitudes.
        The protocols in~\cite{10.1145/3564246.3585248} are examples of proof verification and property testing. Further, each protocol has a single isolated property testing task, relying on non-negative amplitudes. Can the protocols be generalized, so that one may transfer the results of~\cite{10.1145/3564246.3585248} to $\text{QMA}(2)$?
    \end{problem}

    \begin{problem}
        The work in~\cite{zhang2022classical} highlights the number of rounds required for two distinct protocols. The remote state preparation protocol demands a constant number of rounds, whereas the gadget-assisted verification protocol sees a linear increase in the number of rounds. As a result, the overall number of rounds for the classical verification of quantum computation protocol is linear. It is imperative to explore whether there exists a viable solution to overcome this limitation.
    \end{problem}
    \begin{problem}
    \label{problem::compver}
        The previous works (See~\cite{brakerski2018cryptographic} and~\cite{mahadev2018classical}) gave way to a series of protocols for various problems. Can the protocols or techniques by the work in~\cite{zhang2022classical} be used for the above problems? Specifically, using classical verification of quantum computation or remote state preparation with verifiability might help to suggest faster protocols for other problems.
    \end{problem}

    \begin{problem}
    \label{p:QSecurity:3}
        The applications of the random oracle in the construction of the protocol in~\cite{zhang2022classical} are (1) constructing the underlying \emph{symmetric key encryption scheme} used in lookup tables (2) in RO-padded Hadamard tests (as appeared in~\cite{zhang2021succinct}). 
        Is it possible to prove the instantiation of the random oracle or a way of reducing its usage?
    \end{problem}

    \begin{problem}
    \label{p:QSecurity:4}
        Is it possible to use the findings from~\cite{aaronson2023certified} to develop a secure and verifiable certified randomness scheme that can run on current devices? Developing a sampling-based quantum supremacy experiment that can run on a NISQ device and allow for classical verification seems quite challenging.
    \end{problem}

    \begin{problem}
        One may instantiate the \emph{lifting lemma} of Arora et al.~\cite{arora2023quantum} with the proof of quantumness from~\cite{yamakawa2022verifiable}. Consequently, the obtained problem has the property that solutions may publicly be verified. Is it possible to explore the possibility of further pushing this quantum soundness, and obtain verification of BQP with a BPP verifier relative to a random oracle?
    \end{problem}

\subsection{Quantum Machine Learning}
\label{subsec: machine learning}
    \begin{problem}
    \label{knapsackML}
        A hybrid algorithm already exists to solve medium and large instances of the multidimensional knapsack problem~\cite{garcia2021knn}. Is it possible to explore using dynamic parameters instead of fixed parameters defined in~\cite{garcia2021knn} for the optimization process?  The dynamic parameters might be capable of adjusting as the algorithm is executed, resulting in more powerful algorithms. Do reinforcement learning techniques help to adjust the parameters?
    \end{problem}

    \begin{problem}
        \label{li2022-2}
        In~\cite{li2020qubit}, a method is proposed by selecting an initial mapping and searching in a filtered and depth-limited setting. This setting is utilized in the searching step and is a useful SWAP combination that makes two-qubit gates in the logical circuit executable, as many as possible. Are there machine learning and deep learning algorithms to quickly choose the best action in searching and selecting the maximal value of a defined value function in Equation (2) of the original paper~\cite{li2020qubit}, especially when the search depth becomes large?
    \end{problem}

\subsection{Complexity}
\label{subsec: complexity}

    \begin{problem}
    \label{p:QComplexity:3}
        The relationship between the class of problems solvable by quantum algorithms (denoted by BQP) and the well-known classical classes P and NP has already been discussed in the literature~\cite{mohr2014quantum}. Does BQP interact with NP-Complete? 
    \end{problem}

    \begin{problem}
        Consider the time-dependent evolution that appeared in the work in~\cite{brandao2021models}. A long-time linear complexity growth in holographic systems was originally concentrated on time-independent Hamiltonian evolution. Is it possible to attempt to prove something related to the complexity $\mathcal{C}_{\delta}(e^{-iHt})$ for a specific many-body Hamiltonian $H$?
    \end{problem}

    \begin{problem}
        A modular approach of the work in~\cite{harris2023scalable} provides the potential for integrating other test rounds. Under the settings of the latter work, the corresponding subroutines of test rounds, computation rounds, and resource estimation were separated. Is it possible to define a clear procedure for algorithmic resource estimation, using the tests they included?
    \end{problem}

    \begin{problem}
        Consider the binary oracle setting. There exists a matching upper and lower bound up to polylogarithmic factors~\cite{cornelissen2022near}. Can deriving sharp bounds for other norms in the above model be investigated? Is it possible to explore the case of $l_{\infty}\text{-norm}$ as it relates strongly to the quantum algorithm and the applications that appeared in~\cite{cornelissen2022near}?
    \end{problem}

    \begin{problem}
    \label{p:QComplexity:4}
        We already know that
         $$\text{BPP}^{\text{QNC}_{\mathcal{O}(1)}} \subsetneq \text{QNC}^{\text{BPP}}$$, $$\text{QNC}^{\text{BPP}}_{\mathcal{O}(1)} \subsetneq \text{BPP}^{\text{QNC}}$$ and $$\text{BPP}^{\text{QNC}^{\text{BPP}_{\mathcal{O}(1)}}} \subsetneq \text{BPP}^{\text{QNC}} \cup \text{QNC}^{BPP}$$ are all separations of hybrid quantum depth, and are all according to search problems. Can the same separations be proved w.r.t. decision problems~\cite{arora2023quantum}?
    \end{problem}

    \begin{problem}
        Consider the random oracle model. The random oracle is the least structured type of oracle. But, since it is an oracle, it helps establish provable lower bounds. Given an instantiable quantum depth separation. Is it possible to establish the separation under the plain model? Using more structured problems may yield stronger separations.
    \end{problem}

    \begin{problem}
    \label{p:QComplexity:5}
        Given problems solved by polynomial-sized circuits in a hybrid setting for the gates. Denote this as the class $\text{QDepth}_d$. A crucial constraint here is that the length of the longest path connecting quantum gates is at most $d$. The connections are made by quantum wires. Is the union of every $r\text{-level}$ generalizations of $\text{BPP}^{\text{QNC}^{\text{BPP}}_{d}}$~\cite{arora2023quantum} (s.t. $r$ is bounded polynomially) equals $\text{QDepth}_d$? Also, are the separating problems (as well as $\text{d-Rec}[\mathcal{P}]$ in general, for classical query sound $\mathcal{P}$) not in $\text{QDepth}_d$?
    \end{problem}

    \begin{problem}
    \label{p:QComplexity:6}
        Three quantum algorithms for solving 3-SAT problems using quantum P systems are proposed in the work discussed in~\cite{leporati2007three}. What are the computational properties of quantum P systems that contain and process entangled objects? What are the limits of quantum P systems' computational power when tackling problems harder than NP-complete ones? 

    \end{problem}

    \begin{problem}    
    \label{p:QAlgorithm:10}
        A quantum complexity lower and upper bounds for independent set problems in graphs already exists in the literature~\cite{doern2007quantum}. Is it possible to construct an exact algorithm for the maximum independent set problem with time complexity $\mathcal{O}(c^n)$ for some $c < 1.1$? 
   \end{problem}

    \begin{problem}
    \label{p:QComplexity:7}
        It is suggested to limit the time complexity of the guessing tree method used for solving the maximum matching problem. This restriction would enable a comparison between the maximum matching algorithm explained in~\cite{kimmel2021query} and the current quantum maximum matching algorithms that prioritize minimizing time complexity over query complexity.
    \end{problem}

    \begin{problem}
    \label{p:QComplexity:8}

        In consideration of the repeated application of Grover's search, as demonstrated in the study in~\cite{dorn2009quantum}, the question arises as to whether it is possible to achieve a time complexity that is within a logarithmic factor of the query complexity observed in the investigation discussed in~\cite{witter2022query}.
    \end{problem}

    \begin{problem}
    \label{p:QComplexity:9}
        The problem of maximum matching in general graphs has a query complexity of $\mathcal{O}(\sqrt{m}n^\frac{3}{4})$~\cite{witter2022query}. However, is it possible to improve the time complexity of this problem by employing repeated applications of Grover's search, while ensuring a logarithmic factor reduction from the aforementioned query complexity?
    \end{problem}

    \begin{problem}
    \label{p:QComplexity:81}
        Is the $\mathcal{O}(n^2\log^2n)$ quantum algorithm for computing a maximum matching optimal~\cite{dorn2009quantum}? The lower bound for this problem is $\Omega(n^{1.5})$. But is there a better upper bound for computing a maximum matching in bipartite graphs?
    \end{problem}

    \begin{problem}
        The article~\cite{zhang2022classical} presents a noteworthy improvement in the time complexity of the algorithm compared to previous works. The article demonstrates a improvment in time complexity from $\mathcal{O}(\textbf{poly}(\kappa) |C|^3)$ to $\mathcal{O}(\textbf{poly}(\kappa) |C|)$. However, it is essential to note that the $\text{big-}\mathcal{O}$ notation conceals a large constant factor, making the results in~\cite{zhang2022classical} less practical. It should be noted that this constant factor is due to the security analysis, which appears to be not tight enough. Although the explicit bounds for the constants in the above work are loose, it is worth considering the possibility of obtaining tighter bounds.
    \end{problem}

    \begin{problem}
    \label{p:QComplexity:10}
        Using a problem inspired by Brakerski et al.'s proof of quantumness method~\cite{brakerski_et_al:LIPIcs:2020:12067} allows for proving more fine-grained separations between hybrid classes. For instance $$\text{BPP}^{\text{QNC}_{\mathcal{O}(1)}} \subsetneq \text{QNC}^{\text{BPP}}$$ $$\text{QNC}^{\text{BPP}}_{\mathcal{O}(1)} \subsetneq \text{BPP}^{\text{QNC}}$$ and $$\text{BPP}^{\text{QNC}^{\text{BPP}_{\mathcal{O}(1)}}} \subsetneq \text{BPP}^{\text{QNC}} \cup \text{QNC}^{BPP}$$ One might then ask whether these separations yield finer-grained proofs of quantum depth as well. This cannot be immediately concluded from the existing work (for instance, see~\cite{arora2023quantum}).
    \end{problem}

    \begin{problem}
    \label{strEditDist1}
        Is showing a non-trivial lower bound on the quantum computational complexity of computing edit distance possible (See~\cite{boroujeni2021approximating} for more details on the existing results)?
    \end{problem}

    \begin{problem}
    \label{p:QComplexity:11}
        Given a pattern and a text, consider the pattern occurrences in the text with at most $k$ Hamming mismatches. A lower bound of $\Omega(\sqrt{kn})$ is already proved on the quantum query. But, is there a way to improve the quantum query complexity so that it gets closer to the above lower bound? A query complexity of $k^{3/4} n^{1/2 + o(1)}$ already exisits.
    \end{problem}

    \begin{problem}
        A quantum oracle has been developed to generate pseudorandom states, as reported in the literature \cite{https://doi.org/10.4230/lipics.tqc.2021.2}. A pertinent problem in this context is the unitary synthesis problem: whether it is possible to implement every $n$-qubit unitary transformation in $\text{poly}(n)$ time, relative to some classical oracle, remains a topic of active research and discussion.
    \end{problem}

    \begin{problem}
    \label{p:QComplexity:12}
        A quantum oracle already exists, relative to which $BQP = QMA$ \cite{https://doi.org/10.4230/lipics.tqc.2021.2}. Is it possible to construct a classical oracle separation between QCMA and QMA? 
    \end{problem}

    \begin{problem}
        The existence of Hamiltonians with low energy states that exhibit non-trivial complexity, as determined by the quantum circuit depth required to prepare them, is well-established. However, the question of whether the proof techniques employed in the latest research (as outlined in~\cite{anshu2023nlts}) can be extended to yield non-trivial lower bounds for non-commuting Hamiltonians remains open.
    \end{problem}

\subsection{Classification}
\label{subsec: classification}
    \begin{problem}
    \label{de2022}
        Quantum gates on $n$ qubits can encode non-directed graphs with $n$ nodes. An algorithm can handle an exponential number of graphs within polynomial time and provides insight into invariants on graph isomorphism~\cite{de2022quantum}. However, it is unclear under what circumstances non-isomorphic graphs can share the same classification table of subgraphs by the number of edges. Moreover, the algorithm's efficacy in classifying graphs that are considered hard to classify by classical algorithms, such as strong-regular graphs, remains unknown.
    \end{problem}
        
    \begin{problem}
        Mycielski transformation is a graph theory method that also applies to quantum graphs. It affects clique and chromatic numbers. Motzkin-Straus clique number finds the largest possible clique using eigenvalues of the adjacency matrix. The cones that correspond to the clique numbers for quantum graphs can be characterized. An investigation is needed to determine which Motzkin-Straus clique numbers the Mycielski transformation preserves.
    \end{problem}

    \begin{problem}  
        Consider the quantum version of the Stiebitz theorem as follows: Given a classical graph $G$, $n \geq 1$ and $r_j \geq 1$ for $j = 1, \cdots, n$ then let: 
        $$\mu_{\{r_1, \cdots r_n\}}(G) \coloneqq \mu_{r_n - 1}(\cdots \mu_{r_2 - 1}(\mu_{r_1 - 1}(G)) \cdots )$$
        $\mu(G)$ is the resulting graph of applying Mycielski transformation on a given graph $G$. For $n = 0$, the $\{r_1, \cdots , r_n\}$ is identified with $\empty$ and $\mu_{\empty}(G) = G$. For $k \geq 2$ denote 
        $$\mathcal{M}_k = \{ \mu_{\{r_1, \cdots r_{k-2}\}}(K_2) \ | \ r_j \geq 1, \ j=1, \cdots, k-2 \}$$
        that is the set of all generalized Mycielski transformations of $K_2$ obtained from $k-2$ consecutive applications of $\mu_{r-1}(\cdot)$ possibly with different $r$s in every iteration. Let $\chi_{\text{loc}(\mathcal{G})}$ be: 
        $$\chi_{\text{loc}(\mathcal{G})} = \min \{ c \in \mathbb{N} \ | \ \exists \text{c-coloring with dim}(\mathcal{N}) = 1 \}$$
        s.t. $\mathcal{N}$ is a finite von Neumann algebra. For every $G \in \mathcal{M}_k$, it holds that $\chi_{\text{loc}(\mathcal{G})} \geq k$. Let $\mathcal{K}_n$ be the quantum complete graph on $\text{Mat}_n$ equipped with the tracial $\delta\text{-form} \ \psi_n$. Define: 
        $$\mathbb{M}_k = \{ \mu_{\{ r_1 \cdots r_{k-2} \} }(\mathcal{K}_2) \ | \ r_j \geq 1, j = 1, \cdots k-2 \}$$
        For which type of (quantum) chromatic numbers do we have $\chi_{q}(\mathcal{G}) \geq k$ for all $\mathcal{G} \in \mathbb{M}_k$?
    \end{problem}

    \begin{problem}
        Denote $\mathcal{G}(n, p)$ a family of graphs on $n$ vertices whose edges are drawn independently w.p. $p$. What is the likely quantum chromatic number of $G \in \mathcal{G}(n, p)$? Consider the rank-r versions of the quantum chromatic number of a graph~\cite{cameron2006quantum}. For random graphs, it almost certainly holds that $\chi(G) = \chi_q(G)$ s.t. $\chi_q(G)$ is the quantum chromatic number of $G$. It is already known that $\chi(G) \sim \frac{n \log \frac{1}{1-p}}{2 \log n}$ w.h.p. Is it then possible to show that for all $\varepsilon > 0$, $$\chi^{(1)}_{q}(G) \geq (1-\varepsilon)\frac{n \log \frac{1}{1-p}}{2 \log n}?$$
    \end{problem}
        
    \begin{problem}
        Is there a way to find the smallest graph (and the smallest number of colours) exhibiting a separation between classical and quantum chromatic numbers? 
    \end{problem}

    \begin{problem}
         The clustering property of approximate code words first considers constants $c_1, c_2, \delta_0$ exist s.t. $0 \leq \delta < \delta_0$ and every vector $y \in \{0, 1\}^n$. Then, it holds that a $[[n, k, d]]$ CSS code defined by classical codes $(C_x, C_z)$ clusters approximate code words~\cite{anshu2023nlts}. Denote $G^{\delta}_{z}$ as the set of vectors that violate at most a $\delta$-fraction of checks from $C_z$. It holds that:
         \begin{itemize}
             \item If $y \in G_{z}^{\delta}$, then either $|y|_{C_{x}^{\perp}} \leq c_1 \delta n$ or else $|y|_{C_{x}^{\perp}} \geq c_2 n$.
             \item If $y \in G_{x}^{\delta}$ then either $|y|_{C_{z}^{\perp}} \leq c_1 \delta n$ or else $|y|_{C_{z}^{\perp}} \geq c_2 n$.
         \end{itemize} Does the above property hold for every constant-rate and linear-distance quantum code?
    \end{problem}

\subsection{Security}
\label{subsec: security}
    \begin{problem}
    \label{p:QSecurity:5}
        There already exists a quantum oracle that generates pseudorandom states relatively~\cite{https://doi.org/10.4230/lipics.tqc.2021.2}. Still, can the security of such an oracle against adversaries with the power of QMA be guaranteed? 
    \end{problem}

    \begin{problem}
    \label{p:QSecurity:6}
        The algorithm suggested in~\cite{kimmel2021query} solves the maximum matching problem. Is this algorithm optimal? Also, can the lower bound be improved using the general adversary bound?
    \end{problem}

\subsection{Others}

    \begin{problem}

        Given $k$ lists $L_1, \cdots ,L_k$, each consisting of uniformly random $n$-bit strings of size $2^\frac{n}{k}$, the $k\text{-}xor$ problem with lists aims to find a $k$-tuple $x_1, \cdots , x_k \in L_1 \cross \cdots \cross L_k$ such that $x_1 \oplus \cdots \oplus x_k = 0$, if such a tuple exists.

        In the Quantum-Accessible Quantum Memory (QAQM) model of quantum computation, the quantum computation can use as many qubits as needed. The question is whether there are better algorithms to solve the $k\text{-}xor$ problem without QAQM or even in the low-qubits model.
    \end{problem}     
    \begin{problem}
    \label{prob:problemname7}
        There already exists a $(2 + \varepsilon)\text{-approximate}$ polynomial compression for the Connected $\mathcal{F}\text{-Deletion}$ problem when $\mathcal{F}$ contains at least one planar graph~\cite{ramanujan2021approximate}.
           
        Is there a $(1 + \varepsilon)\text{-approximate}$ kernel for this problem (for every $0 < \varepsilon < 1$)? What is the best approximation achievable for this problem in polynomial time?
    \end{problem}
            
    \begin{problem}
        \label{prob:problemname8}
        Is it possible to find a fixed-parameter algorithm for $\text{Connected Planer} \\ \mathcal{F}\text{-Deletion}$ with a single-exponential dependence on $k$ s.t. $k \in \mathbb{N}$? The existing work suggests a factor-$(2+\varepsilon)$ parameterized approximation algorithm for the problem, with the running time of $2^{\mathcal{O}(k \log k)} n^{\mathcal{O}(1)}$ s.t. $0 < \varepsilon < 1$.
    \end{problem}
  
    \begin{problem}

        The problem of maximum cardinality matching can be optimized using 2000Q, a QA hardware from D-Wave. A better version of the QUBO formulation was used for this purpose. As part of their energy gap scale investigations, graphs with up to $22$ edges were fabricated. It is worth exploring whether this analysis can be extended to a $G_2$ graph. Doing so would provide direct insights into the original Graph family and enable verification of its scaling behaviour with the maximum match problem for the $G_n$ graph family.
        
    \end{problem}
    
    \begin{problem}
    \label{strEditDist2}
        An existing quantum algorithm approximates the edit distance within a constant factor of seven~\cite{boroujeni2021approximating}. But is there a sub-quadric algorithm with a smaller constant approximation factor?
    \end{problem} 
    
    \begin{problem}
    \label{strEditDist3}
        The edit distance can be approximated within a constant factor of seven~\cite{boroujeni2021approximating}. Does a quantum algorithm exist that approximates the edit distance of two strings within a constant factor in near-linear time?
    \end{problem}
        
    \begin{problem}
        The Measurement-Based Quantum Computation (MBQC) model is a relevant approach to consider in the field of quantum computation. However, a circuit in the \emph{usual} circuit model transformed to the MBQC model may result in a considerable increase in the circuit width. This is because while the usual circuit model allows for arbitrary wire gate application, the MBQC model only permits interactions between adjacent wires. Furthermore, the use of long-range gates is costly. In light of this, one may inquire whether a protocol can be suggested for both models. Alternatively, one may focus on developing a protocol specifically tailored to the usual circuit model.
    \end{problem}

    \begin{problem}
        The work in~\cite{anshu2023nlts} posits that quantum local testability is not a necessary condition for the construction of the work. The property, as established in the aforementioned work (Property 1), suffices for clustering the classical distributions of low-energy states. However, it is important to note that the property in question is weaker than local testability. Extant literature establishes that local testability implies NLTS. It is worthwhile to investigate the implications codes with Property 1 have for the quantum PCP conjecture~\cite{aharonov2013guest}.
    \end{problem}

    \begin{problem}
        The study in~\cite{anshu2023nlts} has successfully established a property referred to as Property 1. Upon analyzing from the perspective of chain complexes, Property 1 seems to exhibit a relationship with the small-set boundary and co-boundary expansion as specified under Definition 1.2 of~\cite{hopkins2022explicit}. This relationship may introduce a classical analog of the NLTS property. Furthermore,~\cite{hopkins2022explicit} presents a construction technique for classical Hamiltonians that are hard to approximate. The construction technique leverages the sum-of-squares hierarchy.
    \end{problem}

    \begin{problem}
    \label{mariella}
        Consider analyzing the subgraph isomorphism problems on a quantum gate-based computer~\cite{mariella2023quantum}. Is it possible to generalize the problem (e.g., digraph isomorphism, commonly induced subgraphs, labelled edges and nodes), improve the compilation of the resulting circuit, improve classical solvers, and a comparative study of different Ansatze?
    \end{problem}

    \begin{problem}
    \label{p:Error:5}
        Consider \emph{basketing}, a post-selection technique~\cite{harris2023scalable}. This allows for the addressing of time-dependent noise behaviours and reduces overhead. Is exploring another case where the client suffers from state-preparation noise possible? In this case, one may consider more general classes of computation.
    \end{problem}

    \begin{problem}
    \label{prob:problemname2}
        An extant approach has been developed to tackle the maximum independent set (MIS) problem using quantum wires. Specifically, this method offers a solution for non-planar graphs and high-dimensional ones. Quantum wires, a set of auxiliary wire atoms, have been proposed to facilitate strong interactions between distant atoms~\cite{kim2022rydberg}, thus allowing the synthesis of an entire target graph, $G_T$, from a simple first graph, $G_0$. However, the applicability of quantum wires in other optimization problems beyond MIS remains uncertain.
        
   \end{problem}

   \begin{problem}
        There already exists a quantum oracle $\mathcal{O}$ s.t. $\text{BQP}^\mathcal{O} = \text{QMA}^\mathcal{O}$ and pseudorandom unitary transformations (PRUs) (and hence pseudorandom states (PRSs)) exist relative to $\mathcal{O}$. Do either PRUs or PRSs exist relative to $\mathcal{O}$ for a classic oracle? 
    \end{problem}

    \begin{problem}
        The best reduction of the subproblem count is achieved by the limits of the Lov\'asz number, but also at the highest computational costs~\cite{kerger2023asymptotically}. Additionally, the impracticality of applying the adapted implementation on large graphs was not ideal. Is there a faster way(s) of calculating the Lov\'asz number? If yes, the above limits would have been usable.
    \end{problem}

     \begin{problem}
     \label{pf2}
        Is it possible to develop methods for quantum simulation that can be used for traditional models of quantum computation alternately~\cite{kassal2011simulating}?
    \end{problem}

    \begin{problem}
    \label{p:Simulattion:1}
        Most of the dedicated quantum simulators built so far are designed to simulate condensed matter systems~\cite{kassal2011simulating}. Can experimental progress toward simulating chemical systems be developed?
    \end{problem}

    \begin{problem}
    \label{p:QAlgorithm:11}

        The minimum and maximum weight matching problems can be solved in $\mathcal{O}(n(m + n\log n))$ time. There exists an algorithm that finds a maximum weight augmenting path in graph $G$ in $\mathcal{O}(m + n\log n)$ time. The maximum weight matching can be found by enlarging the matching with the path in $G$ in $n$ steps. It is unclear if a quantum algorithm exists for the minimum and maximum weight matching, but if there is, it could improve the running time of the classical algorithm~\cite{dorn2009quantum}.
    \end{problem}
    
    \begin{problem}
    \label{p:QSupremacy:3}
        Is it possible to add iterative parameter tuning and new problem classes, specifically the challenging and low-precision Maximum Independent Set (MIS) problem, to the existing method of generating arbitrary structured problems for D-Wave quantum processing units (QPUs)~\cite{8477865}? This can help compare QPU performance with classical algorithms and identify potential use cases for quantum annealing in real-world applications.
        
   \end{problem}



\section{Papers in Summary}
\label{sec:summary}
Researchers have been exploring innovative ways to harness the power of quantum computing to tackle complex problems. One notable effort is the integration of Quantum Annealing (QA) with Conflict-Driven Clause Learning (CDCL) in solving Boolean satisfiability problems (SAT), as showcased by the HyQSAT approach~\cite{10071022}. By strategically ordering embedded clauses and leveraging quantum parallelism alongside classical algorithms, HyQSAT achieved a remarkable 12.62X quantum speedup on DWave 2000s.

However, the potential of quantum algorithms is not without challenges. The Ohya-Masuda Quantum algorithm \cite{ohya1999quantum}, while theoretically capable of solving 3-SAT problems efficiently, encounters obstacles due to the practical difficulty of executing a crucial non-unitary transformation. Inaccuracies stemming from imperfect measurements undermine the accuracy of results, revealing limitations in realizing this algorithm's promise.

Quantum optimization also enters the realm of combinatorial problems, such as the Maximum Independent Set (MIS) problem. Pichler et al.~\cite{pichler2018quantum} proposed a novel approach utilizing Rydberg atom arrays. By encoding MIS solutions in the ground states of interacting atoms through the Rydberg blockade mechanism, they established a link between 3-SAT instances and MIS problems, showcasing the potential of quantum computing in tackling complex combinatorial challenges.

Building upon this foundation, Ebadi et al.~\cite{ebadi2022quantum} delved into quantum optimization for the Maximum Independent Set problem using programmable arrays of neutral atoms trapped in optical tweezers. Exploring a range of system sizes and coherence properties, their study sheds light on the practical considerations and potential benefits of employing quantum techniques to enhance optimization processes.

In a separate avenue of exploration, researchers examined the relationship between quantum and classical complexity in translationally invariant tiling and Hamiltonian problems. By revealing the quantum and classical complexity bounds for independent set problems, they contribute to our understanding of the power and limitations of quantum algorithms in this context.

While quantum computing holds the promise of exponential speedups for certain problems, its intricacies and challenges are still being unveiled. As researchers navigate the intricate interplay between quantum and classical worlds, they uncover new insights into the potential and boundaries of quantum algorithms in various computational domains.

Especially in recent years, quantum computing has emerged as a promising avenue for addressing complex computational challenges. A series of notable studies have explored the application of quantum algorithms to a variety of combinatorial optimization problems, ranging from graph theory~\cite{de2022quantum, mariella2023quantum, siraichi2019qubit,li2020qubit,bravyi2022hybrid,bochniak2023quantum,salehi2022unconstrained,Ge_2020,cocchi2021graph,pichler2018quantum,kim2022rydberg,doern2007quantum,8477865,pelofske2019solving,kerger2023asymptotically,pelofske2019solving,ramanujan2021approximate,kimmel2021query,witter2022query,dorn2009quantum,gabow1990data,mcleod2022benchmarking,vert2020revisiting,cameron2006quantum,bravyi2022hybrid,bravyi2020obstacles} to chemistry simulations~\cite{kassal2011simulating}.

One intriguing area of exploration involves solving the Maximum Independent Set (MIS) problem using quantum annealing~\cite{8477865}. By mapping MIS to Quadratic Unconstrained Binary Optimization (QUBO) problems, researchers have harnessed the power of commercial quantum processing units (QPUs). These efforts have yielded significant insights into the potential of quantum annealing for handling intricate combinatorial optimization tasks.

In the realm of satisfiability problems, a novel quantum cooperative search algorithm has been introduced to tackle the 3-SAT problem~\cite{10071022}. This innovative approach combines Grover's search algorithm with classical heuristic techniques. The resulting algorithm demonstrates improved performance compared to conventional methods, showcasing the synergy between quantum and classical computing paradigms.
Advancements have also been made in subset-sum problems, where classical and quantum heuristic algorithms have been refined~\cite{bonnetain2020improved,zheng2022quantum}. By leveraging extended representations and quantum search strategies, researchers achieved quantum speedups, contributing to the growing body of evidence supporting the quantum advantage in certain problem domains.

The formidable Traveling Salesman Problem has been targeted with quantum backtracking techniques~\cite{moylett2017quantum}, leading to a quadratic quantum speedup for bounded-degree graphs. This breakthrough suggests that quantum computing could hold the key to more efficient solutions for this notoriously challenging optimization problem.
Furthermore, adiabatic quantum computing has been harnessed to address fundamental challenges in graph theory. The Hamiltonian cycle problem, a classic NP-complete problem, was approached through QUBO formulations suitable for adiabatic quantum computation. These efforts shed light on the feasibility of quantum solutions for graph-related problems.

In the realm of molecular simulations, quantum computing offers promising avenues for simulating chemical systems. Researchers have proposed algorithms for simulating complex biomolecular solutions, leveraging the inherent properties of quantum systems to provide insights into challenging chemical problems. This paves the way for advancements in fields such as drug discovery and material science.

These studies collectively underscore the growing influence of quantum computing on various domains of computational research. From tackling graph-related conundrums to simulating intricate molecular interactions, quantum algorithms are showing great promise in revolutionizing how we approach complex computational problems. As the field continues to evolve, these innovative approaches may reshape our understanding of optimization and simulation in the quantum era.

Quantum complexity and computing involve tricky math called \emph{unitary transformations} for changing quantum systems over time. Imagine the simplest way to do quantum stuff – that's \emph{quantum complexity}. People thought this would grow smoothly with time, but it is complicated. Brandao et al.~\cite{brandao2021models} added to this by showing how quantum complexity grows in different models, using a trick with random stuff called \emph{unitary designs}.

For real-world problems, we want quantum computers to solve puzzles, like putting puzzle pieces into a puzzle. Hamilton and Humble~\cite{hamilton2017identifying} found a better way to do this called \emph{minor set cover}, making it easier. Solving subset sum problems is another puzzle. Bernstein et al.~\cite{bernstein2013quantum} made a quantum trick to solve it faster. Helm and May~\cite{helm2018subset} improved this trick even more.
We can also use light to solve puzzles. Xu et al.~\cite{xu2020scalable} built a machine with light that is good at solving puzzles. It is even faster than supercomputers! Other folks made smart ways to solve puzzles with less computer memory~\cite{delaplace2019improved}.
A big challenge is 3-SAT puzzles. Some smart people figured out how to solve these puzzles using quantum computers~\cite{bravyi2011efficient}. Other folks showed that solving 3-SAT puzzles with quantum stuff is hard~\cite{gosset2016quantum}.

So, people are trying different tricks with quantum computers to solve hard puzzles and make things work better. It sounds like they are finding new ways to solve puzzles using magic math.

Researchers are exploring how quantum computers can solve complex problems. They convert 3-SAT problems into a form that quantum computers can handle better. An improved method called ab-QAOA~\cite{yu2023solution} helps solve hard 3-SAT and Max-3-SAT problems using fewer steps. Quantum annealers, like D-Wave, are tested to solve SAT and MaxSAT problems. A bio-molecular~\cite{chang2021quantum} algorithm is proposed to solve the independent set problem efficiently. QAOA's behaviour on graphs is studied, revealing its performance patterns. A quantum algorithm is developed to speed up solving the Traveling Salesman Problem (TSP). The Grover algorithm is used to tackle the Hamiltonian Cycle Problem (HCP)~\cite{jiang2022quantum}.

Researchers create better formulations for k-SAT and HCP problems, improving scaling. Quantum algorithms for graph problems are analyzed, including Eulerian and Hamiltonian tours. Quantum annealers are applied to graph partitioning and TSP with time windows. An adiabatic quantum algorithm is proposed for TSP. Concepts of graph-induced operators and lattice gauge theory are explored for solving HCP. NISQ circuit compilation is linked to the TSP on a torus. In simpler terms, researchers are finding ways to use quantum computers to solve tough problems in more efficient ways. They're developing new methods and algorithms for various problems, like SAT, TSP, and Hamiltonian cycles, using the power of quantum mechanics.

Researchers have been exploring quantum computing techniques to solve challenging optimization problems, such as the Multidimensional Knapsack Problem. They combined Quantum Particle Swarm Optimization with local search methods, yielding competitive outcomes. Another approach used the Quantum Approximate Optimization Algorithm (QAOA) to enhance solutions for constrained optimization problems, outperforming classical methods. The integration of the k-nearest neighbour technique and quantum cuckoo search led to better results in resource allocation and knapsack problems. A Diversity-Preserving Quantum Particle Swarm Optimization strategy~\cite{lai2020diversity} managed populations efficiently for improved solutions.

Quantum techniques were applied to multi-knapsack problems, displaying superiority over classical methods. Qubit mapping was refined using subgraph isomorphism, reducing extra steps in quantum circuits. An innovative GR3 Algorithm demonstrated fast parallel subgraph isomorphism searching. Moreover, researchers devised an efficient method for subgraph isomorphism in quantum circuits. In the field of biology, an adiabatic quantum algorithm contributed to DNA motif model discovery.

The Quantum Alternating Operator Ansatz~\cite{cook2020quantum} offered approximate solutions for combinatorial problems, while the Quantum Approximate Optimization Algorithm tackled minimum vertex cover problems~\cite{pelofske2019solving,zhang2022applying,wang2023quantum,marsh2019quantum,wang2022classically}. The Grover search algorithm achieved quantum speedup for the vertex cover problem. A Quantum-Walk assisted algorithm targeted bounded NP optimization problems. Notably, quantum algorithms were designed for matching problems, proving their effectiveness in various scenarios~\cite{jeffery2022multidimensional,bonnetain2020improved,mahasinghe2019solving,kempe2003quantum,childs2009universal,lovett2010universal,apers2019unified,marsh2019quantum,belovs2012learning}.

Researchers have been exploring the capabilities of quantum computing in solving complex optimization problems. McLeod and Sasdelli~\cite{mcleod2022benchmarking} benchmarked D-Wave's quantum annealers, Advantage, and 2000Q, using the maximum cardinality matching problem. They found that Advantage outperformed 2000Q for larger problems and analyzed the scaling of diabatic transition probability. Fowler addressed the minimum weight perfect matching problem on a modular qubit array, proving efficient solutions even in the presence of errors.

Witter~\cite{witter2022query} developed a query-efficient quantum algorithm for maximum matching, demonstrating improved quantum upper bounds. Basso et al.~\cite{basso2021quantum} evaluated Quantum Approximate Optimization Algorithm (QAOA) performance on large-girth regular graphs, showing potential quantum speedup. Guerreschi and Matsuura~\cite{guerreschi2019qaoa} simulated QAOA and suggested that practical quantum speedup may require hundreds of qubits. Crooks~\cite{crooks2018performance} optimized QAOA circuits for the MaxCut problem and highlighted its performance relative to classical algorithms.

Boulebnane and Montanaro~\cite{boulebnane2021predicting} analyzed QAOA's performance in the infinite-size limit for MAX-CUT, while Venere et al. formulated and characterized Set Packing on quantum annealers. Ajagekar et al.~\cite{ajagekar2022hybrid} proposed hybrid classical-quantum optimization techniques for solving mixed-integer programming problems in production scheduling, and Chang et al.~\cite{chang2014quantum} introduced a new approach for mixed-integer programming using quantum processors. Bernal et al. presented integer programming techniques for minor embedding in quantum annealers.

Cameron et al.~\cite{cameron2006quantum} studied the quantum chromatic number of graphs, establishing relationships between the clique number and orthogonal representations. They also demonstrated a separation between classical and quantum chromatic numbers for specific cases. These studies collectively contribute to our understanding of quantum computing's potential and limitations in addressing diverse optimization challenges.

Researchers have been exploring quantum algorithms for solving complex optimization and graph-related problems. One approach, the Recursive Quantum Approximate Optimization Algorithm (RQAOA), was introduced by Bravyi et al.~\cite{bravyi2022hybrid} It focuses on approximate vertex k-coloring, comparing its performance with classical and hybrid methods on specific graph types. Bochniak and Pawe{\l}~\cite{bochniak2023quantum} extended classical Mycielski transformations to quantum graphs, studying their effects on chromatic and clique numbers.

Younes~\cite{younes2015bounded} developed a bounded-error quantum algorithm for max-bisection and min-bisection problems, using iterative partial negation and measurement. Wang proposed the Classically-Boosted Quantum Optimization Algorithm (CBQOA), combining classical preprocessing and quantum search to outperform previous approaches for optimization.

Saleem et al.~\cite{saleem2021quantum} introduced a quantum divide and conquer algorithm for distributed computing, enabling optimization on subgraphs using quantum circuits. Le et al. provided subquadratic kernels for connected minor-hitting sets, addressing NP-complete packing/covering problems on graphs.

Chang et al.~\cite{chang2014quantum} investigated bio-molecular solutions' role as oracles in Grover's algorithm for the hitting-set problem, with experimental validation. Ramanujan introduced an approximate compression technique for the Connected $\mathcal{F}-\text{Deletion}$ problem, achieving a 

$(2 + \varepsilon)$-approximation.

Miyamoto et al.~\cite{miyamoto2020quantum} combined quantum search and classical dynamic programming for the minimum Steiner tree problem, achieving subexponential complexity. Kerger et al.~\cite{kerger2023asymptotically} introduced faster quantum algorithms for distributed computing, showing improvements over classical models in specific cases.

Childs et al.~\cite{childs2000finding} explored a quantum adiabatic evolution approach for finding cliques in random graphs, revealing quadratic runtime for small graphs. Pelofske et al.~\cite{pelofske2019solving} investigated solving large Maximum Clique problems on a quantum annealer through decomposition and pruning.
Harris and Kashefi~\cite{harris2023scalable} introduced an error mitigation protocol for BQP-type computations on a quantum computer with time-dependent noise, using verification techniques. Lu et al.~\cite{lu2023approximation} approximated the nearest classical-classical state to a quantum state, proposing a gradient-driven descent flow on Stiefel manifolds to quantify quantumness.

These studies collectively enhance our understanding of quantum algorithms, optimization, and graph-related problems, highlighting quantum computing's potential and limitations in diverse scenarios.

In recent years, quantum computing has garnered significant attention for its potential to revolutionize various fields. Several groundbreaking studies have emerged, shedding light on key aspects of quantum algorithms, complexity theory, and practical applications.

Anshu et al.~\cite{anshu2023nlts} made a remarkable contribution by proving the No Low-Energy Trivial State (NLTS) conjecture. They demonstrated that specific constant-rate and linear-distance Quantum LDPC error-correcting codes correspond to NLTS local Hamiltonians. This finding suggests a deep connection between quantum error correction and complex low-energy states.

Arora et al.~\cite{arora2023quantum} delved into the computational power of shallow quantum circuits within the Random Oracle Model. Their work highlighted intriguing separations between classical and shallow quantum computation classes, providing insights into the potential of adaptive measurements in quantum circuits.

Hhan et al.~\cite{hhan2023quantum} explored the quantum complexity of discrete logarithms and related group-theoretic problems. They introduced a quantum generic group model, presenting lower bounds that contribute to our understanding of the efficiency of quantum algorithms for these problems.
Aaronson's work~\cite{aaronson2018pdqp} bridged the gap between classical and quantum computation by showing that combining quantum advice and non-collapsing measurements allows a quantum computer to solve decision problems in polynomial time, even though each enhancement alone lacks significant computational advantage.

Aaronson and Hung~\cite{aaronson2023certified} introduced certified randomness from quantum supremacy experiments. They proposed a protocol to generate cryptographically certified random bits using quantum circuit sampling, enhancing cryptographic security and applications like proof-of-stake cryptocurrencies.

In the realm of quantum walks, Jeffery and Zur~\cite{jeffery2022multidimensional} improved upper bounds on the time complexity of $k$-distinctness problems, offering insights into the quantum speedup, achieved in these scenarios.

Leverrier and Ze\'mor~\cite{leverrier2022quantum} focused on efficient decoding for quantum Tanner codes, enhancing our understanding of error correction mechanisms in quantum systems.

Cornelissen and Hamoudi~\cite{cornelissen2023sublinear} leveraged the power of quantum algorithms for mean estimation, demonstrating substantial improvements in estimating the mean of vector-valued random variables, with implications for various applications.

Harris and Kashefi~\cite{harris2023scalable} introduced an error mitigation protocol. The protocol is for running BQP-type computations on a quantum computer s.t. the noises are time-dependent. The authors hired verification techniques by Leichtle et al.~\cite{leichtle2021verifying}. This allowed their results to interleave standard and test computation rounds. One should notice that the above technique provides a local correctness guarantee. In particular, it bounds the probability of returning a correct classical output exponentially. The authors further introduced a post-selection technique they named \emph{basketing}. It could address time-dependent noise behaviours and also allow overhead reduction.

Overall, these recent studies contribute significantly to our understanding of quantum algorithms, complexity theory, and their potential applications, bringing us closer to harnessing the power of quantum computing for practical use cases.

\section{Conclusion}
In conclusion, the revelations of quantum mechanics have illuminated a distinct realm, governed by its principles. Quantum computing stands out as a promising avenue to transcend the limitations of classical computing, unlocking solutions to previously insurmountable problems. However, challenges such as uncontrolled errors pose obstacles to its full potential.

This survey has delved into the realm of quantum computing, spotlighting its attempts to tackle computationally hard problems in classical computation. While classical computers rely on bits and analog systems guided by boolean logic gates, quantum computation employs qubits. These quantum systems, capable of existing in multiple states simultaneously, grant quantum computers a unique advantage in processing information in parallel and swiftly arriving at solutions.

The transformative power of quantum computing is evident in its ability to expedite computations deemed challenging for classical computers. Shor's algorithm and Grover's algorithm, for instance, have demonstrated remarkable capabilities in factorizing large integers and searching unstructured datasets, respectively. This computational prowess holds the potential to revolutionize fields such as cryptography, materials science, energy technologies, and quantum chemistry.

The complexity of quantum systems positions them as ideal for processing optimization problems and simulating intricate quantum systems. These simulations, often impractical for classical computers, offer insights into fundamental physics, chemistry, and biology. The mastery of these subjects can drive innovations with far-reaching implications, from the development of new materials to advancements in medicine.

The anticipated supremacy of quantum computation is poised to reshape various industries, including finance, healthcare, logistics, drug discovery, and artificial intelligence. The computational power of quantum systems has already impacted cryptography, necessitating the exploration of quantum-resistant cryptographic methods to ensure the future security of digital communication.

In summary, the journey into the quantum world holds the promise of unlocking unparalleled computational capabilities, paving the way for groundbreaking advancements across scientific, technological, and industrial domains.


Various research groups have undertaken endeavours to tackle classical computational problems using quantum devices. In our exploration, we have compiled papers that delve into this approach and pinpoint key open problems demanding attention.

Nevertheless, these initiatives have encountered substantial challenges. Grasping the limitations inherent in current quantum computing methods is crucial for surmounting these obstacles. Below, we outline ten of the most formidable challenges confronting researchers in this field:

    \begin{enumerate}
        \item \textbf{Quantum Supremacy:}
        The question of whether quantum computers can surpass classical computers (\emph{quantum supremacy}) is a common one, and is explored in problems~\ref{p:QSupremacy:1},~\ref{p:QSupremacy:2}, and~\ref{p:QSupremacy:3}. The concept of quantum supremacy is multifaceted and raises several questions: What precisely is quantum supremacy? Is it restricted to certain types of problems or is it all-encompassing? How can we demonstrate that quantum computers can solve problems that are practically infeasible tasks for classical computers? 

        To demonstrate that quantum computers can solve problems that are practically impossible for classical computers to handle, one must establish a method of proof. This can be accomplished through the use of formal proof or through empirical evidence gathered from experimental data. As quantum computers continue to develop and improve, the demonstration of quantum supremacy will likely become increasingly feasible.

        \item \textbf{Quantum Error Correction:}
        The management and control of qubits presents a formidable challenge in the context of quantum computing, given the system's susceptibility to noise and errors during computation. The detection and correction of these errors is an essential task to enhance the reliability and scalability of such computing systems. Collectively, problems~\ref{p:Error:1},~\ref{p:Error:2},~\ref{p:Error:3},~\ref{p:Error:4}, and ~\ref{p:Error:5} represent exemplars of the challenges that arise in this arena, and underscore the importance of developing effective strategies for addressing them.

        \item \textbf{Quantum Algorithms:}
        To employ quantum hardware instead of classical alternatives, it is necessary to develop algorithms that are congruent with these devices. To fully harness the supremacy of quantum computing, it is imperative that we design algorithms that are more efficient than their classical counterparts. Examples of such algorithms are presented in problems: ~\ref{p:QAlgorithm:1},~\ref{p:QAlgorithm:3},~\ref{p:QAlgorithm:4},~\ref{p:QAlgorithm:5},~\ref{p:QAlgorithm:6},~\ref{p:QAlgorithm:7},~\ref{p:QAlgorithm:9},~\ref{p:QAlgorithm:10}, and~\ref{p:QAlgorithm:11}.

        \item \textbf{Hybrid Quantum-Classical Computing:} 
        The integration of classical and quantum computers can be a potent approach to solving intricate problems. This is typically achieved by allocating specific tasks to each type of computer. Two exemplary research papers that utilized hybrid computation in their work are Problems~\ref{p:Hybrid:1} and~\ref{p:Hybrid:2}. In light of the complex nature of contemporary problems, the amalgamation of classical and quantum computing has emerged as a promising avenue for addressing hitherto intractable computational challenges. The judicious allocation of tasks between classical and quantum computers can lead to significant improvements in computational efficiency and accuracy.

        \item \textbf{Quantum Simulation:}
        Quantum systems are characterized by their complexity and multi-dimensional nature, rendering their simulation on classical computers challenging. Nonetheless, the simulation of a quantum system utilizing another quantum system may offer a more efficient alternative. This avenue of approach could potentially unlock novel possibilities in the fields of materials science, as well as chemistry. Two illustrative examples of problems germane to this subject matter are problems~\ref{p:Simulation:1} and~\ref{p:Simulation:2}.

        \item \textbf{Quantum Cryptography:} 
        Quantum computation has the potential to significantly impact cryptography due to the unique properties of quantum mechanics, such as entanglement. However, the extent to which quantum computing can enhance security protocols, such as quantum key distribution and secure communication, remains unclear. Notwithstanding, numerous problems in quantum cryptography have been identified, as enumerated by problems~\ref{p:QSecurity:1},~\ref{p:QSecurity:2},~\ref{p:QSecurity:3},~\ref{p:QSecurity:4},~\ref{p:QSecurity:5}, and~\ref{p:QSecurity:6}.

        \item \textbf{Scalability:} 
        Quantum technologies have the potential to be revolutionary, but their scalability remains a challenge. Currently, controlling quantum systems and reducing noise is not fully feasible, resulting in a limited number of qubits in quantum computers. As problems~\ref{p:Scalability:1},~\ref{p:Scalability:2},~\ref{p:Scalability:3},~\ref{p:Hybrid:1}, and~\ref{p:Scalability:6} highlight, building larger and more stable quantum systems is essential for their advancement.

        \item \textbf{Quantum Software \& Programming Languages:}
        As previously stated, quantum computers operate using distinctive algorithms. Consequently, the integration of quantum computers into practical applications necessitates the development of user-friendly programming languages and tools. It is highly desirable to have a quantum processor capable of executing any program to obtain the solution. However, the development of such a precisely programmable processor appears to be an insurmountable challenge. This issue is discussed in depth in Problem~\ref{p:QSoftware:1}.

        \item \textbf{Quantum Annealing:} 
        Quantum annealing is a computational technique that aims to find the state with the minimum energy in a given system. Unlike gate-based computation, quantum annealing is distinct and cannot be readily compared with it. More details are discussed in Problem~\ref{p:QA:1}.

        \item \textbf{Quantum Complexity Theory:}
        Quantum computational complexity classes have a unique definition that differs from the classical one. The relationship between these classes and classical complexity classes is unknown. Problems~\ref{p:QComplexity:2} to~\ref{p:QComplexity:12} are related to this topic. Problems:~\ref{p:QComplexity:2},~\ref{p:QComplexity:3},~\ref{p:QComplexity:4},~\ref{p:QComplexity:5},~\ref{p:QComplexity:6},~\ref{p:QComplexity:7},~\ref{p:QComplexity:8},~\ref{p:QComplexity:9},~\ref{p:QComplexity:10},~\ref{p:QComplexity:11}, and~\ref{p:QComplexity:12} are about this subject.
    \end{enumerate}

    It is imperative to acknowledge that the unresolved problems in quantum computation not only serve to advance our understanding of the field but also play a pivotal role in unlocking its full potential. The intersection of quantum computing and classical computation presents a vast and uncharted research landscape. Researchers have the opportunity to generate innovative solutions and expand the scope of what is achievable. By addressing these open questions, we can pave the way for further advancements in the field of quantum computation, ultimately leading to a greater understanding of complex computational problems.

 \section{Declarations}
 
\subsection{Data and Materials}
It is a theoretical article, and we have not used any specific data material.

 \subsection{Competing Interests}
 There is no competing personal or financial interest with other people or organizations.

\subsection{Funding}
There is no source of funding for this research.
\bibliography{sample-base.bib}
\bibliographystyle{acm.bst}

\end{document}